\documentclass{article}
\usepackage{amsfonts}
\usepackage{amssymb}
\usepackage{amscd}
\newcommand{\C}{\mathbb C}
\newcommand{\bp}{{\mathbb P}}
\newcommand{\R}{\mathbb R}
\newcommand{\bt}{{\mathbb T}}
\newcommand{\Z}{{\mathbb Z}}
\newcommand{\la}{\langle}
\newcommand{\ra}{\rangle}
\newcommand{\ca}{{\mathcal A}}
\newcommand{\cb}{{\mathcal B}}
\newcommand{\cc}{{\mathcal C}}
\newcommand{\ce}{{\mathcal E}}
\newcommand{\cf}{{\mathcal F}}
\newcommand{\cg}{{\mathcal G}}
\newcommand{\ch}{{\mathcal H}}
\newcommand{\cj}{{\mathcal J}}
\newcommand{\cl}{{\mathcal L}}
\newcommand{\rp}{{\mathcal P}}
\newcommand{\cm}{{\mathcal M}}
\newcommand{\cn}{{\mathcal N}}
\newcommand{\co}{{\mathcal O}}
\newcommand{\cs}{{\mathcal S}}
\newcommand{\ct}{{\mathcal T}}
\newcommand{\cv}{{\mathcal V}}
\newcommand{\U}{{\mathcal {U}}}
\newcommand{\cw}{{\mathcal W}}
\newcommand{\cz}{{\mathcal Z}}
\newcommand{\cp}{\C{\bp}^1_x}
\newcommand{\cx}{\C{\bp}^1_{x_0}}
\newcommand{\ag}{{\mathop{\mbox{Aut}}\nolimits G}}
\newcommand{\im}{{\mathop{\mbox{Im}}\nolimits\,}}
\renewcommand{\ker}{{\mathop{\mbox{Ker}}\nolimits\,}}
\newcommand{\m}{{\mathop{\mbox{Map}}\nolimits}}
\newcommand{\di}{{\mathop{\mbox{Diff}}\nolimits\,}}

\newcommand{\fc}{{\mathfrak C}}
\newcommand{\frg}{{\mathfrak G}}
\newcommand{\fh}{{\mathfrak H}}
\newcommand{\fj}{{\mathfrak j}}
\newcommand{\fg}{{\mathfrak g}}
\newcommand{\fo}{{\mathfrak O}}
\newcommand{\fs}{{\mathfrak S}}
\newcommand{\fu}{{\mathfrak U}}
\newcommand{\fx}{{\mathfrak X}}
\newfont{\eu}{eusm10}
\newfont{\ef}{eufb10}
\newfont{\esb}{eusb10}
\newcommand{\bg}{{\esb G}}
\newcommand{\bm}{{\esb M}}
\newcommand{\bn}{{\esb N}}

\newcommand{\gc}{{\ef C}}
\newcommand{\gh}{{\ef H}}
\newcommand{\vv}{{\eu V}}
\newcommand{\ww}{{\eu W}}
\def\a{\alpha}
\def\b{\beta}
\def\g{\gamma}
\def\d{\delta}
\def\e{\eta}
\def\l{\lambda}
\def\o{\omega}
\def\p{\partial}
\def\ve{\varepsilon}
\def\vp{\varphi}
\def\r{\rho}
\def\s{\sigma}
\def\t{\triangle}
\def\up{\Upsilon}
\def\th{\Theta}
\def\bl{\backslash}
\def\hra{\hookrightarrow}
\def\lra{\longrightarrow}

\begin{document}
\begin{flushright}
 hep-th/9803183
\end{flushright}
\vskip 3cm

\begin{center}
{\LARGE \bf Self-Dual Yang-Mills: Symmetries and Moduli Space}
\vskip 1.5cm
{\Large A.D.Popov}\footnote {E-mail: popov@thsun1.jinr.dubna.su}\\
\vskip 1cm
{\em Bogoliubov Laboratory of Theoretical Physics\\
JINR, 141980 Dubna, Moscow Region, Russia}\\
\end{center}
\vskip 1.1cm
\begin{quote}
\noindent
{\bf Abstract}. Geometry of the solution space of the self-dual
Yang-Mills (SDYM) equations in Euclidean four-dimensional space
is studied. Combining the twistor and group-theoretic approaches,
we describe the full infinite-dimensional symmetry group of the SDYM
equations and its action on the space of local solutions to the field
equations. It is argued that owing to the relation to a holomorphic
analogue of the Chern-Simons theory,  the SDYM theory may be as solvable
as 2D rational conformal field theories, and successful nonperturbative
quantization may be developed. An algebra acting on the space of self-dual
conformal structures on a 4-space (an analogue of the Virasoro algebra)
and an algebra acting on the space of self-dual connections (an analogue of
affine Lie algebras) are described. Relations to problems of topological
and $N=2$ strings are briefly discussed.
\end{quote}

\vskip 0.9cm

\tableofcontents

\newpage
\section{Introduction}
\label{1}

In the past two decades significant progress in understanding
integrable, conformal and topological quantum field theories in two
dimensions has been achieved. In many respects this progress was
related to the existence of an infinite number of symmetries making
it possible not only to describe the space of classical solutions to
2D field equations, but also to advance essentially the
nonperturbative quantization of 2D theories. Among symmetry algebras
of 2D models, the most important role is played by the Virasoro and
affine Lie algebras (see e.g.~\cite{kyoto}-\cite{New}). The use of
transformation groups, their orbits and representations forms the
basis of the dressing transformation method~\cite{ZS}-\cite{FT} and
of the Kyoto's school approach~\cite{Sato,kyoto,UT} to solvable
equations of 2D and 3D field theories.

\smallskip

In four dimensions there also exist integrable, conformal and
topological field theories, and naturally the following question
arises:  Can the methods and results of 2D theories be transferred to
4D theories?  On the whole, the answer is positive for 4D integrable
and topological field theories. At the same time, the knowledge of
4D conformal field theories (CFT) beyond the trivial case
of free field theories is much less explicit, and not so many exact
results are obtained (see e.g.~\cite{HPS}-\cite{HW} and references
therein).  Usually, one connects this with the fact that, unlike the
2D case, the conformal group in four dimensions is finite-dimensional,
and constraints arising from conformal invariance
are not sufficient for a detailed description of 4D CFT's. One of our
aims is to demonstrate that for a special subclass of CFT's --
integrable 4D CFT's -- this is a wrong impression based on the
consideration of only local (manifest) symmetries.

\smallskip

There actually exists only one nonlinear integrable model in 4D
described by the self-dual Yang-Mills (SDYM) equations defined on a
4-manifold with the self-dual Weyl tensor~\cite{Pen}-\cite{AHS}. This
unique theory is conformally invariant, and it is usually considered
as a 4D analogue of the 2D WZNW theory. It is expected that many
results of 2D rational CFT's can be extended to the SDYM theory. This
was discussed, for instance, in~\cite{NS,LMNS}, where the
quantization of the SDYM theory on  K\"ahler manifolds was
considered.  We shall give additional arguments in favour of the
conjecture that the SDYM model is a good starting point for the
development of 4D quantum CFT's.

\smallskip

The main purpose of our paper is to describe {\it all} symmetries of
the SDYM equations and, in particular, algebras generalizing the
Virasoro and affine Lie algebras to the 4D case. In contrast with the
WZNW model, most symmetries of the SDYM model are nonlocal. These
symmetries are local symmetries of a holomorphic analogue of the
Chern-Simons theory on a 6D twistor space, and the SDYM theory is
connected with this model via the nonlocal Penrose-Ward transform.
The use of this correspondence makes it possible to simplify
considerably the investigation of symmetries of the SDYM equations.
The lift from a 4D self-dual space to its 6D twistor space is useful for
understanding correct degrees of freedom and  correct symmetries
of the SDYM theory. We show that just as in the case of the 2D WZNW
theory these symmetries {\it completely} define the space of local
solutions to the field equations and therefore  the quantization of
the SDYM theory is connected with the construction of representations
of a symmetry algebra.

\smallskip

Roughly speaking, the symmetry group of a system of differential
equations is the group that maps solutions of this system into one
another. From this point of view the transformation groups of type
$\m(X_3; G)$ (maps: space $X_3\to$ group $G$, $dim_{\R} X_3=3$)
considered in~\cite{NS,LMNS} are not symmetry groups, since in
general their action does not preserve the solution space.  The
above-mentioned groups $\m(X_3; G)$ can be considered as ``off-shell"
symmetry groups reflecting only the field content of the theory and
acting on free  fields. These groups are not connected with the
integrability and can be introduced in a space-time of an arbitrary
dimension (see e.g.~\cite{BGH}).

\smallskip

Study of ``on-shell" infinitesimal symmetries of the SDYM equations
(in 4D Euclidean space) began from the papers ~\cite{Prasad} and was
continued in ~\cite{Chau}-\cite{tai}. In ~\cite{Chau, Dolan}  it has
been shown that the obtained infinitesimal symmetries form the affine
Lie algebra $\fg\otimes\C\; [\l,\l^{-1}]$ when the gauge potential
$A=A_\mu dx^\mu$ takes values in the Lie algebra $\fg$ of a group
$G$. Ueno and Nakamura~\cite{UN} have shown that on the solution
space of the SDYM equations it is possible to define an infinitesimal
action of a larger Lie algebra of holomorphic maps from a domain on
the twistor space $\cz$ of $\R^4$ into the algebra $\fg$.
Takasaki~\cite{Tak} described this algebra in terms of Sato's
approach to soliton equations. But Crane~\cite{Crane} showed that in
general the group corresponding to the Ueno-Nakamura algebra does not
preserve the solution space  of the SDYM equations and indicated a
vagueness of geometrical meaning of these transformations. The
above-mentioned infinitesimal symmetries do not exhaust all
symmetries of the SDYM equations, which has been shown in the papers
~\cite{Taka}-\cite{ita} where  Virasoro-type symmetries were
described.  In this paper, we describe the {\it full} symmetry group
of the SDYM equations.

\newpage
\section{Self-duality and manifest symmetries}
\label{2}

\subsection{Definitions and notation}
\label{2.1}
We consider the Euclidean space ${\R}^4$ with the metric
$\d_{\mu\nu}$, a gauge potential  $A=A_\mu dx^\mu$ and the Yang-Mills
{}field $F=dA+A\wedge A$ with components
$F_{\mu\nu}=\p_\mu A_\nu - \p_\nu A_\mu + [A_\mu , A_\nu ]$,
where $\mu , \nu ,...=1,...,4$, $\p_\mu := \p / \p x^\mu$.  The
{}fields $A_\mu$ and $F_{\mu\nu}$ take values in the Lie algebra $\fg$
of an arbitrary semisimple compact Lie group $G$.  We suppose that
$G$ is a matrix group $G\subset GL(n, \C)$.

\smallskip

The SDYM equations have the form
$$
\frac{1}{2} \varepsilon _{\mu\nu\rho\sigma}F_{\rho\sigma}=F_{\mu\nu},
\eqno(2.1)
$$
where $\varepsilon _{\mu\nu\rho\sigma}$ is the completely antisymmetric
tensor in $\R^4$ and $\varepsilon _{1234}=1$. Here, and throughout the
paper, we use the Einstein summation convention, unless otherwise stated.

\smallskip

In this paper,  we study the space of smooth {\it local} solutions
to the SDYM
equations (2.1). More precisely, we suppose $A_\mu$ to be smooth on
an arbitrary open ball $U$ in $\R^4$, and we do not fix boundary
conditions for $A_\mu$. As $U$ we shall also consider open subsets in
$\R^4$ which can be dense subsets in $\R^4$ and can even coincide
with $\R^4$. We shall consider smooth solutions of the
SDYM equations on $U$ lying in an open neighbourhood of some
fixed solution $A^0_\mu$, for instance, in a neighbourhood of the
vacuum $A^0_\mu =0$ (local solutions).
The set of local solutions is an infinite-dimensional space and
contains finite-dimensional moduli spaces of global solutions
(instantons, monopoles etc.) as subspaces.

\subsection{Gauge symmetries}
\label{gs}

Equations (2.1) are manifestly invariant under the group of gauge
transformations
$$
A_\mu \mapsto A_\mu^g = g^{-1} A_\mu g + g^{-1} \p_\mu g,
\  F_{\mu\nu} \mapsto F_{\mu\nu}^g = g^{-1} F_{\mu\nu} g,
\eqno(2.2)
$$
where $g=g(x)\in G, x\in U\subset \R^4$. For infinitesimal gauge
transformations we have
$$
\d_{\vp} A_\mu =D_\mu \vp\equiv\p_\mu \vp + [A_\mu , \vp ],
\eqno(2.3)
$$
where $\vp (x)\in\fg, x\in U$.

\smallskip

The fields $A_\mu$ and $A^g_\mu$ differing by the gauge transformation
(2.2) are considered to be equivalent. That is why
gauge transformations are ``trivial" symmetries.

\subsection{Conformal symmetries}
\label{cs}

It is well-known that the SDYM equations (2.1) are invariant with
respect to (w.r.t.) the group of conformal transformations of the
space $\R^4$.  This group is locally isomorphic to the group
$SO(5,1)$.  On the coordinates $x^\mu$ and components $A_\mu$ of the
gauge potential $A$ the group of conformal transformations acts in the
{}following way:\\

$${\mathrm {translations:}}\quad x^\mu \mapsto \tilde x^\mu =
x^\mu + a^\mu ,
\ A_\mu (x^\nu )\mapsto \tilde A_\mu = A_\mu (x^\nu + a^\nu ),
\eqno(2.4a)$$

$$ {\mathrm {rotations:}} \qquad x^\mu \mapsto \tilde x^\mu =
a^\mu_\nu x^\nu ,
\ A_\mu (x^\nu )\mapsto \tilde A_\mu = (a^{-1})^\sigma_\mu A_\sigma
(a^\nu_\rho x^\rho ),
\eqno(2.4b)$$

$${\mathrm {dilatations:}} \qquad x^\mu \mapsto
\tilde x^\mu = e^\a x^\mu ,
\ A_\mu (x^\nu )\mapsto \tilde A_\mu = e^{-\a} A_\mu (e^\a x^\nu ),
\eqno(2.4c)$$

 $${{\mathrm {special\ conformal}}\atop {\mathrm {transformations:}}}\quad
x^\mu \mapsto \tilde x^\mu =
\frac{x^\mu +\a^\mu x^2}{1+2\a_\nu x^\nu +\a^2 x^2},\
 A_\mu (x^\nu )\mapsto \tilde A_\mu = \frac{\p x^\sigma}{\p\tilde x^\mu}
 A_\sigma (\tilde x^\nu ),
\eqno(2.4d)$$\\
where $a=(a^\mu_\nu )\in SO(4)$, $a^\mu , \a ,\a^\mu\in \R$,
$\a ^2:=\a _\nu \a ^\nu$, $x^2:=x_\nu x^\nu$.

\smallskip

{}For infinitesimal conformal transformations we have
$$
\d_N A_\mu ={\cl}_NA_\mu\equiv N^\nu\p_\nu A_\mu + A_\nu\p_\mu N^\nu,
\eqno(2.5)
$$
where ${\cl}_N$ is the Lie derivative along a vector field $N$
and $N=N^\nu\p_\nu$ is any generator of the 15-parameter conformal group,
$$
X_a = \d_{ab}\eta^b_{\mu\nu} x_\mu\p_\nu ,\
Y_a = \d_{ab}\bar\eta^b_{\mu\nu} x_\mu\p_\nu ,\  P_\mu =\p_\mu ,
$$
$$
K_\mu =\frac{1}{2}x^2 \p_\mu - x_\mu x^\nu\p_\nu ,
\quad D=x^\nu\p_\nu .
\eqno(2.6)
$$
Here $\{X_a\}$ and $\{Y_a\},\ a,b,...=1,2,3,$ generate two commuting
$SO(3)$ subgroups in $SO(4)$, $P_\mu$ are the translation generators            ,
$K_\mu$ are the generators of special conformal transformations, $D$ is
the dilatation generator,
$\eta^a_{\mu\nu}=\{\epsilon^a_{bc}, \mu =b, \nu =c;\ \delta^a_\mu , \nu =4;
-\delta^a_\nu , \mu =4 \}$ are the self-dual 't~Hooft tensors and
$\bar\eta^a_{\mu\nu}=\{\epsilon^a_{bc}, \mu =b, \nu =c;\ -\delta^a_\mu ,
\nu =4; \delta^a_\nu , \mu =4 \}$ are the anti-self-dual 't~Hooft tensors.

\vspace{0.2cm}

{\bf Remark.} It is well-known that for semisimple structure group $G$
there are no {\it local} symmetries of the SDYM equations differing from
the gauge and conformal symmetries described above.

\section{Complex geometry of twistor spaces}
\label{3}

\subsection{Complex structure on $\R^4$}
\label{csr4}

To write down a linear system for eqs.(2.1) and to clarify its
geometrical meaning, it is necessary to introduce a complex structure
$J$  on $\R^4$ (and thus on any open subset $U\subset\R^4$).  This means
that we must introduce on $\R^4$ a tensor $J^\nu_\mu$ such that
$J^\sigma_\mu J_\sigma^\nu = -\delta^\nu_\mu$.  It is well-known that
all constant complex structures on $\R^4$ are parametrized by the
two-sphere $S^2\simeq SO(4)/U(2)$, and the most general form of
$J^\nu_\mu$ is
$$
J_{\mu}^{\nu} = s_a \bar\eta_{\mu\sigma}^{a}\delta^{\sigma\nu},
\eqno(3.1)
$$
where real numbers $s^a$ parametrize $S^2$: $s_as^a=1$. Using the
identities for the 't~Hooft tensors
$$
\bar\eta^{a}_{\mu\sigma} \bar\eta^{b}_{\nu\sigma} =
\delta^{ab}\delta_{\mu\nu} +\epsilon^{abc}\bar\eta_{\mu\nu}^c,
\eqno(3.2)
$$
it can be shown that $J^2=-1$. The other admissible choice of the
complex structure $\tilde J_{\mu}^{\nu} = s_a \eta_{\mu\sigma}^{a}
\delta^{\sigma\nu}$ corresponds to choosing the opposite orientation
on $\R^4$ and  transition from self-duality to anti-self-duality
equations.

\smallskip

Eigenvalues of the operator $J=(J^\nu_\mu )$ (applied to vectors)
are $\pm i$, and we can introduce two subspaces in $\C^4=\R^4\otimes \C$,
$$
V^{1,0}=\{V\in\C^4: J^\mu_\sigma V^\sigma = i V^\mu\},\
V^{0,1}=\{V\in\C^4: J^\mu_\sigma V^\sigma = - i V^\mu\}.
\eqno(3.3a)
$$
As a basis in $V^{1,0}$ and $V^{0,1}$ one may take vectors with
the components
$$
\{V^{(1)\mu }_1\}=\{\frac{1}{2}, -\frac{i}{2}, -\frac{1}{2}\bar\l ,
\frac{i}{2}\bar\l\},\
\{V^{(1)\mu }_2\}=\{\frac{1}{2}\bar\l , \frac{i}{2}\bar\l , \frac{1}{2},
\frac{i}{2}\},
\eqno(3.3b)
$$
$$
\{\bar V^{(1)\mu }_1\}=\{\frac{1}{2}, \frac{i}{2}, -\frac{1}{2}\l ,
-\frac{i}{2}\l\},\
\{\bar V^{(1)\mu }_2\}=\{\frac{1}{2}\l , -\frac{i}{2}\l , \frac{1}{2},
-\frac{i}{2}\},
\eqno(3.3c)
$$
where $\l$ and $\bar\l$ are local holomorphic and antiholomorphic
coordinates on the sphere $S^2$, $\l =(s^1+is^2)/(1+s^3)$.

\smallskip

Using $J$, one can introduce vector fields $V^{(1)}_A=V^{(1)\mu}_A\p_\mu$
of the type $(1,0)$ and vector fields $\bar V^{(1)}_A=\bar
V^{(1)\mu}_A\p_\mu$ of the type $(0,1)$ w.r.t. $J$, where $A,B,...=1,2$.
We have
$$
\bar V^{(1)}_1= \bar V^{(1)\mu}_1\p_\mu =\frac{1}{2}(\p_1+i\p_2)
-\frac{\l}{2}(\p_3+i\p_4)=\p_{\bar y^1} - \l\p_{y^2},
\eqno(3.4a)
$$
$$
\bar V^{(1)}_2 = \bar V^{(1)\mu}_2\p_\mu =\frac{1}{2}(\p_3-i\p_4)
+\frac{\l}{2}(\p_1-i\p_2)=\p_{\bar y^2} + \l\p_{y^1},
\eqno(3.4b)
$$
where
$$
y^1=x^1 + ix^2,\ y^2=x^3-ix^4,\ \bar y^1=x^1 - ix^2,\
\bar y^2=x^3 + ix^4
\eqno(3.5)
$$
are the canonical complex coordinates on $\R^4\simeq\C^2$.

\subsection{Riemann sphere $\C{\bp}^1$}
\label{3.2}

In (3.3) we have introduced the complex coordinate $\l$ on
$S^2\simeq\C{\bp}^1$, parametrizing  complex structures on $\R^4$.
Using the stereographic projection $S^2\to \R^2$, one can introduce two
coordinate patches $\Omega_1\simeq \R^2$ and $\Omega_2\simeq \R^2$ of
the sphere with the coordinates
$$
\o^1_1=\frac{s^1}{1+s^3}\ {\mbox{and}}\
\o^2_1=\frac{s^2}{1+s^3}\ {\mbox{on}}\ \Omega_1,\
\o^1_2=\frac{s^1}{1-s^3}\ {\mbox{and}}\
\o^2_2=\frac{s^2}{1-s^3}\ {\mbox{on}}\ \Omega_2,
\eqno(3.6)
$$
in which the metric on $S^2$ is conformally flat.

\smallskip

We introduce the standard complex structure $\fj$ on $S^2$
with the components
$$
\fj = (\fj^A_B),\   \fj^A_C\fj^C_B=-\d^A_B,\
\fj^1_2=-\fj^2_1=-1
\eqno(3.7)
$$
in the coordinates $\{\o^A_1\}$. Now we can introduce vector fields,
holomorphic and antiholomorphic  w.r.t. $\fj$,  on $\Omega_1$ as
$$
V^{(1)}_3=\frac{1}{2}(\p_{\o^1_1}-i\p_{\o^2_1})=\p_\l ,\qquad
\fj^A_B V^{(1)B}_3=iV^{(1)A}_3,
\eqno(3.8)
$$
$$
\bar V^{(1)}_3=\frac{1}{2}(\p_{\o^1_1}+i\p_{\o^2_1})=\p_{\bar\l },\qquad
\fj^A_B \bar V^{(1)B}_3=-i\bar V^{(1)A}_3,
\eqno(3.4c)
$$
where $\l =\o^1_1+i\o^2_1$ is the complex coordinate on $\Omega_1\simeq\C$.
Analogously, we introduce the complex coordinate $\zeta =\o^1_2-i\o^2_2$
on $\Omega_2\simeq \C$ and vector fields $V^{(2)}_3=\p_\zeta$,
$\bar V^{(2)}_3=\p_{\bar\zeta}$ on $\Omega_2$.

\smallskip

So the sphere $S^2$ can be covered by two coordinate patches $\Omega_1$,
$\Omega_2$, with $\Omega_1$, the neighbourhood of $\l =0$, and $\Omega_2$,
the neighbourhood of $\l =\infty$. Let us fix $\a_1,
\a_2$: $0\le \a_1<1<\a_2\le\infty$ and put
$$
\Omega_1=\{\l\in\C: |\l |<\a_2\},\
\Omega_2=\{\l\in\C\cup\infty : |\l |>\a_1\}.
\eqno(3.9)
$$
The sphere $S^2$, considered as a complex projective line $\C{\bp}^1=
\Omega_1\cup\Omega_2$, is the complex manifold obtained by patching
together $\Omega_1$ and $\Omega_2$ with the coordinates $\l$ and $\zeta$
related by $\zeta =\l^{-1}$ on $\Omega_1\cap\Omega_2$. For example,
if $\Omega_1=\{\l\in\C: |\l |<\infty\}$ and
$\Omega_2=\{\l\in\C\cup\infty: |\l |>0\}$,
$\Omega_1\cap\Omega_2$ is the multiplicative group
$\C^{\mathbf{\ast}}$ of complex numbers $\l\ne\{0,\infty\}$.

\subsection{Twistor space}
\label{3.3}

We consider an open subset $U$ in $\R^4$. As a smooth manifold the
twistor space ${\rp}\equiv{\rp}(U)$ of $U$ is a direct product
of the spaces $U$ and $\C{\bp}^1$: ${\rp}=U\times\C{\bp}^1$
and is the bundle of complex structures on $U$~\cite{AHS}. This space
can be covered by two coordinate patches:
$$
{\rp}=\U_1\cup\U_2,\ \U_1=U\times\Omega_1,\ \U_2=U\times\Omega_2,
\eqno(3.10a)
$$
with the coordinates $\{x^\mu, \l ,\bar\l \}$ on $\U_1$ and $\{x^\mu,
\zeta , \bar\zeta \}$ on $\U_2$. The two-set open cover $\fo =\{\Omega_1,
\Omega_2\}$ of the Riemann sphere $\C{\bp}^1$ was described in
\S\,3.2.  We shall consider the intersection $\U_{12}$ of $\U_1$
and $\U_2$
$$
\U_{12}:=\U_1\cap\U_2=U\times(\Omega_1\cap\Omega_2)
\eqno(3.10b)
$$
with the coordinates $x^\mu\in U$, $\l ,\bar\l\in \Omega_{12}:=
\Omega_1\cap\Omega_2$. Thus, the twistor space $\rp$ is a trivial bundle
$\pi : {\rp}\to U$ over $U$ with the fibre $\C{\bp}^1$, where
$\pi : \{x^\mu , \l , \bar\l \}\to \{x^\mu\}$ is the canonical
projection.

\smallskip

We shall also consider the twistor space ${\cz}\equiv{\cz}(\R^4)$
of $\R^4$ which as a smooth manifold is a direct product
${\cz}=\R^4\times\C{\bp}^1$. The twistor space $\rp$ is an open
subset of $\cz$. In its turn, ${\cz}\simeq\C{\bp}^3 -\C{\bp}^1$
is an open subset in the space $\C{\bp}^3$ which is the twistor space
of the sphere $S^4$. Formally, $\rp$ concides with $\cz$ if
we take $U=\R^4$; that is why we denote the cover of $\cz$
by the same letters $\U_1=\R^4\times\Omega_1, \U_2=\R^4\times\Omega_2$.
Since $\rp$ is an open subset of $\cz$, a
complex structure will be discussed for $\cz$.

\smallskip

Having the complex structure $J$ on $\R^4$ and the complex structure
$\fj$ on $S^2$, we can introduce a complex structure ${\cj}=
(J, {\fj})$ on $\cz$. The vector fields $\{\bar V^{(1)}_a\}$ on $\U_1$,
introduced in (3.4), are vector fields of
the type (0,1) w.r.t. the complex structure $\cj$. Vector fields
$\{\bar V_a^{(2)}\}$ of the type (0,1) on $\U_2$ have the form
$$
\bar V_1^{(2)}=\zeta\p_{\bar y^1}-\p_{y^2},\quad
\bar V_2^{(2)}=\zeta\p_{\bar y^2}+\p_{y^1},\quad
\bar V_3^{(2)}=\p_{\bar\zeta},
\eqno(3.11a,b,c)
$$
and we have
$$
\bar V^{(1)}_1=\l\bar V^{(2)}_1,\quad
\bar V^{(1)}_2=\l\bar V^{(2)}_2,\quad
\bar V^{(1)}_3=-\bar\l^2\bar V^{(2)}_3
\eqno(3.12a,b,c)
$$
on $\U_{12}=\U_1\cap\U_2$.

\smallskip

Now we can introduce complex coordinates $\{z^a_1\}$ on $\U_1$
and $\{z^a_2\}$ on $\U_2$ as solutions of the equations
$\bar V^{(1)}_a(z^b_1)=0$ and $\bar V^{(2)}_a(z^b_2)=0$. We have
$$
z^1_1=y^1-\l \bar y^2,\quad
z^2_1=y^2+\l \bar y^1,\quad z^3_1=\l ,
\eqno(3.13a)
$$
$$
z^1_2=\zeta y^1-\bar y^2,\quad
z^2_2=\zeta y^2+\bar y^1,\quad z^3_2=\zeta
\eqno(3.13b)
$$
and on the intersection $\U_{12}$ these coordinates
are connected by the holomorphic transition function $f_{12}$
$$
z^a_1=f^a_{12}(z^b_2)\quad \Leftrightarrow \quad
z^1_1=f^1_{12}(z^b_2)=\frac{z^1_2}{z^3_2},\
z^2_1=f^2_{12}(z^b_2)=\frac{z^2_2}{z^3_2},\
z^3_1=f^3_{12}(z^b_2)=\frac{1}{z^3_2}.
\eqno(3.13c)
$$
From (3.13) it is not difficult to derive the formulae
$$
\frac{\p}{\p\bar z^1_1}=\g_1\bar V^{(1)}_1,\quad
\frac{\p}{\p\bar z^2_1}=\g_1\bar V^{(1)}_2,\quad
\frac{\p}{\p\bar z^3_1}=\bar V^{(1)}_3 + \bar y^2\g_1
\bar V^{(1)}_1 - \bar y^1\g_1\bar V^{(1)}_2,
\eqno(3.14a)
$$
where $\g_1=1/(1+\l\bar\l )$. Analogously, on $\U_2$
$$
\frac{\p}{\p\bar z^1_2}=\g_2\bar V^{(2)}_1,\quad
\frac{\p}{\p\bar z^2_2}=\g_2\bar V^{(2)}_2,\quad
\frac{\p}{\p\bar z^3_2}=\bar V^{(2)}_3 - \bar y^1\g_2\bar V^{(2)}_1-
\bar y^2\g_2\bar V^{(2)}_2,
\eqno(3.14b)
$$
where $\g_2=1/(1+\zeta\bar\zeta )$.

\smallskip

It is easy to check that the local basis (0,1)-forms w.r.t. $\cj$  are
$$
\bar\theta^1_{(1)}=\gamma_1(d\bar y^1-\bar\l dy^2),\quad
\bar\theta^2_{(1)}=\gamma_1(d\bar y^2+\bar\l dy^1),\quad
\bar\theta^3_{(1)}=d\bar\l \quad \mbox{on}\quad \U_1,
\eqno(3.15a)
$$
$$
\bar\theta^1_{(2)}=\gamma_2(\bar\zeta d\bar y^1- dy^2),\quad
\bar\theta^2_{(2)}=\gamma_2(\bar\zeta d\bar y^2+ dy^1),\quad
\bar\theta^3_{(2)}=d\bar\zeta \quad \mbox{on}\quad \U_2.
\eqno(3.15b)
$$
The exterior derivative $d$ on $\cz$ splits into $\p$ and
$\bar\p$: $d=\p +\bar\p $, where
$$
\bar\p =d\bar z^a_1\frac{\p}{\p\bar z^a_1}=\bar\theta^a_{(1)}
\bar V^{(1)}_a  \quad \mbox{on}\quad \U_1,
\eqno(3.16a)
$$
$$
\bar\p =d\bar z^a_2\frac{\p}{\p\bar z^a_2}=\bar\theta^a_{(2)}
\bar V^{(2)}_a  \quad \mbox{on}\quad \U_2,
\eqno(3.16b)
$$
and the operator $\p$ is connected with $\bar\p$ by means of complex
conjugation. As usual $d^2=\p^2=\bar\p^2=\p\bar\p +\bar\p\p =0$.

\smallskip

It follows from (3.12), (3.13) and (3.15) that as a complex manifold
$\cz$ is not a direct product $\C^2\times\C{\bp}^1$, but is a
nontrivial holomorphic vector bundle $p:\ {\cz}\to\C{\bp}^1$.
Moreover, from (3.12), (3.13) and (3.15) it follows that $\cz$
coincides with a total space of the rank 2 holomorphic vector bundle
$L^{-1}\oplus L^{-1}$ over $\C{\bp}^1$,
$$
p:\ {\cz}=L^{-1}\oplus L^{-1}\ \longrightarrow\ \C{\bp}^1,
\eqno(3.17)
$$
where $L$ is the tautological complex line bundle over $\C{\bp}^1$
with the transition function  $\l^{-1}$, and the first Chern class
$c_1(L)$ equals $-1$: $c_1(L)=-1$. Its dual $L^{-1}$ is isomorphic to
the hyperplane bundle  (Chern class $c_1(L^{-1})=1$) over $\C{\bp}^1$.
The twistor space $\rp$ of $U\subset \R^4$ is an open subset of $\cz$
 and ${\cz}=L^{-1}\oplus L^{-1}\simeq \C{\bp}^3 - \C{\bp}^1$ is an
open subset of $\C{\bp}^3$. Holomorphic sections of the bundle (3.17)
are projective lines
$$
\C{\bp}^1_y =\left\{
\l\in\Omega_1:
z^1_1=y^1+\l \tilde y^2,\ z^2_1=y^2+\l \tilde y^1\atop
\zeta\in\Omega_2:
z^1_2=\zeta y^1+ \tilde y^2,\ z^2_2=\zeta y^2+\tilde y^1\right\}
\eqno(3.18)
$$
parametrized by the points $y=\{y^1,y^2,\tilde y^1,\tilde y^2\}\in\C^4$.

\subsection{Real structure on twistor space}
\label{3.4}

A real structure on the complex twistor space $\cz$ is defined as
an antiholomorphic involution $\tau :\ {\cz}\to{\cz}$,
defined by the antipodal map $\l\mapsto -1/\bar \l$ on the
$\C{\bp}^1$ factor,
$$
\tau (x^\mu ,\l ) = (x^\mu , -1/\bar \l ), \ \tau^2 =1.
\eqno(3.19)
$$
This involution takes the complex structure $\cj$ on $\cz$ to its
conjugate $-\cj$, i.e., it is antiholomorphic. It is obvious from the
definition (3.19) that $\tau$ has no fixed points on $\rp\subset \cz$
but does leave the fibres $\cp\ ,\ x\in U,$ of the bundle ${\rp}\to
U$ invariant. The same is true for the fibres $\cp$ of the bundle
${\cz}\to \R^4$.  Fibres $\cp$ of the bundle ${\rp}\to U$ are also
{\it real} holomorphic sections of the bundle (3.17) for which we
have $\tilde y^1=\bar y^1,\ \tilde y^2=-\bar y^2$ in (3.18), i.e.,
they are parametrized by $\{x^\mu\}=\{y^A,\bar y^A\}\in U$.

\smallskip

An extension of the involution $\tau$ to complex functions
$f(x^\mu ,\l )$ has the form~\cite{Ma}
$$
\tau : f(x,\l )\mapsto \tau (f(x,\l ))\equiv f_\tau (x,\l ):=
\overline{f(\tau (x,\l ))}=\overline{f(x,-\bar\l^{-1})}.
\eqno(3.20)
$$
In particular, for the complex coordinates $\{z^a_1\}$ and $\{z^a_2\}$
on $\cz$ we have
$$
\tau (z^1_1)=z^2_2,\ \tau (z^2_1)=-z^1_2,\ \tau (z^3_1)=-z^3_2,\
\Leftrightarrow
$$
$$
\tau(z^a_1)=B^a_bz^b_2,\ B^1_2=1,\ B^2_1=-1,\ B^3_3= -1.
\eqno(3.21)
$$
All the rest components of the constant matrix $B=(B^a_b)$ are equal
to zero.

\smallskip

Using (3.21), it is not difficult to verify that for the transition
function (3.13c) compatible with the real structure $\tau$, we have
$$
\tau (f^a_{12})=B^a_b\tilde f^b_{12},
\eqno(3.22)
$$
where $\tilde f_{12}$  is the transition function inverse to $f_{12}$
$$
z^1_2=\tilde f^1_{12}(z^b_1)=\frac{z^1_1}{z^3_1},\quad
z^2_2=\tilde f^2_{12}(z^b_1)=\frac{z^2_1}{z^3_1},\quad
z^3_2=\tilde f^3_{12}(z^b_1)=\frac{1}{z^3_1}.
\eqno(3.23)
$$
So all the holomorphic data are compatible with $\tau$.

\section{The Penrose-Ward correspondence}
\label{pwc}

\subsection{Complex vector bundles over $U$ and $\rp$}
\label{cvbo}

Let us consider a principal $G$-bundle $P=P(U,G)=U\times G$
over $U\subset \R^4$. Then, a gauge potential $A=A_\mu dx^\mu$
(a connection 1-form) defines a connection $D:=d+A=dx^\mu (\p_\mu
+A_\mu )$ on the bundle $P$, and the 2-form $F=dA+A\wedge A=
\frac{1}{2}F_{\mu\nu}dx^\mu\wedge dx^\nu$ is the curvature of the
connection form $A$. We shall consider irreducible connections.
Suppose a representation of $G$ in the complex vector space $\C^n$
is given. In the standard manner we associate with $P$ the complex
vector bundle
$$
E=P\times_G\C^n\simeq U\times\C^n,
$$
which is topologically trivial.

\smallskip

Using the projection $\pi : {\rp}\to U$ of the twistor space $\rp$ on
$U$, we can pull back $E$ to a bundle $E':=\pi^{\mathbf{\ast}}E$ over
$\rp$, and the pulled back bundle $E'$ is trivial on
 the fibres $\cp$ of the bundle ${\rp}\rightarrow U$.
We can set components of $\pi^{\mathbf{\ast}}A$ along the fibres
equal to zero and then the pulled back connection $D'$ will have
the form $D'= D+d\l\p_\l+ d\bar\l\p_{\bar\l}$ (on $\U_1$)$=
D+d\zeta\p_\zeta + d\bar\zeta\p_{\bar\zeta}$ (on $\U_2$).

\subsection{Self-duality $\Rightarrow$ holomorphy}
\label{sdh}

The twistor space $\rp$ of the space $U\subset \R^4$ is a complex
three-dimensional manifold with the coordinates $\{z^a_1\}$ on
$\U_1\subset \rp$ and $\{z^a_2\}$ on $\U_2\subset \rp$,
${\rp}=\U_1\cup\U_2$.  Using the (0,1)-forms (3.15), we introduce the
(0,1) components $B_a$ of the connection 1-form
$\pi^{\mathbf{\ast}}A=A_\mu dx^\mu =B^{1,0} + B^{0,1}\equiv \bar B+B$  by
the formulae
$$
\{ B^{(1)}_1:= A_{\bar y^1} - \l A_{y^2},\ B^{(1)}_2:=
A_{\bar y^2} + \l A_{y^1},\ B^{(1)}_3:=0\}\quad \mbox{on}\quad \U_1,
\eqno(4.1a)
$$
$$
\{ B^{(2)}_1:= \zeta A_{\bar y^1} - A_{y^2},\ B^{(2)}_2:=
\zeta A_{\bar y^2} + A_{y^1},\ B^{(2)}_3:=0\}\quad \mbox{on}\quad \U_2.
\eqno(4.1b)
$$
Notice that $B^{(1)}_a=\l B^{(2)}_a$ on $\U_{12}$. One can also
introduce the components $B_{\bar z^a_{1,2}}$ of $B$ along the
antiholomorphic vector fields $\p_{\bar z^a_{1,2}}$ from (3.14),
$$
\{B_{\bar z^1_1}: = \g_1 B^{(1)}_1,\
B_{\bar z^2_1}: = \g_1 B^{(1)}_2,\
B_{\bar z^3_1}: =y^2 \g_1 B^{(1)}_1 - y^1\g_1B^{(1)}_2\}
\quad \mbox{on}\quad \U_1,
\eqno(4.2a)
$$
$$
\{B_{\bar z^1_2}: = \g_2 B^{(2)}_1,\
B_{\bar z^2_2}: = \g_2 B^{(2)}_2,\
B_{\bar z^3_2}: =-\bar y^1 \g_2 B^{(2)}_1 - \bar
y^2\g_2B^{(2)}_2\}
\quad \mbox{on}\quad \U_2.
\eqno(4.2b)
$$
Then we have $\pi^{\mathbf{\ast}}A=\bar B+B$ and
$$
B\equiv B^{0,1}=B_{\bar z^a_1}d\bar z^a_1=B^{(1)}_a\bar\theta^a_{(1)}
\quad \mbox{on}\quad \U_1,
\eqno(4.3a)
$$
$$
B\equiv B^{0,1}=B_{\bar z^a_2}d\bar z^a_2=B^{(2)}_a\bar\theta^a_{(2)}
\quad \mbox{on}\quad \U_2.
\eqno(4.3b)
$$

\smallskip

Now we can introduce components of the connection $D'$ on the
complex vector bundle $E'$ which are (0,1) components w.r.t.
the complex structure $\cj$ on $\rp$,
$$
D':=\p_{\bar B}+\bar\p_B,\quad \bar\p_B=\bar\p + B,
\eqno(4.4)
$$
where the operator $\bar\p$ was introduced in (3.16), the (0,1)-form
$B$ was introduced in (4.3) and the operator $\p_{\bar B}=\p +\bar B$
is the (1,0) component of the operator $D'$.

\vskip 2mm

{\bf Remark.} In most cases we shall further write down
 formulae and equations in the trivialization over
$\U_1\subset \rp$.

\smallskip

Let us consider the equations
$$
\bar\p_B\xi =0
\eqno(4.5)
$$
on a smooth local section $\xi$ of the bundle $E'$.
The local solutions of these equations are by definition
 the local holomorphic sections of the complex vector
bundle $E'$. The bundle $E'\to \rp$ is said to be
holomorphic if eqs.(4.5) are compatible, i.e., $\bar\p^2_B =0
\ \Rightarrow$ the (0,2) components of the curvature of
$D'$ are equal to zero.

\smallskip

In the trivialization over $\U_1$, eqs. (4.5) are equivalent to the
equations
$$
[(D_1+iD_2)-\l (D_3+iD_4)]\xi_1(x, \l , \bar\l )=0,
\eqno(4.6a)
$$
$$
[(D_3-iD_4)+\l (D_1-iD_2)]\xi_1(x, \l , \bar\l )=0,
\eqno(4.6b)
$$
$$
\p_{\bar\l }\xi_1(x, \l , \bar\l )=0,
\eqno(4.6c)
$$
and analogously in the trivialization over $\U_2$. Equation (4.6c)
simply means that $\xi_1$ is a function of $x^\mu$ and $\l$
(does not depend on $\bar\l$). If eq.(4.6c) is solved,  the
remaining two equations (4.6a,b) for $\xi_1(x,\l )$ are usually called
the {\it linear system} for the SDYM equations ~\cite{BZ}. It is readily
seen that the compatibility conditions $\bar\p^2_B=0$ of eqs.(4.6)
are identical to the SDYM equations (2.1), which in the coordinates
$\{y^1, y^2, \bar y^1, \bar y^2\}$ have the form
$$
F_{y^1y^2}=0,\ F_{\bar y^1\bar y^2}=0,\ F_{y^1\bar y^1}+F_{y^2\bar y^2}=0,
\eqno(4.7)
$$
i.e., eqs. (4.7) follow from the equations $\bar\p^2_B=0$. Therefore,
if a gauge potential $A=A_\mu dx^\mu$ is a smooth solution of eqs.(4.7)
on a domain $U$ in $\R^4$, there exist solutions of eqs.(4.5), and
the bundle $E'\to \rp$ is holomorphic.

\smallskip

For the cover $\fu=\{\U_1, \U_2\}$ of ${\rp}=\U_1\cup \U_2$, eqs.(4.5)
have a local solution $\xi_1$ over $\U_1$, a local solution $\xi_2$
over $\U_2$ and $\xi_1=\xi_2$ on the overlap $\U_{12}=\U_1\cap \U_2$
(i.e., it is a section over $\rp$). We can always represent
$\xi_1, \xi_2$ in the form $\xi_1=\psi_1\chi_1$, $\xi_2=\psi_2\chi_2$,
where $G^{\C}$-valued functions $\psi_1$ and $\psi_2$  nonsingular
on $\U_1$ and $\U_2$ satisfy the equations
$$
\bar\p_B\psi_1=0,\quad \bar\p_B\psi_2 =0
\eqno(4.8)
$$
on $\U_1$ and $\U_2$, respectively. The vector-functions $\chi_{1,2}
\in\C^n$ are holomorphic on $\U_{1,2}$,
$$
\bar V^{(1)}_a\chi_1 =0,\quad  \bar V^{(2)}_a\chi_2 =0.
\eqno(4.9)
$$
It follows from (4.8) that
$$
(\p_{\bar y^1}\psi_1 - \l \p_{y^2}\psi_1)\psi_1^{-1}=
(\p_{\bar y^1}\psi_2 - \l \p_{y^2}\psi_2)\psi_2^{-1}=
-(A_{\bar y^1} - \l A_{y^2}),
\eqno(4.10a)
$$
$$
(\p_{\bar y^2}\psi_1 + \l \p_{y^1}\psi_1)\psi_1^{-1}=
(\p_{\bar y^2}\psi_2 + \l \p_{y^1}\psi_2)\psi_2^{-1}=
-(A_{\bar y^2} + \l A_{y^1}),
\eqno(4.10b)
$$
$$
\p_{\bar\l}\psi_1=\p_{\bar\l}\psi_2=0.
\eqno(4.10c)
$$
Moreover, the vector-functions $\chi_1$ and $\chi_2$ are related by
$$
\chi_1 = {\cf}_{12}\chi_2
\eqno(4.11)
$$
on $\U_{12}$, i.e.,
$$
{\cf}_{12}:=\psi_1^{-1} \psi_2
\eqno(4.12)
$$
is the transition matrix in the bundle $E'$ and
${\cf}_{21}:=\psi_2^{-1} \psi_1={\cf}_{12}^{-1}$.
From eqs.(4.8), (4.10) it follows that ${\cf}_{12}$
is the holomorphic $G^{\C}$-valued function on $\U_{12}$
with nonvanishing determinant.

\smallskip

{\bf Remarks}

\smallskip

{\bf 1.} The matrices $\psi_1$ and $\psi_2$ are matrix fundamental
solutions, i.e., the columns of $\psi_1, \psi_2$ form frame fields for
$E'$ over $\U_1$, $\U_2$. In other words, matrix-valued functions
$\psi_1, \psi_2$ define a trivialization of the bundle $E'$ over
$\U_1$, $\U_2$. At the same time, $\chi_1=\chi_1(z^a_1)$ and
$\chi_2=\chi_2(z^a_2)$ are \v{C}ech fibre coordinates of the bundle
$E'$ over $\U_1$ and $\U_2$. The representation of $\xi_{1,2}$ in the
form $\xi_1=\psi_1\chi_1$, $\xi_2=\psi_2\chi_2$ is simply an
expansion of the sections $\xi_{1,2}$ in the basis sections
$\psi_{1,2}$ with the components $\chi_{1,2}$ (see e.g.~\cite{GH}).

\smallskip

{\bf 2.} The matrix-valued functions $\psi_{1,2}$ are
$\C^{\infty}$-functions on $\U_{1,2}$, and any transition matrix of
the form (4.12) defines a bundle $E'$, which is topologically
trivial, but holomorphically nontrivial, since $\psi_{1,2}$ are not
holomorphic functions on $\U_{1,2}$.  On the other hand, eqs.(4.10c)
mean that the restriction of $E'$ to any real projective line $\cp$
($x\in U$) is holomorphically trivial: $E'|_{\cp}\simeq \cp\times
\C^n$.

\subsection{Gauge transformations and holomorphic equivalence}
\label{4.3}

It is easy to see that the local gauge transformations (2.2)
of the gauge potential $A$ are induced by the transformations
$$
\psi_1\mapsto \psi^g_1:=g^{-1}(x)\psi_1,\
\psi_2\mapsto \psi^g_2:=g^{-1}(x)\psi_2,
\eqno(4.13)
$$
and the transition matrix ${\cf}_{12}=\psi_1^{-1}\psi_2$ is
invariant under these transformations because $(\psi^g_1)^{-1}
\psi^g_2 = \psi^{-1}_1\psi_2$.

\smallskip

On the other hand, the components $\{A_\mu\}$ of the gauge potential $A$
in (4.10) will not change after transformations
$$
\psi_1\mapsto \psi^{h_1}_1:=\psi_1h^{-1}_1,\
\psi_2\mapsto \psi^{h_2}_2:=\psi_2h^{-1}_2,
\eqno(4.14)
$$
where $h_1$ is any regular holomorphic $G^{\C}$-valued function on
$\U_1$ and $h_2$ is any regular holomorphic $G^{\C}$-valued function
on $\U_2$.  This means that a class of holomorphically equivalent
bundles over the twistor space $\rp$ corresponds to a self-dual
connection on $U$.  Recall that holomorphic bundles with the
transition matrices $\hat{\cf}_{12}$ and ${\cf}_{12}$ are called
holomorphically equivalent if
$$
\hat{\cf}_{12}=h_1 {\cf}_{12}h_2^{-1}
\eqno(4.15)
$$
for some regular matrices $h_1, h_2$ such that
$h_1$ is holomorphic on $\U_1$ and $h_2$ is holomorphic on $\U_2$.

\subsection{Unitarity conditions}
\label{4.4}

It follows from eqs.(4.10) that in the general case the components
$\{A_\mu\}$ of the gauge potential $A$ will take values in the Lie
algebra $\fg^{\C}$, because  $\psi_{1,2}$ are $G^{\C}$-valued. This
is equivalent to the consideration of $A_\mu$ with values in the Lie
algebra $\fg$, but with complex components $A_\mu^k$ in the expansion
$A_\mu =A_\mu^kT_k$ in the generators $\{T_k\}$ of the Lie group $G$.
If we want to consider real gauge fields,  we have to impose
additional reality conditions on the bundle $E'$ induced by the real
structure $\tau$ on $\rp$ (see \S\,3.4) and by an automorphism
$\tilde\sigma$ of the Lie algebra $\fg^{\C}$ such that $\fg=\{ a\in
\fg^{\C}: \tilde\sigma (a)=a, \tilde\sigma^2 =id\}$.
Such a reality structure in
the bundle $E'$ exists for any compact Lie group $G$~\cite{AHS}, and we
shall describe it for the case $G=SU(n)$, $\fg=su(n)$.

\smallskip

Namely, in the case $\fg=su(n)$ we have $A_\mu^\dagger =-A_\mu$
and therefore
$$
A^\dagger_{y^1} = - A_{\bar y^1},\quad
A^\dagger_{y^2} = - A_{\bar y^2},
\eqno(4.16a)
$$
where $\dagger$ denotes Hermitian conjugation. Then the matrices
${\cf}_{12}\in SL(n,\C)$ and $\psi_1, \psi_2\in SL(n,\C)$
have to satisfy on $\U_{12}$ the following unitarity conditions
(see e.g.~\cite{Crane}):
$$
{\cf}_{12}^\dagger (\tau (\bar z^a_1))={\cf}_{12}(z^a_1),
\eqno(4.16b)
$$
$$
\psi^\dagger_1(\tau (x,\l  ))=\psi_2^{-1} (x,\l ),
\eqno(4.16c)
$$
where the action of $\tau$ on the coordinates of the space $\rp$
was described in \S\,3.4.

\vskip 2mm

{\bf Remark.} For simplicity, we shall always consider the case
$G=SU(n)$ when discussing real gauge fields.

\smallskip

Thus, starting from a bundle $E$ over $U\subset\R^4$ with a
self-dual connection, we have constructed a topologically trivial
holomorphic vector bundle $E'$ over $\rp$ satisfying the
conditions: (1) $E'$ is holomorphically trivial on each real
projective line $\cp$, $x\in U$, in $\rp$;
(2) $E'$ has a real structure.

\subsection{Riemann-Hilbert problems}
\label{4.5}

Suppose we have a nonsingular matrix-valued function
${\cf}(x,\l )\in SL(n,\C )$ on $\Omega_1\cap\Omega_2\subset
\C{\bp}^1$ (see \S\,3.2) depending holomorphically on $\l$
and smoothly on some parameters $\{x^\mu\}$.  Then a parametric
Riemann-Hilbert problem is to find matrix-valued functions $\psi_1,
\psi_2\in SL(n,\C)$ on $\Omega_1\cap\Omega_2$ such that $\psi_1$ can
be extended continuously to a regular (i.e., holomorphic with a
non-vanishing determinant) matrix-valued function on $\Omega_1$,
$\psi_2$ can be extended to a regular matrix-valued function on
$\Omega_2 $ and
$$
{\cf}(x,\l )=\psi_1^{-1}(x,\l )\psi_2(x,\l )
\eqno(4.17)
$$
on $\Omega_1\cap\Omega_2$.

\smallskip

It follows from the Birkhoff decomposition theorem (see
e.g.~\cite{Pres}) that for a fixed $x$ {\em any} holomorphic on
$\Omega_1\cap\Omega_2$ nonsingular matrix-valued function $\cf$
admits a decomposition
$$
{\cf}=\psi_1\Lambda\psi_2,
\eqno(4.18)
$$
where $\psi_1, \psi_2$ are defined above and $\Lambda$ is a diagonal
matrix whose entries are integral powers $k_i\in {\Z}$ of $\l$,
$k_1+...+k_n=0$. The $k_i$'s are unique up to permutation and are
Chern classes of the holomorphic line bundles over $\C{\bp}^1$
which occur in the decomposition of the holomorphic vector bundle over
$\C{\bp}^1$ with $\cf$ as a transition matrix (Grothendieck's theorem).

\smallskip

If $\Lambda$ is the identity matrix, the decomposition (4.18)
is called a solution to the Riemann-Hilbert problem. For these matrices
$\cf$, the factorization is unique up to a transformation
$$
\psi_1(x,\l )\mapsto \psi_1^g=g^{-1}(x)\psi_1(x,\l ),\
\psi_2(x,\l )\mapsto \psi_2^g=g^{-1}(x)\psi_2(x,\l ),
\eqno(4.19)
$$for some matrix $g(x)\in SL(n,\C)$. So the Riemann-Hilbert problem
can only be solved `generically' and (4.17) may not have a solution
{}for all values of the parameters $x^\mu$. But if a factorization
(4.17) exists at some $x_0^\mu$, then it exists in an open
neighbourhood $U$ of $x_0^\mu$. Usually, $\Lambda\ne 1$ on a
submanifold of codimension 1 (or more) of the parameter space.  The
points $x^\mu$ for which $\Lambda\ne 1$ are called jumping points,
and projective lines $\cp$ corresponding to these points $x$ are
called jumping lines.  In the twistor construction the jumping points
$x\in\R^4$ give rise to singularities in the SDYM potential $A$. For
details see e.g.~\cite{WW,MW}.

\subsection{Holomorphy $\Rightarrow$ self-duality}
\label{4.6}

Suppose we have a topologically trivial holomorphic vector
bundle $E'$ over $\rp$ with the cover $\fu=\{\U_1, \U_2\}$
and a transition matrix ${\cf}_{12}$ satisfying the unitarity condition
(4.16b). Considering ${\cf}_{12}$ for fixed $x^\mu\in U$,
we obtain a parametric Riemann-Hilbert problem on  $\C{\bp}^1$.
Then in a set of all possible transition matrices we choose
those for which a solution of the Riemann-Hilbert problem exists.

\smallskip

After finding a Birkhoff decomposition (4.17) for ${\cf}_{12}$
we consider $(\bar V_a^{(1)}\psi_1)\psi_1^{-1}$ and
$(\bar V_a^{(1)}\psi_2)\psi_2^{-1}$ as functions on $\U_1$
and $\U_2$ with values in the Lie algebra $su(n)$. For definitions
of the (0,1) vector fields $\bar V_a^{(1)}$, $\bar V_a^{(2)}$
see \S\,3. From the holomorphy of ${\cf}_{12}$
it follows that
$$
(\bar V_a^{(1)}\psi_1)\psi_1^{-1}=(\bar V_a^{(1)}\psi_2)\psi_2^{-1}
\eqno(4.20)
$$
on $\U_{12}$. Notice that as functions on $\C{\bp}^1$ the
matrices $\psi_1$ and $\psi_2$ are regular on $\Omega_1$ and
$\Omega_2$, respectively. Hence, $\psi_{1,2}$ can be expanded
on $\Omega_1\cap\Omega_2$ in powers of $\l$:
$$
\psi_1(x,\l )=\sum_{n=0}^{\infty}\l^n\psi_1^n (x),\
\psi_2(x,\l )=\sum_{n=0}^{\infty}\l^{-n}\psi_2^n (x).
\eqno(4.21)
$$
If we substitute the expansion of $\psi_{1,2}$ in powers of $\l$
into (4.20), both the sides of (4.20) must be linear in $\l$,
and we have
$$
(\p_{\bar y^1}\psi_1 - \l \p_{y^2}\psi_1)\psi_1^{-1}=
(\p_{\bar y^1}\psi_2 - \l \p_{y^2}\psi_2)\psi_2^{-1}=
-(A_{\bar y^1}(x) - \l A_{y^2}(x)),
\eqno(4.22a)
$$
$$
(\p_{\bar y^2}\psi_1 + \l \p_{y^1}\psi_1)\psi_1^{-1}=
(\p_{\bar y^2}\psi_2 + \l \p_{y^1}\psi_2)\psi_2^{-1}=
-(A_{\bar y^2}(x) + \l A_{y^1}(x)),
\eqno(4.22b)
$$
where
$$
A_{y^1}:=- \mathop{Res}\limits_{\l =0}\l^{-2}
(\bar V_2^{(1)}\psi_2)\psi_2^{-1}
\equiv -\oint_{S^1}\frac{d\l}{2\pi i\l^2}
(\bar V_2^{(1)}\psi_2)\psi_2^{-1}=
-(\p_{y^1}\psi^0_2)(\psi^0_2)^{-1},
\eqno(4.23a)
$$
$$
A_{y^2}:=\mathop{Res}\limits_{\l =0}\l^{-2}
(\bar V_1^{(1)}\psi_2)\psi_2^{-1}
\equiv \oint_{S^1}\frac{d\l}{2\pi i\l^2}
(\bar V_1^{(1)}\psi_2)\psi_2^{-1}=
-(\p_{y^2}\psi^0_2)(\psi^0_2)^{-1},
\eqno(4.23b)
$$
$$
A_{\bar y^1}:=- \mathop{Res}\limits_{\l =0}\l^{-1}
(\bar V_1^{(1)}\psi_1)\psi_1^{-1}
\equiv -\oint_{S^1}\frac{d\l}{2\pi i\l}
(\bar V_1^{(1)}\psi_1)\psi_1^{-1}=
-(\p_{\bar y^1}\psi^0_1)(\psi^0_1)^{-1},
\eqno(4.23c)
$$
$$
A_{\bar y^2}:=- \mathop{Res}\limits_{\l =0}\l^{-1}
(\bar V_2^{(1)}\psi_1)\psi_1^{-1}
\equiv -\oint_{S^1}\frac{d\l}{2\pi i\l}
(\bar V_2^{(1)}\psi_1)\psi_1^{-1}=
-(\p_{\bar y^2}\psi^0_1)(\psi^0_1)^{-1}.
\eqno(4.23d)
$$
Here, the contour $S^1=\{\l\in \C{\bp}^1: |\l|=1\}$
circles once around $\l =0$ and  the contour integral determines
 residue $Res$ at the point $\l =0$.

\smallskip

The components $\{A_\mu\}$ of the gauge potential defined by (4.23)
satisfy the SDYM equations on $U$ which are the compatibility
conditions of eqs.(4.22). Thus, starting from a holomorphic
matrix-valued function ${\cf}_{12}$ which is a transition
matrix of a holomorphic vector bundle $E'$ over the twistor
space $\rp$, we have completed the procedure of reconstructing a
gauge potential $A$ which defines a self-dual connection on a
complex vector bundle $E$ over $U\subset \R^4$. As it was explained
in \S\,4.3, the transformations (4.14), (4.15) of ${\cf}_{12}$ into a
holomorphically equivalent transition matrix $h_1{\cf}_{12}h_2^{-1}$
do not change $A_\mu$, and gauge transformations $A_\mu\mapsto
A^g_\mu$ inducing the transformations (4.13) do not change ${\cf}_{12}$.
It follows from the twistor construction that a self-dual gauge
potential $A$ is real-analytic.

\smallskip

To sum up, we have described a one-to-one correspondence between
gauge equivalence classes of solutions to the SDYM equations on an
open subset $U$ of the Euclidean 4-space and equivalence classes of
holomorphic vector bundles $E'$ over the twistor space $\rp$
satisfying the conditions: (i) bundles $E'$ are holomorphically
trivial on each real projective line  $\cp$, $x\in U$, in $\rp$, (ii)
each $E'$ has a real structure. This is the Euclidean version of
Ward's theorem~\cite{AW, WW}.

\vskip 2mm

{\bf Remark.} A twistor correspondence between self-dual gauge fields
and holomorphic bundles also exists in a more general
situation~\cite{AHS}.  Let us consider a real oriented four-manifold
$M$ with a metric $g$ of signature $(++++)$. The 4-manifold $M$ is
called self-dual if its Weyl tensor is self-dual. In~\cite{AHS}
it was proved that the twistor space ${\cz}\equiv {\cz}(M)$ for a
self-dual manifold $M$ is a complex analytic 3-fold. There is a
natural one-to-one correspondence between self-dual bundles $E$ over
$M$ (in particular, over $\R^4, S^4, T^4,...$) and holomorphic vector
bundles $E'$ over the twistor space $\cz$. In the general case,
bundles $E$ and $E'$ are not topologically trivial, as it takes place
in the case of Euclidean space $\R^4$, when
$\rp\subset{\cz}(\R^4)=\R^4\times\C{\bp}^1.$

\section{Holomorphic bundles in the \v{C}ech approach}
\label{5}

We are going to analyse the twistor correspondence between
self-dual complex vector bundles $E$ over $U\subset\R^4$
and holomorphic vector bundles $E'$ over $\mathcal  P$ from the
group-theoretic point of view, i.e., we want to describe groups
acting on the space of transition matrices $\cf_{12}$
of the bundles $E'$, on the space of self-dual gauge potentials
$A$ and on the moduli space of self-dual gauge fields. In our
discussion, we shall use the notion of {\it local groups},
(local) {\it actions} of (local) groups on sets, {\it germs},
{\it sheaves} and {\it \v{C}ech cohomology}, definitions of which are
recalled in Appendices A,B and C.

\smallskip

In this section, we shall describe symmetries and the moduli space of
{\it all} holomorphic vector bundles over $\rp$. This means that we
shall consider holomorphic bundles over $\rp$ which are not
necessarily holomorphically trivial over $\cp\hra{\rp}$, $x\in U$,
and do not satisfy the unitarity condition (4.16b).  As recalled in
Appendices B and C, there is a one-to-one correspondence between the
set of isomorphism classes of holomorphic bundles over a complex
space $X$ and the \v{C}ech 1-cohomology set $H^1(X, {\ch})$ of the
space $X$ with values in the sheaf ${\ch}={\co}^{G^{\C}}$ of germs of
holomorphic maps from $X$ into the complex Lie group $G^{\C}$. We
shall consider this correspondence for our case of the complex
twistor space $\rp$ and the group $G^{\C}=SL(n,\C )$ and describe
it from the group-theoretic point of view.

\subsection{Moduli space of holomorphic bundles over the twistor
            space $\rp$}
\label{ms}

We consider the two-set open cover $\fu=\{\U_1, \U_2\}$ of $\rp$
(see \S\,3.3), where $\U_1, \U_2$ are Stein manifolds. For this
cover we have the following q-simplexes
$\la\U_{\a_0},...,\U_{\a_q}\ra$:
$\la\U_1\ra $, $\la\U_2\ra$, $\la\U_1,\U_2\ra$,
$\la\U_2,\U_1\ra$, supports $\U_1, \U_2, \U_{12}:= \U_1\cap\U_2$
of which are nonempty sets.  Further, a {\it q-cochain} of the
cover $\fu$ with the coefficients in the sheaf $\ch=\co^{SL(n,\C)}$
is a map $f$,  which associates with any q-simplex
$\la\U_{\a_0},...,\U_{\a_q}\ra$ a section of the sheaf $\ch$
over $\U_{\a_0}\cap...\cap\U_{\a_q}$: $\ f_{\a_0...\a_q}
\equiv$ $f(\U_{\a_0}\cap...\cap\U_{\a_q})\in{\ch}
(\U_{\a_0}\cap...\cap\U_{\a_q})$. In other words, a q-cochain
of the cover $\fu$ with values in $\ch$ is a collection
$f=\{f_{\a_0\ldots\a_q}\}$ of sections of the sheaf $\ch$
over nonempty intersections $\U_{\a_0}\cap...\cap\U_{\a_q}$.
The set of q-cochains is denoted by $C^q(\fu , \ch)$ (see Appendix C).
In the considered case we have the sets of 0-cochains $C^0(\fu , \ch)$
and 1-cochains $C^1(\fu , \ch)$.

\smallskip

The set  $C^0(\fu , \ch)$ is a {\it group} under a
pointwise multiplication. For $h=\{h_1, h_2\}$, $f=\{f_1, f_2\}\in
C^0(\fu , \ch)$ we have
$$
hf=\left\{(hf)_1, (hf)_2\right\}:=\left\{h_1f_1, h_2f_2\right\},
\eqno(5.1)
$$
where $h_\a , f_\a\in\ch(\U_\a )\equiv\Gamma(\U_\a ,\ch),
\a =1,2.$  The set $C^1(\fu , \ch)$ of all 1-cochains forms
a {\it group} under the following operation:
if $h=\{h_{12}, h_{21}\}, f=\{f_{12}, f_{21}\} \in C^1(\fu , \ch)$,
then
$$
hf=\left\{(hf)_{12}, (hf)_{21}\right\}:=\{h_{12}f_{12}, h_{21}f_{21}\},
\eqno(5.2)
$$
where $h_{12}, h_{21}, f_{12}, f_{21}\in \ch(\U_{12})\equiv\Gamma(\U_{12},
\ch)$. Notice that $h_{12}$ and $h_{21}$ ($f_{12}$ and $f_{21}$) are
elements of two different groups $\ch(\U_{12}): \{h_{12}, h_{21}\}
\in\ch(\U_{12})\times\ch(\U_{12})$.

\smallskip

{}For the two-set open cover $\fu$,  sets of {\it 0-} and
{\it 1-cocycles} are defined by the formulae
$$
Z^0(\fu,\ch)=\left\{\{h_1,h_2\}\in C^0(\fu,\ch): h_1=h_2 \
\mbox{on}\ \U_{12}  \right\},
\eqno(5.3)
$$
$$
Z^1(\fu,\ch)=\left\{\{h_{12},h_{21}\}\in C^1(\fu,\ch):
h_{12}=h_{21}^{-1} \right\},
\eqno(5.4)
$$
and the space $Z^0(\fu,\ch)$ coincides with the {\it group}
$H^0(\rp,\ch)\equiv\Gamma (\rp,\ch)$ of global sections of the sheaf $\ch$.
The set $Z^1(\fu,\ch)$ is not a group for the non-Abelian sheaf $\ch$.

\smallskip

{}Finally, two cocycles $\cf, \hat\cf\in Z^1(\fu,\ch)$ are said to be
equivalent, $\hat\cf\sim\cf$, if
$$
\hat\cf_{12}=h_1\cf_{12} h_2^{-1},
\eqno(5.5)
$$ for some element $h=\{h_1,h_2\}\in C^0(\fu,\ch)$ restricted to
$\U_{12}$.  A
set of equivalence classes of 1-cocycles $\cf$ with respect to the
equivalence relation (5.5) is called a {\it \v{C}ech 1-cohomology set}
and denoted by $H^1(\fu, \ch)$.  In the general case we should take the
direct limit of these sets $H^1(\fu, \ch)$ over successive refinement
of cover $\fu$ of $\rp$ to obtain $H^1(\rp, \ch)$, the \v{C}ech
1-cohomology set of $\rp$ with coefficients in $\ch$. But in our case
$\U_1,\U_2$ are Stein manifolds and therefore  $H^1(\fu, \ch)=
H^1(\rp, \ch)$. The cohomology set $H^1(\rp, \ch)$ is identified with
the set of all holomorphic vector bundles over $\rp$ with the group
$SL(n,\C)$ which are considered up to equivalence (5.5), i.e., with
the {\it moduli space} of holomorphic vector bundles $E'$.

\subsection{Action of the group $C^0(\fu,\ch)$ on the space $Z^1(\fu,\ch)$}
\label{5.2}

Suppose that we are given a cover $\{\U_\g\}$ of the space $\rp$, $\g
=1,2,...$, and the groups $C^0(\{\U_\g\},\ch)$ and
$C^1(\{\U_\g\},\ch)$ of 0-cochains and 1-cochains. Let us define
the following action of the group $C^0$ on the group $C^1$
(automorphism $\s_0(h,.)$):
$$
\s_0(h,f)_{\a\b}=h_\b f_{\a\b}h_\b^{-1}\ \mbox{(no summation)},
\eqno(5.6)
$$
where $h=\{h_\a\}\in C^0(\{\U_\g\},\ch),
\ f=\{f_{\a\b}\}\in C^1(\{\U_\g\},\ch)$.
Now we can define a twisted homomorphism
$\d^0:\ C^0\to C^1$ of the group $C^0$ into the group $C^1$
by the formula~\cite{Oni}
$$
\d^0(h)_{\a\b}=h_\a h^{-1}_\b ,
\eqno(5.7a)
$$
where $\d^0(h)=\{\d^0(h)_{\a\b}\}\in C^1(\{\U_\g\},\ch)$. It is not
difficult to see that
$$
\d^0(hg)=\d^0(h)\s_0(h,\d^0(g)),
\eqno(5.7b)
$$
i.e., the homomorphism $\d^0$ is ``twisted" by $\s_0$. The twisted
homomorphism $\d^0$ permits one to define an action $\r_0$ of the group
$C^0$ on $C^1$ as on a set. The corresponding transformations act on
$C^1$ by the formula~\cite{Oni}
$$
\r_0(h,f)=\d^0(h)\s_0(h, f)\ \Leftrightarrow \
\r_0(h,f)_{\a\b}=h_\a f_{\a\b}h_\b^{-1}\ \mbox{(no summation)},
\eqno(5.8a)
$$
$$
\r_0(gh,f)=\r_0(g,\r_0(h,f)),
\eqno(5.8b)
$$
where $h,g\in C^0(\{\U_\g\},\ch),\ f\in C^1(\{\U_\g\},\ch)$.
Of course, in (5.6)-(5.8) it is implied that the components $h_\a$ of the
element $h\in C^0$ are restricted to $\U_{\a\b}$. It is not difficult to
verify that the action (5.8) preserves the space of 1-cocycles
$Z^1(\{\U_\g\},\ch) \subset C^1(\{\U_\g\},\ch)$.

\smallskip

{}For a two-set open cover $\fu=\{\U_1,\U_2\}$ of $\rp$, the action $\r_0$
of the group $C^0$ on the space $Z^1(\fu,\ch)$ of 1-cocycles has the form
$$
\r_0(h,\cf)_{12}=h_1\cf_{12}h_2^{-1},
\eqno(5.9)
$$
where $h\in C^0(\fu,\ch), \cf\in Z^1(\fu,\ch)$. As already said,
the action (5.9) (the special case of (5.8)) preserves the space $Z^1$, and
the quotient space $\r_0(C^0)\bl Z^1$ ($\r_0(C^0)$ acts on $Z^1$
on the left),
i.e., the space of orbits of the group $C^0$ in $Z^1$,
$$
H^1(\rp,\ch)=  H^1(\fu,\ch):= \r_0(C^0(\fu,\ch))\bl Z^1(\fu,\ch),
\eqno(5.10)
$$
is the \v{C}ech 1-cohomology set.

\subsection{Action of the group $C^1(\fu,\ch)$ on the space $Z^1(\fu,\ch)$}
\label{5.3}

{}For a two-set open cover $\fu$ of $\rp$ one may define an automorphism
$\s (h,.): C^1\to C^1, h\in C^1$, of the group of 1-cochains by
the formula
$$
\s (h,f)_{12} =h_{21}f_{12}h_{21}^{-1},
\eqno(5.11)
$$
where $h,f\in C^1(\fu,\ch)$, and a twisted homomorphism $\d : C^1\to C^1$
by the formula
$$
\d (h)=\{\d (h)_{12},\
\d (h)_{21}\}=\{h_{12}h_{21}^{-1}, h_{21}h_{12}^{-1}\},
\eqno(5.12)
$$
where $h,\d (h)\in C^1(\fu,\ch)$. With the help of the homomorphisms
$\s$ and $\d$
one can define the action of the group $C^1$ on itself as follows:
$$
\r (h,f)=\d (h)\s(h,f)\ \Leftrightarrow \
\r (h,f)_{12} =h_{12}f_{12}h_{21}^{-1},
\eqno(5.13a)
$$
$$
\r (gh,f)=\r (g,\r (h,f)),
\eqno(5.13b)
$$
where $g, h, f\in C^1(\fu,\ch)$. This action preserves the set
$Z^1(\fu,\ch)$ of 1-cocycles, and for a cocycle $\cf\in Z^1(\fu,\ch)$
we have
$$
\tilde \cf_{12}:=\r (h,\cf)_{12}= h_{12}\cf_{12}h_{21}^{-1}.
\eqno(5.14)
$$
It is easy to see that $\tilde\cf_{21}:=h_{21}\cf_{21}h_{12}^{-1}=
(h_{12}\cf_{12}h_{21}^{-1})^{-1}=\tilde\cf^{-1}_{12}$, i.e., $\tilde\cf$
is a 1-cocycle.

\smallskip

{}For $h=\{h_{12}, h_{21}\}\in  C^1(\fu,\ch)$, the matrices $h_{12},
h_{21} \in SL(n,\C)$ are arbitrary holomorphic matrix-valued functions on
$\U_{12}$ and therefore  with the help of the action (5.14) one can
obtain any cocycle from $Z^1(\fu,\ch)$. In other words, the action of
$C^1(\fu,\ch)$ on $Z^1(\fu,\ch)$  is transitive, and the set $Z^1$
can be identified with a homogeneous space $C^1/C^1_\t$,
$$
Z^1(\fu, \ch)=C^1(\fu, \ch)/C^1_\t (\fu, \ch),
\eqno(5.15a)
$$
where
$$
C^1_\t (\fu, \ch)= \left\{\{h_{12},h_{21}\}\in C^1(\fu, \ch):
h_{21}=h_{12}\right\}
\eqno(5.15b)
$$
is the stability subgroup of the trivial cocycle $\cf_{12}^0=1$.
The group $C^1_\t (\fu, \ch)$ is the kernel of the homomorphism (5.12).
Thus, the group $C^1(\fu, \ch)$ acts {\it transitively} on the space
$Z^1(\fu, \ch)$ of holomorphic bundles $E'$ over $\rp$.

\vskip 2mm

{\bf Remark.} The description of the group $C^1$ and of its action
on the space $Z^1$ of cocycles in terms of matrix-valued functions
depends on a cover of the space $\rp$. For a general system of local
trivializations with an open cover $\{\U_\g\}$, $\g\in I$, the
elements $\cf$ of $Z^1(\{\U_\g\}, \ch)$ must satisfy the conditions
$$
\cf_{\a\a}=1\ \mbox{(no summation)}\quad \mbox{on}\quad \U_\a ,
\quad \cf_{\b\a}=\cf^{-1}_{\a\b}
\quad \mbox{on}\quad \U_{\a\b}:=\U_\a\cap\U_\b ,
\eqno(5.16a)
$$
$$
\cf_{\a\b}\cf_{\b\g}\cf_{\g\a}=1\ \mbox{(no summation)}\quad
\mbox{on}\quad \U_{\a\b\g}:=
\U_\a\cap\U_\b\cap\U_\g\ne\varnothing.
\eqno(5.16b)
$$
Then $C^1(\{\U_\g\}, \ch)$ acts on $\cf\in Z^1(\{\U_\g\}, \ch)$
as follows:
$$
\cf_{\a\b}\ \mapsto \ \tilde\cf_{\a\b}:=\r (h, \cf )_{\a\b}=
h_{\a\b} \cf_{\a\b} h^{-1}_{\b\a}\ \mbox{(no summation)}.
\eqno(5.17)
$$
It is easily checked that the conditions (5.16a) for $\tilde\cf$ are
satisfied, and from the conditions (5.16b) imposed on $\tilde\cf_{\a\b}$
it follows that
$$
h_{\a\b}|_{\U_{\a\b\g}} = h_{\a\g}|_{\U_{\a\b\g}}.
\eqno(5.18)
$$
It simply means that $h_{\a\b}$ are defined on
$$
{\mathop{\cup}\limits_{\a,\b\in I}}\ \U_{\a\b},
\eqno(5.19)
$$
and we denote by $\bar C^1(\{\U_\g\}, \ch)$  the subgroup of all
elements $h=\{h_{\a\b}\}\in C^1(\{\U_\g\}, \ch)$ satisfying (5.18).
Thus, we obtain
$$
Z^1(\{\U_\g\}, \ch)=\bar C^1(\{\U_\g\}, \ch)/\bar C^1_\t (\{\U_\g\}, \ch),
\eqno(5.20a)
$$
where
$$
\bar C^1_\t (\{\U_\g\}, \ch)=\left\{\{h_{\a\b}\}\in
\bar C^1(\{\U_\g\}, \ch): h_{\b\a}=h_{\a\b}\right\}
\eqno(5.20b)
$$
is the stability subgroup of the trivial cocycle $\cf^0_{\a\b}=1$. For
a two-set open cover $\fu=\{\U_1,\U_2\}$ we have $\bar C^1(\fu, \ch)=
C^1(\fu, \ch)$.

\smallskip

It follows from the definitions that the groups $C^0(\fu, \ch)$ and
$C^1(\fu, \ch)$ are direct products
$$
C^0(\fu, \ch)=
\ch(\U_1)\times \ch(\U_2)\equiv \Gamma (\U_1, \ch)\times
\Gamma (\U_2, \ch)\ni\{h_1, h_2\},
\eqno(5.21a)
$$
$$
C^1(\fu, \ch)=
\ch(\U_{12})\times \ch(\U_{12})\equiv \Gamma (\U_{12}, \ch)\times
\Gamma (\U_{12}, \ch)\ni\{h_{12}, h_{21}\},
\eqno(5.21b)
$$
of the groups $\ch(\U_1), \ch(\U_2)$ and $\ch(\U_{12})$ of sections
over $\U_1, \U_2$ and $\U_{12}$ of the sheaf $\ch$ . Respectively,
$C^1_\t(\fu ,\ch)$ coincides with the diagonal subgroup in the group
$\ch(\U_{12})\times \ch(\U_{12})$, and  $Z^1(\fu,\ch)$ coincides with
the subset of elements $h=\{h_{12}, h_{12}^{-1}\}$ from the group
$C^1(\fu, \ch)$.

\smallskip

Collating formulae (5.10) and (5.15), we obtain that
$$
H^1(\rp,\ch)=\r_0(C^0)\setminus C^1/C^1_\t ,
\eqno(5.22)
$$
i.e., the moduli space of holomorphic bundles $E'$ over $\rp$ is
parametrized by the {\it double coset space} (5.22). It is not
difficult to see that the 1-cohomology set (5.22) is isomorphic to

\smallskip

(i) the set of $C^1_\t$-orbits in $Y^1:=\r_0(C^0)\setminus C^1$,

(ii) the set of $C^0$-orbits in $Z^1=C^1/C^1_\t $,

(iii) the set of $C^1$-orbits in $Y^1\times Z^1$,

\smallskip

\noindent where an action of $h\in C^1$ on $(y,z)\in Y^1\times Z^1$
is defined by the formula
$$
C^1\times (Y^1\times Z^1)\ni (h, (y,z)): \
(y,z)\mapsto (yh, \r(h^{-1}, z))\in Y^1\times Z^1.
$$

\smallskip

To sum up, for the {\it space} $Z^1(\fu,\ch)$ of holomorphic bundles
$E'$ over $\rp$, the group $C^1(\fu, \ch)$ of 1-cochains  for the
cover $\fu$ with values in the sheaf $\ch$ of non-Abelian groups acts
on the transition matrices $\cf_{12}$ of bundles $E'$ by the left
multiplication on matrices $h_{12}\in\ch(\U_{12})$ and by the right
multiplication on  matrices $h_{21}^{-1}\in\ch(\U_{12})$.  This group
acts on $Z^1$ transitively, and the space $Z^1$ is the {\it coset
space} (5.15a) (or (5.20a) for an arbitrary cover of $\rp$).  So
$C^1(\fu, \ch)$ is the symmetry group of the space of holomorphic
bundles $E'$ in the \v{C}ech approach. The {\it moduli space} $H^1(\rp,
\ch)$ of bundles $E'$ is the {\it double coset space} (5.22).

\subsection{The group $\fh(\rp)$ of automorphisms of the
            complex manifold $\rp$}
\label{ga}

Let  $X$ be a compact smooth manifold, $G$ a compact simple connected
Lie group  and $\ag$ a group  of automorphisms of the group $G$.
Consider the group $\m\,(X;G)$ of smooth maps from $X$ into $G$ and
the connected component of the unity $\m_0\,(X;G)$ of the group
$\m\,(X;G)$.  It is well-known that the group of automorphisms of the
group $\m_0\,(X;G)$ is a semidirect product $$ \di
(X)\ltimes\m\,(X;\ag) \eqno(5.23) $$ of the diffeomorphism group
$\di(X)$ of the manifold $X$ and the group of automorphisms
$\m\,(X;\ag)$ (for proof see \S~3.4 in ~\cite{Pres}).

\smallskip

As a set the space $Z^1(\fu,\ch)$ considered above coincides with the
group $\m\,(\U_{12}; SL(n,\C))$ of holomorphic maps from $\U_{12}$
into $SL(n,\C)$ and it is an analogue of the group $\m_0\,(X;G)$.
The group $C^1(\fu,\ch)$  acting on the space $Z^1(\fu,\ch)$ is
respectively an analogue of the group of automorphisms $\m\,(X;\ag)$.
It is clear that there should be an analogue  of the  diffeomorphism
group from (5.23), i.e., some group of transformations of the
coordinates of the space $\rp$ acting on the set $Z^1(\fu,\ch)$.

\smallskip

Remember that as a smooth manifold the twistor space is $\rp=U\times
S^2$. At the same time, $\rp$ is a complex 3-manifold, and in \S\,3.3 we
have introduced the complex coordinates $z_1:\ \U_1\to\C^3,\ z_2:\
\U_2\to\C^3$ on $\rp$ and the holomorphic transition function
$f_{12}$ connecting $z_1$ and $z_2$ on $\U_{12}$.  Let $\e :\
\rp\to\rp$ be an arbitrary transformation from the group $\di(\rp)$
of diffeomorphisms of the twistor space $\rp$.  Let us denote by
$\tilde\U_1:=\e (\U_1),\ \tilde\U_2:=\e (\U_2)$ the images of the
open sets $\U_1, \U_2$ in $\rp$. We have
$$
\e (\rp)=\e (\U_1 \cup \U_2)= \e (\U_1) \cup \e(\U_2)=
\tilde\U_1\cup\tilde\U_2,
\eqno(5.24a)
$$
$$
\e (\U_{12})=\e (\U_1 \cap \U_2)= \e (\U_1) \cap \e(\U_2)=
\tilde\U_1\cap\tilde\U_2,
\eqno(5.24b)
$$
since the map $\e$ is a bijection.

\smallskip

Let us consider the restriction of the map $\e$ to $\U_{12}$, i.e.,
the {\em local} diffeomorphism $\e\mid_{\U_{12}}: \U_{12}\to\rp$.  On
$\tilde\U_{12}=\e (\U_{12})$ one can always introduce complex
coordinates $\hat z_1:\ \tilde\U_{12}\to\C^3,\ \hat z_2:\
\tilde\U_{12}\to\C^3$ related by a holomorphic transition function
$\hat f_{12}$ such that the map $\e\mid_{\U_{12}}:\
\U_{12}\to\tilde\U_{12}$ will be holomorphic in the chosen
coordinates.  In other words, domains $\U_{12}$ and $\tilde\U_{12}$
are biholomorphic and there exist holomorphic functions $\e_1$,
$\e_2$ such that~\cite{Bou}
$$
\hat z^a_1\circ\e =\e^a_1(z^b_1),\
\hat z^a_2\circ\e =\e^a_2(z^b_2),\
\hat z^a_1=\hat f^a_{12}(\hat z^b_2).
\eqno(5.25)
$$
These maps form the (local) group $\fh(\U_{12})$.

\smallskip

Having the group $\fh(\U_{12})$ of local holomorphic maps
$\e\mid_{\U_{12}}:\ \U_{12}\to\rp$, one can define its action on
transition matrices $\cf$ of holomorphic bundles over $\rp$.
But in this connection the following questions arise:
\begin{enumerate}
\item Is it possible to introduce on $\tilde \U_1\cup\tilde\U_2$
      complex coordinates $\tilde z_1:\ \tilde\U_1\to\C^3,\
      \tilde z_2:\ \tilde\U_2\to\C^3$ related by a holomorphic
      transition function $\tilde f_{12}$?
\item  Can the coordinates $\hat z_1, \hat z_2$ on $\tilde\U_{12}$
       be extended to $\tilde \U_1$, $\tilde\U_2$ and will they
       be equivalent to the coordinates $\tilde z_1$, $\tilde z_2$?
\end{enumerate}

The diffeomorphism group $\di(\rp)$ acts not only on  transition
matrices of bundles $E'$ over $\rp$, but also on the complex
structure of the space $\rp$.  But a change of the complex
structure of the space $\rp$ leads to a change of the conformal
structure  and a metric on $U\subset\R^4$ by virtue of the twistor
correspondence ~\cite{Pen, AHS}. If we are interested in symmetries
of the SDYM equations on the space $U$ with a conformally flat metric,
then we have to consider only those diffeomorphisms $\e\in\di(\rp)$
which preserve the complex structure of $\rp$.  These maps
$\e\,:\rp\to\rp$ form the group of biholomorphic transformations of the
space $\rp$ which we shall denote by $\fh (\rp)$. It is a subgroup of
the diffeomorphism group:  $\fh (\rp)\subset \di(\rp)$.

\smallskip

In the coordinates $z_1, z_2, \tilde z_1, \tilde z_2$  transformations
$\e\in\fh (\rp)$ are defined by the holomorphic functions
$$
\tilde z_1^a\circ\e =\e^a_1(z_1),\
\tilde z_2^a\circ\e =\e^a_2(z_2),\
\tilde z_1^a =\tilde f_{12}^a(\tilde z_2^b).
\eqno(5.26)
$$ Formulae (5.26) are not always convenient because there the
coordinates $z_\a$ are calculated at points $p\in\rp$, and the
coordinates $\tilde z_\a$ are calculated at points $q=\e
(p)\in\rp$. It is often more convenient to define $\e$ by
transition functions $\e_{\a\b}$ from $z_\a$ to $\tilde z_\b$ in the
domains $\ \U_\a\cap\tilde\U_\b$ (if
$\U_\a\cap\tilde\U_\b\ne\varnothing$), then  $z_\a$ and $\tilde z_\b$
are calculated at the same points $p\in\U_\a\cap\tilde\U_\b$.  For
example, the conformal transformations (2.4) of the space $\R^4$ induce
such holomorphic transformations of coordinates $\{z^a_1\}\mapsto
\{\tilde z^a_1\}$ of the twistor space $\cz =\cz(\R^4)$ that on
$\U_1\cap\tilde\U_1$ we have

$${\mathrm {translations:}}\quad
\tilde z^1_1=z^1_1+a_1-\bar a_2z^3_1,\quad
\tilde z^2_1=z^2_1+a_2+\bar a_1z^3_1,\quad  \tilde z^3_1=z^3_1,
$$

$$
\begin{array}{l}
{\mbox{rotations}}\\
{{\mbox{induced by}\ }\{X_a\}}
\end{array}
:\quad
\left (
\begin{array}{c}
\tilde z^1_1\\
\tilde z^2_1\\
\tilde z^3_1
\end{array}
\right )=
\left (
\begin{array}{ccc}
c&d&0\\
-\bar d&\bar c&0\\
0&0&1
\end{array}
\right )
\left (
\begin{array}{c}
z^1_1\\
z^2_1\\
z^3_1
\end{array}
\right ),\quad
\left (
\begin{array}{cc}
c&d\\
-\bar d&\bar c
\end{array}
\right )
\in SU_L(2),
$$

$$
\begin{array}{l}
{\mbox{rotations}}\\
{{\mbox{induced by}\ }\{Y_a\}}
\end{array}
:\quad
\tilde z^1_1=\frac{z^1_1}{a-\bar b z^3_1},\
\tilde z^2_1=\frac{z^2_1}{a-\bar b z^3_1},\
\tilde z^3_1=\frac{b+\bar a z^3_1}{a-\bar b z^3_1},\
\left (
\begin{array}{cc}
a&b\\
-\bar b&\bar a
\end{array}
\right )
\in SU_R(2),
$$

$${\mathrm {dilatations:}}\quad \tilde z^1_1=e^\a z^1_1,\
\tilde z^2_1=e^\a z^2_1,\ \tilde z^3_1= z^3_1,
$$

$${{\mathrm {special\ conformal}}\atop {\mathrm {transformations}}}:\quad
\tilde z^1_1=\frac{z^1_1}{1+\a_1z^1_1+\a_2z^2_1},\
\tilde z^2_1=\frac{z^2_1}{1+\a_1z^1_1+\a_2z^2_1},\
\tilde z^3_1=\frac{z^3_1-\bar\a_1z^1_1+\bar\a_2z^2_1}
{1+\a_1z^1_1+\a_2z^2_1}.
$$\\
Here $a_1, a_2, \a_1, \a_2\in\C, \a\in\R$.

\subsection{Action of the group $\fh(\rp)$ on the space $Z^1(\fu,\ch)$}
\label{5.5}

Action of the group $\fh(\rp)$ of complex-analytic diffeomorphisms
of the space $\rp$ on transition matrices of holomorphic bundles $E'$
over $\rp$ is defined in the following  way. We consider a two-set open
cover $\fu =\{\U_1, \U_2\}$ of $\rp$ and a transition matrix
$\cf\in Z^1(\fu,\ch)$ of a bundle $E'$. After a transformation
$\fh(\rp)\ni\e : \rp\to\rp$ we have a new cover $\tilde\fu =
\{\tilde\U_1, \tilde\U_2\},\ \tilde\U_1=\e (\U_1),\ \tilde\U_2=\e (\U_2).$
Let us consider the common refinement both of the covers. Denote
$$
\hat\U_1:=\U_1\cap\tilde\U_1,\
\hat\U_2:=\U_1\cap\tilde\U_2,\
\hat\U_3:=\U_2\cap\tilde\U_1,\
\hat\U_4:=\U_2\cap\tilde\U_2,
\eqno(5.27a)
$$
to give the refined cover $\hat\fu =\{\hat\U_1, \hat\U_2, \hat\U_3,
\hat\U_4\}$.

\smallskip

The cocycle $\cf\in Z^1(\fu,\ch)$ induces the following 1-cocycle
$\hat\cf\in Z^1(\hat\fu,\ch)$:
$$
\hat\cf=\{
\hat\cf_{12}, \hat\cf_{13}, \hat\cf_{14}, \hat\cf_{23},
\hat\cf_{24}, \hat\cf_{34}\}
:=\{1, \cf_{12}(z^a_1), \cf_{12}(z^a_1),
\cf_{12}(z^a_1), \cf_{12}(z^a_1), 1\},
\eqno(5.27b)
$$
where $\hat\cf_{\a\b}$ is defined in $\hat\U_{\a\b}:=
\hat\U_{\a}\cap\hat\U_{\b}$  and
$$
\hat\U_{12}:=\hat\U_1\cap\hat\U_2=\U_1\cap\tilde\U_{12},\
\hat\U_{13}:=\hat\U_1\cap\hat\U_3=\U_{12}\cap\tilde\U_{1},\
\hat\U_{14}:=\hat\U_1\cap\hat\U_4=\U_{12}\cap\tilde\U_{12},\
$$
$$
\hat\U_{23}:=\hat\U_2\cap\hat\U_3=\U_{12}\cap\tilde\U_{12},\
\hat\U_{24}:=\hat\U_2\cap\hat\U_4=\U_{12}\cap\tilde\U_{2},\
\hat\U_{34}:=\hat\U_3\cap\hat\U_4=\U_2\cap\tilde\U_{12}.
\eqno(5.28)
$$
The cocycle $\hat\cf$ is equivalent to the cocycle $\cf$, and
the group $\fh(\rp)$ acts on $\cf\in Z^1(\fu,\ch)$ as follows:
$$
\fh(\rp)\ni\e :\ \cf\mapsto\hat\cf\mapsto\r (\e , \cf )\equiv
\hat\cf^\e =\{\hat\cf_{12}^\e , \hat\cf_{13}^\e , \hat\cf_{14}^\e ,
\hat\cf_{23}^\e , \hat\cf_{24}^\e , \hat\cf_{34}^\e \},
$$
$$
\hat\cf_{12}^\e :=1,\ \hat\cf_{13}^\e :=\cf_{12}(\e^a_1(z_1)),
\ \hat\cf_{14}^\e :=\cf_{12}(\e^a_1(z_1)),
$$
$$
\hat\cf_{23}^\e :=\cf_{12}(\e^a_1(z_1)),\
\hat\cf_{24}^\e :=\cf_{12}(f^a_{12}(\e^b_2(z_2))),\
\hat\cf_{34}^\e :=1.
\eqno(5.29)
$$
In the general case cocycles $\hat\cf$ and $\hat\cf^\e$ are not equivalent
and therefore  the group $\fh(\rp)$ of biholomorphic transformations
of the twistor space $\rp$ acts nontrivially on the space $Z^1(\fu,\ch)$.
This action includes refining of the cover and a transition to an equivalent
cocycle.

\smallskip

It is usually considered that elements $\e\in\fh(\rp)$ which are close
to the identity do not move the covering sets. That is, if $\e$ is close to
the identity, it is possible to define the action of such $\e\in
\fh(\rp)$  as follows:
$$
\r (\e , .):\ \cf_{12}\mapsto\r (\e ,\cf)_{12}=\cf^\e_{12}=\cf_{12}(\e_1(z_1)),
\eqno(5.30)
$$
i.e., without using the refined cover $\hat\fu$. In other words, the action
of a neighbourhood of unity of the group $\fh(\rp)$ maps $Z^1(\fu,\ch)$
into itself. In what follows we shall study just this case.

\smallskip

Returning to \S\,\ref{5.3} and to the beginning of \S\,\ref{ga}, we come to
the conclusion that the full group of continuous symmetries acting on
the space $Z^1(\fu,\ch)$ of holomorphic bundles $E'$ over $\rp$ is a
semidirect product
$$
\fh(\rp)\ltimes C^1(\fu,\ch)
\eqno(5.31)
$$
of the group $\fh(\rp)$ of holomorphic automorphisms of the
space $\rp$ and of the group  $C^1(\fu,\ch)$ of 1-cochains for the
cover $\fu$ with values in the  sheaf $\ch$ of
holomorphic maps of the space $\rp$ into the Lie group $SL(n,\C)$.

\section{Symmetries in holomorphic setting}
\label{6}

\subsection{Germs of sets and groups}
\label{6.1}

Let $\cb$ be a set with a marked point $e\in\cb$.
The element $e$ is called the unity . If $\cb$ and
$\cc$ are sets with the marked points which we denote
by the same letter $e$,
then a homomorphism of the set $\cb$ into the set $\cc$ is such
a map $\vp :\cb\to\cc$ that $\vp (e)=e$. The homomorphism $\cb\to\cc$
is said to be the isomorphism if it maps $\cb$ onto $\cc$ bijectively.
The set $\ker\vp=\vp^{-1}(e)$ with the marked point $e$ is called
the kernel of the homomorphism $\vp$.

\smallskip

Let $X$ be a set  with a marked point $e$, and let $Y_1, Y_2$ be
two subsets of the set $X$ also containing the point $e$.  The sets
$Y_1, Y_2$ are called {\it equivalent} at the point $e$ if there
exists such a neighbourhood $Y_3$ of this point that $Y_1\cap
Y_3=Y_2\cap Y_3$.  The class of all sets equivalent to the set $Y_1$
is called the {\it germ} of this set at the point $e$ and denoted by
$\bf Y$~\cite{GR}.  The sets $Y_1, Y_2, Y_3$ are {\it
representatives} of the germ $\bf Y$ of sets.

\smallskip

In Appendix A, a notion of {\it group germs} {\bg}~\cite{Hir}
based on the definition of germs of sets is introduced. {\it
Representatives} of the group germ {\bg} are {\it local groups},
i.e., open neighbourhoods $\cg$ of the identity $e\equiv 1$,  which
are closed under all group operations (multiplication, operation of
inverse etc). In particular, we shall consider the germs {\gc} and
{\gh} of the groups $C^1(\fu,\ch)$ and $\fh(\rp)$ described in
\S\,\ref{5}.

\subsection{Holomorphic triviality of bundles $E'$ on $\cp\hra\rp$}
\label{6.2}

Let us consider the twistor space $\cz\equiv\cz(\R^4)$ of $\R^4$ and
the moduli space $H^1(\cz,\ch)$ of holomorphic bundles $E'$ over the
space $\cz$. With the sheaf $\ch=\co^{SL(n,\C)}$ (of germs) of
holomorphic maps from $\cz$ into the group $SL(n,\C)$ one associates
the sheaf $\co^{sl(n,\C)}$ (of germs) of holomorphic maps from $\cz$
into the Lie algebra $sl(n,\C)$. The Abelian group (by addition) of
cohomologies $H^1(\cz, \co^{sl(n,\C)})$ of the space $\cz$ with
values in the sheaf $\co^{sl(n,\C)}$ parametrizes infinitesimal
deformations of the trivial bundle $E_0'=\cz\times\C^n$ and
$\mbox{dim}\,H^1(\cz, \co^{sl(n,\C)})=\infty$, i.e., in an
arbitrarily small neighbourhood of the trivial bundle $E_0'$ there
exists an infinite number of holomorphically nontrivial bundles $E'$.

\smallskip

Let us fix an arbitrary point $x_0\in\R^4$ and consider a real
projective line $\cx$ embedded into $\cz$. Now   we consider the
restriction $\co^{sl(n,\C)}_{x_0}:=\co^{sl(n,\C)}\!\!\mid_{\cx}$ of
the sheaf $\co^{sl(n,\C)}$ to $\cx$ and the cohomology  group
$H^1(\cx ,\co^{sl(n,\C)}_{x_0})$ parametrizing infinitesimal
deformations of the trivial holomorphic bundle
$E_{0x_0}':=\cx\times\C^n$. It is easily seen that
$$
H^1(\cx ,\co^{sl(n,\C)}_{x_0})=0,
\eqno(6.1)
$$
because $H^1(\C\bp^1,\co)=0$, where $\co$ is the sheaf of germs of
holomorphic functions on $\C\bp^1$. The equality (6.1) means that
there exists a sufficiently small open neighbourhood $U\subset\R^4$
of the point $x_0$ and an open subset $\cm\ni e$ of the
space $H^1(\rp,\ch)$, where $\rp\subset\cz$ is the twistor space
of $U$, such that for the bundles $E'$, representing points $[E']$
of the space $\cm\subset H^1(\rp,\ch)$,  their restriction $E_x'$
to $\cp\hra\rp$ will be holomorphically trivial for any $x\in U$
(version of the Kodaira theorem). In other words, small enough
deformations  do not change the trivializability of the bundle $E'$
over real projective lines in a neighbourhood of a given projective
line $\cx$ (for discussion see e.g.~\cite{WW, MW}).

\smallskip

Projective lines $\{\cp\}_{x\in U}$ form a family of complex
1-manifolds parametrized by $x\in U$, and $\cp$ coincides with
$\C\bp^1\times\{x\}$ in the direct product $\C\bp^1\times
U\simeq\rp$.  We consider holomorphic bundles $E'$ over $\rp$ with
transition matrices $\cf$ from $Z^1(\fu,\ch)$ and their restriction
to $\cp\hra\rp$, $x\in U$. Then a family of holomorphic maps
$\cf_{12}(x,\l)$ from $U\times\Omega_{12}$ into $SL(n,\C)$ determines
a family of vector bundles $E_x':=E'\mid_{\cp}$ over $\cp$, labelled
by the parameters $x\in U\subset\R^4$. In this family there exists a
marked family of holomorphically trivial bundles
$E_{0x}'=\cp\times\C^n$.  Finally, we have introduced an open subset
$\cm$ of the set $H^1(\rp,\ch)$ (being an open neighbourhood of the
marked point $e\in H^1(\rp,\ch)$) of moduli of those bundles $E'$
{}from $H^1(\rp,\ch)$, which are holomorphically trivial on
$\cp\hra\rp$ for all $x\in U$.  With each point $m\equiv [E']\in\cm$
one can associate a bundle $\ce_m:=E'(m)$ over $\rp$. Then we have a
{}family $\{\ce_m\}_{m\in\cm}$ of holomorphic bundles over $\rp$,
parametrized by $m\in\cm$.  The marked point in this family is the
trivial bundle $\ce_0:=E'(e)$ (the isomorphism class of the bundle
$E_0'$).

\smallskip

Let $X$ be a complex space. Consider a family of holomorphic
vector  bundles of rank $n$ with the base $X$ and a family  of complex
parameters $T$, i.e., a holomorphic vector bundle $\ce$ of rank $n$
over $X\times T$. The space $T$ is called the base of deformation.
{}For $t\in T$, we denote by $\ce_t$ a bundle over $X$ which is induced
by restriction of $\ce$ to $X\times\{t\}$ with a natural
identification $X\leftrightarrow X \times\{t\}$ ~\cite{Pal}.  In our
case, we have a holomorphic vector bundle $\ce$ of rank $n$ over
$\rp\times\cm$, $\cm\subset H^1(\rp,\ch)$.

\smallskip

Using the definitions of \S\,\ref{6.1}, one can consider sets
equivalent to the set $\cm$, and a class of all open subsets in
$H^1(\rp,\ch)$, equivalent to the set $\cm$, defines  the {\it germ}
{\bm} of this set at the point $e$. Of course, the notion of
equivalence is supplemented here by the demand that all {\it
representatives} $\cm, \cm',...$ of the germ {\bm} should be moduli
spaces of those bundles from $Z^1(\fu, \ch)$ which are
holomorphically trivial on $\cp$, $x\in U$. Let us stress that a
choice of a concrete representative $\cm, \cm',...$ of the germ {\bm}
is not essential since a different choice gives {\it equivalent
deformations} of the bundle $E_0'$.  That is why in the modern
deformation theory of complex spaces and holomorphic bundles as a
base of deformation one takes not a set with a marked point $e$ but
the germ of this set at the marked point (see e.g.~\cite{Pal}).

\smallskip

Now we take a point $m=[E']\in\cm$ and the transition matrix $\cf(m)\in
Z^1(\fu,\ch)$ in the bundle $E'$ representing this point. Acting on
$\cf(m)$ by all possible elements of $C^0(\fu,\ch)$ by formulae
(5.8), (5.9), we obtain an orbit $\r_0(C^0)(\cf(m))$ of the point
$\cf(m)\in Z^1(\fu,\ch)$ under the action $\r_0$ of the group
$C^0(\fu,\ch)$. This orbit coincides with the space
$\cc(\fu,\ch):=C^0(\fu,\ch)/H^0(\rp,\ch )$, and we denote
it by $\cc_m(\fu,\ch)$. Consider the union of orbits
$$
\cn = {\mathop{\cup}\limits_{m\in\cm}}\ \cc_m(\fu,\ch).
\eqno(6.2)
$$
The space $\cn\subset Z^1(\fu,\ch)$
is a bundle over $\cm$ associated with the principal fibre
bundle $P(\cm,C^0)$,
$$
\cn=P(\cm,C^0(\fu,\ch))\times_{C^0(\fu,\ch)} \cc (\fu,\ch),
\eqno(6.3)
$$
and  the group $C^0$ acts on $\cn$ on the left. The space $\cn$ is a
neighbourhood of the unity  $\cf^0=1$ in the space $Z^1(\fu,\ch)$.
We consider an open subset $\cn'\subset Z^1(\fu,\ch)$ equivalent to
$\cn$ and such that for all transition matrices $\cf$ from $\cn'$
there exists a solution of the Riemann-Hilbert problem (4.17) on
$\cp$ and $\cf^0\in\cn'$. Then  we can introduce the germ {\bn} of
the set $\cn$ at the point $\cf^0$ as a class of sets equivalent to
$\cn$.

\smallskip

The group $C^0(\fu,\ch)$ acts on any representative $\cn$ of the germ
{\bn},  and we have
$$
\cm=\r_0(C^0)\bl\cn,
\eqno(6.4)
$$
i.e.,  $\cm$ is a set of orbits of the group $C^0$ in the space $\cn$
(cf. (5.10)). By virtue of the Penrose-Ward correspondence described
in \S\,\ref{pwc}, there is a bijection between the space $\cm$ and
the moduli space of real-analytic solutions to the SDYM equations on
an open set $U\subset\R^4$ which are sufficiently close to the
trivial solution $A_0=0$.  A set of all such solutions is called the
space of {\it local solutions} (a small open neighbourhood of the
point $A_0=0$).  So, $\cm$ is bijective to the moduli space of local
solutions to the SDYM equations with the marked point $A_0=0$.
However, as a marked point in $Z^1(\fu,\ch)$ one can choose a
transition matrix $\hat\cf$ of a bundle $\hat E'$ over $\rp$,
holomorphically trivial on $\cx$, which corresponds to a solution
$\hat A$ of the SDYM equations. Then one can consider bundles
(trivial on $\cp$, $x\in U$) with transition matrices from an open
neighbourhood $\hat\cn\subset Z^1(\fu,\ch)$ of the point $\hat\cf$
and the moduli space $\hat\cm=\r_0(C^0)\bl\hat\cn$ of these bundles.
This space $\hat\cm$ will be bijective to the space of local
solutions to the SDYM equations that are near the solution $\hat A$.

\subsection{Jumping points and jumping lines}
\label{jpjl}

Let us consider a holomorphic bundle $E'$ over the twistor
space $\cz=L^{-1}\oplus L^{-1}\simeq\R^4\times\C\bp^1$ such that
its restriction to $\rp\subset\cz$ belongs to the space $\cn\subset
Z^1(\fu,\ch)$ introduced in \S\,\ref{6.2}. In general the bundle
$E'$ will be holomorphically trivial on real projective lines $\cp$
parametrized not by $x$ from $U$ but by $x$ from a ``wider"
open set $U'\supset U$. Those points $x$ from $\R^4$, for which
$E'|_{\cp}$ are not holomorphically trivial, are called the {\it jumping
points}, and projective lines $\cp$ corresponding to them  are called
the {\it jumping lines}. In the Ward construction the jumping points give
rise to singularities in the gauge potential $A$. The set $\R^4-U'$
of jumping points has codimension 1 (hypersurface) or more, i.e., the
set $U'$ is an open dense subset in $\R^4$. Lines $\cp$ with $x\in U'$
are called generic lines, and semi-stability of the bundle $E'$ is
equivalent to being  trivial on the generic line. For more details
see e.g. ~\cite{WW, MW}.

\smallskip

Now   we consider a holomorphic bundle $E''$ over $\cz$ such that its
restriction $E''|_{\rp}$ to $\rp$ belongs to $\cn$, and $E''$ is
nonequivalent to the bundle $E'$ considered above. So,  $E'|_{\rp}$
and $E''|_{\rp}$ correspond to different points from the moduli space
$\cm$. The bundle $E''$ will be holomorphically trivial on $\cp$ with $x$
{}from an open set $U''\supset U$ and in the general case $U'\ne U''$. In other
words, subsets of jumping points for the bundles $E'$ and $E''$ do not
coincide. At last, one can consider bundles $E_{inst}'$ over $\cz$
which have no jumping points in $\R^4\subset S^4$. The restriction
of $E_{inst}'$ to $\rp$ belongs to $\cn$, and instantons are parametrized
by a subset $\cn_{inst}$ in the set $\cn$.  It is clear that $\cn_{inst}
\subset\cn$ is a ``small" subset of $\cn$, and for a fixed topological
charge  the dimension of the moduli space $\cm_{inst}$ is finite.

\subsection{Representatives $\cm_0$ and $\cn_0$ of the
            germs {\bm} and {\bn}}
\label{rgmn}

In \S\,\ref{6.2} the germ {\bm} at the point $e$ of the set $\cm$
and the germ {\bn} at the point $\cf^0$ of the set $\cn$ have been
introduced. As an example, we shall describe some
representatives $\cm_0$ and $\cn_0$ of these germs using the standard
$\ve$-$\d$ language.

\smallskip

Consider the twistor space $\rp$ for an open ball $U=\{x\in\R^4:
(x-x_0)^2<r^2_0\}$  of the radius $r_0$ with a center at the point
$x_0\in\R^4$, the cover $\fu=\{\U_1,\U_2\}$ of $\rp$ and the
space  $Z^1(\fu,\ch)$ of holomorphic vector bundles over $\rp$.
{}For the cover $\fo =\{\Omega_1, \Omega_2\}$ of $\C\bp^1$ from
\S\,3.2, we consider the closure $\overline\Omega_{12}:=\{\l\in\C:
\a_1\le|\l|\le\a_2\}$ of the open set $\Omega_{12}= \Omega_1\cap\Omega_2$.
Let $\overline U$ be the closure of the open set $U$:
$\overline U=\{x\in\R^4: (x-x_0)^2\le r^2_0\}$. Then  the closure of
the open set $\U_{12}=U\times\Omega_{12}$ is
$$
\overline\U_{12}=\overline U\times\overline\Omega_{12},
\eqno(6.5)
$$
and $\overline\U_{12}$ is a compact subset of the set $\rp$.

\smallskip

We assume that matrix-valued transition functions $\cf_{12}$ of
bundles $E'$ are not only holomorphic on $\U_{12}$, but
also smooth on $\overline \U_{12}$. This mild assumption can be
replaced by the condition of holomorphy of $\cf_{12}$ in an open
$\d$-neighbourhood of the set $\overline \U_{12}$ with sufficiently
small $\d >0$ ~\cite{GR}.  Length $|\xi |$ of a vector $\xi
=(\xi_1,...,\xi_n)\in \C^n$ is given by the formula $|\xi
|^2=\mathop{\sum}\limits_i |\xi_i |^2= \mathop{\sum}\limits_i \xi_i
\overline\xi_i$. We consider complex $n\times n$ matrices
$A=(a_{ij})$ defining a linear transformation $A:  \xi\mapsto A\xi$.
For the matrices $A$ we define a norm $|A|$ by setting (see e.g.
~\cite{GR}):
$$ |A|:=\mathop{\max}\limits_{\xi\ne 0}\frac{|A\xi
|}{|\xi |}= \mathop{\max}\limits_{|\xi |=1} |A\xi | \eqno(6.6a)
$$
Now let us introduce a norm $\parallel\cdot\parallel$ on the space
$Z^1(\fu,\ch)$ setting
$$
\parallel\cf\parallel = \mathop{\max}\limits_{z_1\in\overline\U_{12}}
|\cf_{12}(z_1)| \eqno(6.6b)
$$
for $\cf\in Z^1(\fu,\ch)$.
Then $Z^1(\fu,\ch)$ turns into a topological space.

It follows from the equality (6.1) discussed in \S\,\ref{6.2}, that
there exists such a positive number $r_1(x)$ depending on
$x\in\overline U$ that the bundle $E_x'=E'|_{\cp}$ will
be holomorphically trivial if its transition matrix satisfies
the condition
$$
\mathop{\max}\limits_{\l\in\overline\Omega_{12}}
|\cf_{12}(x,\l )-1|< r_1(x).
\eqno(6.7a)
$$
The function $r_1(x): \overline U\to \R$ can always be chosen smooth.
It maps the compact space $\overline U$ into $\R$ and therefore
$$
r_1(x)\ge r_1:= \mathop{\min}\limits_{x\in\overline U} r_1(x),
\eqno(6.7b)
$$
i.e., it is bounded from below.
Moreover, one can always choose such a radius $r_0$ of an open ball
$U$ that $r_1$ will be positive: $r_1 > 0$.

\smallskip

We fix the radius $r_0$ of an open ball $U$
and consider all $\cf\in Z^1(\fu,\ch)$ such that
$$
\parallel\cf-1\parallel \equiv \mathop{\max}\limits_{z_1\in
\overline\U_{12}}|\cf_{12}(z_1)-1|<r_1,
\eqno(6.8)
$$
i.e., we consider the transition matrices $\cf\in Z^1(\fu,\ch)$
close to the identity in the norm (6.6b). By virtue of (6.7b), all
such transition matrices will satisfy the condition (6.7a)
for any $x\in\overline U$ and therefore  holomorphically
nontrivial bundles $E'$ over $\rp$, associated with them,
will be holomorphically trivial on $\cp\hra \rp$ for all
$x\in U$.

\smallskip

Notice that in the general case the action
(5.9) of the group $C^0(\fu,\ch)$ does not preserve the condition
(6.8) on $\cf\in Z^1(\fu,\ch)$, but it preserves the condition of
holomorphic triviality of bundles $E'$ on $\cp$. As such,
we can act by the group  $C^0(\fu,\ch)$ on the space of all $\cf$'s
satisfying inequality (6.8) and ``spread" this space over the
space $Z^1(\fu,\ch)$. As usual, two matrices $\cf$ and $\hat\cf$
satisfying the condition (6.8) are considered to be equivalent if
they are connected by formula (5.9). Factorizing the space
of all transition matrices satisfying (6.8) by this equivalence
relation, we get a moduli space $\cm_0$. The space $\cm_0$
is one of representatives of the germ {\bm} at the point
$e=[E_0']$ of the moduli space of holomorphic bundles introduced in
\S\,\ref{6.2}.

\smallskip

Now,  following \S\,\ref{6.2}, we introduce the space
$$
\cn_0=\mathop{\cup}\limits_{m\in \cm_0} \cc_m (\fu,\ch),
\eqno(6.9)
$$
obtained by the ``spread" of $\cf(m)$ over the space $Z^1(\fu,\ch)$
with the help of the action of the group $C^0(\fu,\ch)$. We have
(cf.(6.4))
$$
\cm_0=\r_0(C^0)\bl \cn_0,
\eqno(6.10)
$$
i.e., $\cm_0$ is the space of orbits of the group $C^0(\fu,\ch)$ in
the space $\cn_0$. The space $\cn_0$ is an open neighbourhood of
$\cf^0=1$ in the set $Z^1(\fu,\ch)$ and is one of representatives
of the germ {\bn} at the point $\cf^0=1$ of the space of holomorphic
bundles described in \S\,\ref{6.2}. So, for transition matrices
$\cf_{12}$ from $\cn_0$ the Birkhoff decomposition (4.17) exists for
all $x\in U$.

\subsection{Symmetries of local solutions in the \v{C}ech approach}
\label{6.5}

We consider the space  $Z^1(\fu,\ch)$ of holomorphic bundles
$E'$ over $\rp$ and the open subset $\cn$ in $Z^1(\fu,\ch)$
introduced in \S\,\ref{6.2}. In \S\S\,\ref{5.3}---\,\ref{5.5}
we have defined the group $\frg(\rp,\ch):=\fh(\rp)\ltimes C^1(\fu,\ch)$
and described its action $\r$ on the space  $Z^1(\fu,\ch)$.
This action, of course, does not map $\cn$ into itself
(or into another representative of the germ {\bn}), and one should
consider a {\it local action} of the group $\frg(\rp,\ch)$.

\smallskip

Let us consider an open neighbourhood $\fh$ of the identity  of the
group $\fh(\rp)$, an open  neighbourhood $\fc$ of the identity  of
the group $C^1(\fu,\ch)$ and an open  neighbourhood
$\frg:=\fh\ltimes\fc$ of the identity  of the group  $\frg(\rp,\ch)$.
As explained in the Appendix A and \S\,\ref{6.1}, the local groups
 $\fh$, $\fc$ and $\frg$ are representatives of the germs {\gh},
{\gc}, {\ef G} at the identity of the groups $\fh(\rp)$,
$C^1(\fu,\ch)$  and $\frg(\rp,\ch)$, respectively.  As local groups,
$\fh$, $\fc$ and $\frg$ are isomorphic to the groups $\fh(\rp)$,
$C^1(\fu,\ch)$  and  $\frg(\rp,\ch)$.

\smallskip

The above-mentioned representatives of the germs {\gh} and {\gc} can
always be chosen so that the local group $\frg$ will map the set
$\cn$ into itself. In more detail, there exists a subset $\cn'$ of
the set $\cn$ ($\cn'$ is another representative of the germ {\bn})
such that we have a map $\r:\frg\times\cn'\to\cn$. The map
$\r:\cn''\to\cn$, where $\cn''=\{(a,\cf)\in \frg\times\cn: \r
(a,\cf)\in\cn\}$ is an open subset in $\frg\times\cn$ containing
$\{e\}\times\cn$, is also defined.  In this case, the properties $\r
(e, \cf)=\cf$, $\r (a, \r (b,\cf)) =\r (ab, \cf)$ etc. are fulfilled
{}for all $(a,\cf)\in\cn''$.  In particular, the local group $\fh$ of
biholomorphisms acts on the space $\cn$ by formula (5.30) from
\S\,\ref{5.5}.

\smallskip

{}For the matrix local group $\fc$ we introduce the diagonal subgroup
$$
\fc_{\t}:=\fc\cap C^1_{\t}(\fu, \ch),
\eqno(6.11)
$$
which is the local stability subgroup of the marked cocycle
$\cf^0\in\cn$. For the definition of the group $C^1_{\t}(\fu, \ch)$
see (5.15b). Then,  by repeating all the arguments of \S\,\ref{5.3}
in terms of the local groups, we have
$$
\cn\simeq\fc/\fc_{\t},
\eqno(6.12a)
$$
i.e., $\cn$ is a coset space. In other words, for each representative
$\cn$ of the germ {\bn} of the space of bundles, holomorphically trivial
on $\cp\hra\rp$, one can always choose a representative $\fc$
of the germ {\gc} of the group of 1-cochains such that (6.12a) will
take place. In fact, (6.12a) is a consequence of an isomorphism of
germs
$$
\mbox{\bn}\simeq \mbox{\gc}/\mbox{\gc}_{\t}.
\eqno(6.12b)
$$

\smallskip

Combining (6.12a) and (6.4), we obtain
$$
\cm\simeq\r_0(C^0)\bl\fc/\fc_{\t},
\eqno(6.13a)
$$
i.e., the moduli space of local solutions to the SDYM equations
is a {\it double coset space}. Again, (6.13a) is a consequence of the
isomorphism of germs
$$
\mbox{\bm}\simeq {\bf \r_0(C^0)\bl}\mbox{\gc}/\mbox{\gc}_{\t}.
\eqno(6.13b)
$$

\smallskip

Thus, the {\it full group of continuous symmetries} acting on the
space $\cn$   is a semidirect product
$$
\frg=\fh\ltimes\fc
\eqno(6.14)
$$
of the local group $\fh$ of holomorphic automorphisms of the space
$\rp$ and of the local group $\fc$ of 1-cochains of the cover $\fu$
with values in the sheaf $\ch=\co^{SL(n,\C)}$ of
holomorphic maps of the space $\rp$ into the group $SL(n,\C)$.

\subsection{Unitarity conditions}
\label{6.6}

As it was discussed in \S\,\ref{4.4}, the transition matrices
$\cf_{12}$ in holomorphic bundles $E'\to\rp$ which are compatible
with the real structure $\tau$ on $\rp$ have to satisfy the
additional condition (4.16b). Denote by
$$
Z^1_{\tau}(\fu,\ch):=\left\{\cf\in  Z^1(\fu,\ch):
\cf^{\dagger}_{12}(\tau(\bar z_1))=\cf_{12}(z_1)\right\}
\eqno(6.15)
$$
a subset of transition matrices satisfying this unitarity conditions.

\smallskip

We should next define subgroups $C^0_{\tau}$ in $C^0$ and  $C^1_{\tau}$
in $C^1$ such that their action, described by formulae (5.9) and (5.14),
will preserve $Z^1_{\tau}(\fu,\ch)$. It is not hard to see that
$$
C^0_{\tau}(\fu,\ch)=\left\{ \{h_1,h_2\}\in C^0(\fu,\ch):
h^{\dagger}_1 (\tau(\bar z_1))=h_2^{-1}(z_1)\right\},
\eqno(6.16)
$$
$$
C^1_{\tau}(\fu,\ch)=\left\{ \{h_{12},h_{21}\}\in C^1(\fu,\ch):
h^{\dagger}_{12} (\tau(\bar z_1))=h_{21}^{-1}(z_1)\right\}.
\eqno(6.17)
$$
Actions of these groups on $Z^1_{\tau}$ have the form
$$
\cf_{12}\mapsto \hat\cf_{12}(z_1):=\r_0(h,\cf)_{12}=
h_1(z_1)\cf_{12}(z_1)h^{\dagger}_1(\tau (\bar z_1)),
\quad h\in C^0_{\tau},
\eqno(6.18)
$$
$$
\cf_{12}\mapsto \tilde\cf_{12}(z_1):=\r (h,\cf)_{12}=
h_{12}(z_1)\cf_{12}(z_1)h^{\dagger}_{12}(\tau (\bar z_1)),
\quad h\in C^1_{\tau}.
\eqno(6.19)
$$
By the definitions (6.16) and (6.17), $C^0_{\tau}(\fu,\ch)$ and
$C^1_{\tau}(\fu,\ch)$ are real subgroups in $C^0(\fu,\ch)$
and $C^1(\fu,\ch)$, respectively.

\smallskip

The cocycles $\cf_{12}$ and $\hat\cf_{12}$ from (6.18) define
equivalent bundles $E'\sim \hat E'$, $\cf_{12} \sim\hat\cf_{12}$,
and one can introduce a 1-cohomology set $H^1_{\tau}(\fu,\ch)$
as a set of orbits of the group $\r_0( C^0_{\tau})$ in the space
$Z^1_{\tau}(\fu,\ch)$ of transition matrices compatible with the
real structure $\tau$ on $\rp$,
$$
H^1_{\tau}(\fu,\ch):=\r_0( C^0_{\tau}(\fu,\ch))\bl Z^1_{\tau}(\fu,\ch)
\subset  H^1(\fu,\ch).
\eqno(6.20)
$$
\noindent For the cover $\fu=\{\U_1, \U_2\}$
we have $H^1_{\tau}(\rp,\ch)=H^1_{\tau}(\fu,\ch)$. So, the real
structure $\tau$ on $\rp$ induces a real structure on
$H^1(\rp,\ch)$, and $H^1_{\tau}(\rp,\ch)$ is a set of real
``points" of the space $H^1(\rp,\ch)$ corresponding to the bundles
$E'$ with the unitary structure (6.15).

\smallskip

Consider the action of the group $C^1_{\tau}$ on $Z^1_{\tau}$. As a
stability subgroup of the element $\cf^0=1$ compatible with the
real structure we have the group
$$
C^1_{\tau\t}:=C^1_{\tau}\cap C^1_{\t}=\left\{\{h_{12}, h_{21}\}\in
C^1_{\tau}(\fu,\ch): h_{12}=h_{21}\right\},
\eqno(6.21)
$$
and the space $Z^1_{\tau}(\fu,\ch)$ can be identified with the
quotient space
$$
Z^1_{\tau}(\fu,\ch)=C^1_{\tau}(\fu,\ch)/C^1_{\tau\t}(\fu,\ch).
\eqno(6.22)
$$
The moduli space $H^1_{\tau}(\rp,\ch)$ of holomorphic bundles
$E'$ with the unitary structure coincides with the double coset space
$$
H^1_{\tau}:=\r_0(C^0_{\tau})\bl C^1_{\tau}/C^1_{\tau\t},
\eqno(6.23)
$$
and this set is isomorphic to (i) the set of $C^1_{\tau\t}$-orbits in
$Y^1_{\tau}:=\r_0(C^0_{\tau})\bl C^1_{\tau}$, (ii) the set of
$C^0_{\tau}$-orbits in $Z^1_{\tau}$, (iii) the set of $C^1_{\tau}$-orbits
in $Y^1_{\tau}\times Z^1_{\tau}$.

\smallskip

As to the group $\fh(\rp)$, the action of which on $Z^1(\fu,\ch)$
was described
in \S\,\ref{5.5}, one should choose in it a subgroup $\fh_{\tau}(\rp)$
of those transformations $\eta\in\fh(\rp)$ which are compatible with the
real structure $\tau$ on $\rp$. In terms of the functions $\e_1$ and $\e_2$
from (5.26) representing $\e$ in the chosen coordinates it means that
$$
\overline{\e^a_1(\tau (\bar z_1))}=B^a_b\e^b_2(z_2),
\eqno(6.24)
$$
 where the coefficients $B^a_b$ are written down in (3.21).
Thus, the symmetry group acting on the space $Z^1_{\tau}(\fu,\ch)$
of holomorphic bundles $E'$ satisfying the unitarity conditions is the group
$$
\frg_{\tau}(\rp,\ch):=\fh_{\tau}(\rp)\ltimes C^1_{\tau}(\fu,\ch).
\eqno(6.25)
$$
This group is a real subgroup in the group (5.31).

\smallskip

{}Further, going over to local solutions, we introduce a subset
$\cn_{\tau}$ of those transition matrices from $\cn$ which
satisfy the condition (4.16b), i.e., $\cn_{\tau}:=\cn\cap
Z^1_{\tau}(\fu,\ch)$. One analogously introduces the moduli space
$\cm_{\tau}:=\cm\cap H^1_{\tau}(\rp,\ch)$, the real local groups
$\fh_{\tau}:=\fh\cap\fh_{\tau}(\rp)$, $\fc_{\tau}:=\fc\cap
C^1_{\tau}(\fu,\ch)$ and the germs {\gh}$_{\tau}$, {\gc}$_{\tau}$
corresponding to them. Then  one obtains isomorphisms
$$
\cn_{\tau}\simeq \fc_{\tau}/\fc_{\tau\t}, \quad
\cm_{\tau}\simeq \r_0(C^0_{\tau})\bl \fc_{\tau}/\fc_{\tau\t},
\eqno(6.26)
$$
corresponding to the isomorphisms (6.12), (6.13). At last, as the
symmetry group of the space of real local solutions in the \v{C}ech
approach one gets the local group
$$
\frg_{\tau}=\fh_{\tau}\ltimes\fc_{\tau},
\eqno(6.27)
$$
which is a semidirect product of the local groups $\fh_{\tau}$ and
$\fc_{\tau}$.

\section{Holomorphic bundles: the Dolbeault description}
\label{7}

\subsection{Some definitions}
\label{7.1}

The well-known  Dolbeault theorem reduces a computation of cohomology
spaces of a manifold $X$ with the coefficients in a sheaf of germs of
holomorphic maps from $X$ into a complex {\it Abelian} group $\bt$ to
problems of calculus of $\bt$-valued differential forms of the type
(0,q) on the manifold $X$ (isomorphism between \v{C}ech and Dolbeault
cohomology groups)~\cite{GH}. We want to
describe an analogue of the Dolbeault theorem for the
sheaf $\ch$ of germs of holomorphic maps of the space $\rp$ into the
non-Abelian group $SL(n,\C)$, following mainly the papers ~\cite{Oni}.
This will
permit us to describe symmetries of the space of local solutions to
the SDYM equations on $U\subset\R^4$.  But first, let us recall some
definitions for objects which will be considered below.

\smallskip

Let $K$ be a sheaf of groups and $\ca$ a sheaf of sets on $X$.  We
shall say that $K$ acts on $\ca$ if for any $x\in X$ the group $K_x$
acts on $\ca_x$, and also this action is continuous in the topology
of the sheaves $K$ and $\ca$. It is said that $K$  transitively acts on
$\ca$, if $K_x$ transitively acts on $\ca_x$ for each $x\in X$. In
this case $\ca$ can be identified with a quotient sheaf $K/K'$,
where $K'$ is a sheaf of  stability subgroups $K_x'$ and
stalks of the sheaf $K/K'$  are quotient  spaces $K_x/K_x'$.
Conversely, if $K'$ is a subsheaf of subgroups in $K$, the sheaf
$K/K'$ can be considered as a sheaf of  sets with marked section
$x\mapsto K_x'$, $x\in X$, on which $K$ transitively acts on the
left.

\subsection{The sheaves $\hat\cs$, $\hat\cb^{0,q}$ and $\hat\cb$}
\label{7.2}

Consider the sheaf $\hat\cs$ of germs of {\it smooth} maps from $\rp$
into the group $SL(n,\C)$. The sheaf $\ch$ of germs of {\it
holomorphic} maps $\rp\to SL(n,\C)$ is a subsheaf of the sheaf
$\hat\cs$, and there exists a canonical embedding $i: \ch\to\hat\cs$.
Consider also the sheaf $\hat\cb^{0,q}$ ($q=1,2,...$) of germs of {\it
smooth}  (0,q)-forms on $\rp$ with values in the Lie algebra
$sl(n,\C)$. Let us define a map $\bar\delta^0:
\hat\cs\to\hat\cb^{0,1}$ given for any open set $\U$ of the space
$\rp$ by the formula
$$
\bar\d^0 \hat\psi = -(\bar\p \hat\psi )\hat\psi^{-1},
\eqno(7.1)
$$
where $\hat\psi\in\hat\cs(\U)$, $\bar\d^0 \hat\psi \in \hat\cb^{0,1}(\U)$,
$d=\p +\bar\p$. Let us also introduce an operator $\bar\d^1:  \hat\cb^{0,1}
\to\hat\cb^{0,2}$, defined for any open set $\U\subset\rp$ by the formula
$$
\bar\d^1\hat B =\bar\p\hat B + \hat B\wedge\hat B,
\eqno(7.2)
$$
where $\hat B\in \hat\cb^{0,1}(\U)$, $\bar\d^1\hat B\in
\hat\cb^{0,2}(\U)$. In other words, the maps of sheaves $\bar\d^0:
\hat\cs\to \hat\cb^{0,1}$ and $\bar\d^1: \hat\cb^{0,1} \to\hat\cb^{0,2}$
are defined by means of localizations. In particular, on $\U_1\subset\rp$
we have
$$
(\bar\d^0 \hat\psi_1)_a = -(\bar V^{(1)}_a \hat\psi_1)\hat\psi_1^{-1}
\eqno(7.1')
$$
$$
(\bar\d^1\hat B^{(1)})_{ab}= \bar V^{(1)}_a\hat B^{(1)}_b -
\bar V^{(1)}_b\hat B^{(1)}_a + [\hat B^{(1)}_a, \hat B^{(1)}_b].
\eqno(7.2')
$$
The sheaf $\hat\cs$ acts on the sheaves $\hat\cb^{0,q}$ ($q=1,2,...$)
with the help of the adjoint representation.  In particular, for any
open set $\U\subset\rp$ we have
$$
\hat B\mapsto \mbox{Ad}(\hat\psi, \hat B)=
\hat\psi^{-1}\hat B \hat\psi + \hat\psi^{-1}\bar\p\hat\psi ,
\eqno(7.3a)
$$
$$
\hat F\mapsto \mbox{Ad}(\hat\psi, \hat F)=
\hat\psi^{-1}\hat F \hat\psi ,
\eqno(7.3b)
$$
where $\hat\psi\in\hat\cs(\U)$, $\hat B\in \hat\cb^{0,1}(\U)$,
$\hat F\in \hat\cb^{0,2}(\U)$.

\smallskip

Denote by $\hat\cb$ the subsheaf in  $\hat\cb^{0,1}$ consisting of
germs of (0,1)-forms $\hat B$ with values in $sl(n,\C)$ such that
$\bar\d^1\hat B=0$, i.e., sections $\hat B$ over any open set $\U$
of the sheaf $\hat\cb=\ker\bar\d^1$ satisfy the equations
$$
\bar\p\hat B + \hat B\wedge\hat B=0,
\eqno(7.4)
$$
where $\hat B\in \hat\cb^{0,1}(\U)$. So the sheaf $\hat\cb$ can be identified
with the sheaf of (0,1)-connections $\bar\p_{\hat B}=\bar\p+\hat B$ in
the holomorphic bundle $E'$ over $\rp$.

\subsection{The sheaves $\cs$, $\cb^{0,q}$ and $\cb$}
\label{7.3}

Recall that $\rp$ is the fibre bundle with fibres $\cp$ over  the
points $x$ from $U\subset\R^4$, and the canonical projection $\pi :
\rp\to U$ is defined. The typical fibre $\C\bp^1$ has the $SU(2)$-invariant
complex structure $\fj$ (see \S\,3.2), and the vertical distribution
$V=\ker\pi_*$ inherits this complex structure. A restriction of $V$
to each fibre $\cp$, $x\in U$, is the tangent bundle to that fibre.
The (flat) Levi-Civita connection on $U$ generates the splitting of the
tangent bundle $T(\rp)$ into a direct sum
$$
T(\rp)=V\oplus H
\eqno(7.5)
$$
of the vertical distribution $V$ and the horizontal distribution $H$.

\smallskip

Using the complex structures $\fj$, $J$ and $\cj$ on $\C\bp^1$, $U$
and $\rp$ respectively, one can split the complexified tangent bundle
of $\rp$ into a direct sum
$$
T^{\C}(\rp)=(V^{1,0}\oplus H^{1,0})\oplus (V^{0,1}\oplus H^{0,1})
\eqno(7.6)
$$
of vectors of type (1,0) and (0,1). So we have the integrable
distribution $V^{0,1}$ of antiholomorphic vector fields with
the basis $\bar V_3^{(1)}=\p_{\bar\l}$ on $\U_1\subset\rp$ and
$\bar V_3^{(2)}=\p_{\bar\zeta}$ on $\U_2\subset\rp$. The vector fields
(3.4a), (3.4b) and (3.11a), (3.11b) form a basis in the {\it normal}
{\it bundle} $H^{0,1}$ of a line $\cp\hra\rp$.

\smallskip

Having the canonical distribution $V^{0,1}$ on the space $\rp$,
we introduce the sheaf $\cs$ of germs of {\it partially holomorphic}
maps $\psi : \rp\to SL(n,\C)$, which are annihilated by vector
{}fields from $V^{0,1}$.  In other words, sections of the sheaf $\cs$
over open subsets $\U\subset\rp$ are $SL(n,\C)$-valued functions $\psi$
on $\U$, which satisfy the equations
$$
\p_{\bar\l}\psi =0 \ \ \mbox{on}\ \ \U\cap\U_1,\quad
\p_{\bar\zeta}\psi =0 \ \ \mbox{on}\ \ \U\cap\U_2,
\eqno(7.7)
$$
i.e., they are holomorphic along $\cp\hra\rp$, $x\in U$.
It is obvious that the sheaf $\ch$ of holomorphic
maps from $\rp$ into $SL(n,\C)$, i.e., smooth maps which are
annihilated by vector fields from $V^{0,1}\oplus H^{0,1}$,
is a subsheaf of $\cs$ and $\cs$ is a subsheaf of $\hat\cs$.

\smallskip

Consider now the sheaves $\hat\cb^{0,q}$, introduced in \S\,\ref{7.2}.
Let $\cb^{0,1}$ be the subsheaf of (0,1)-forms from $\hat\cb^{0,1}$
vanishing on the distribution $V^{0,1}$. In components this means that
for any open set $\U\subset\rp$
$$
B^{(1)}_3=0 \ \ \mbox{on}\ \ \U\cap\U_1,\quad
B^{(2)}_3=0 \ \  \mbox{on}\ \ \U\cap\U_2,
\eqno(7.8)
$$
where $B^{(1)}$ belongs to the section of the sheaf $\cb^{0,1}$ over $\U_1$,
and $B^{(2)}$ belongs to the section of the sheaf $\cb^{0,1}$ over $\U_2$.
So $\cb^{0,1}$  is the subsheaf of  $\hat\cb^{0,1}$.

The map $\bar\d^0$, introduced in \S\,\ref{7.2}, induces a map
$\bar\d^0: \cs\to\cb^{0,1}$, defined for any open set $\U$ of the
space $\rp$ by the formula
$$
\bar\d^0\psi = - (\bar\p\psi )\psi^{-1},
\eqno(7.9a)
$$
where $\psi\in \cs(\U)$, $\bar\d^0\psi\in \cb^{0,1}(\U)$. Analogously,
the operator $\bar\d^1$ induces a map $\bar\d^1:  \cb^{0,1}\to
\hat\cb^{0,2}$, given for any open set $\U\subset\rp$ by the formula
$$
\bar\d^1 B=\bar\p B+B\wedge B,
\eqno(7.10a)
$$
where $B\in \cb^{0,1}(\U)$,  $\bar\d^1 B\in\hat\cb^{0,2}(\U)$.
In particular, on $\U_1\subset\rp$ we have
$$
\bar\d^0\psi_1=-\left\{(\bar V^{(1)}_a\psi_1)\psi_1^{-1}\right\}
\bar\theta_{(1)}^a ,
\eqno(7.9b)
$$
$$
\bar\d^1B^{(1)}=\frac{1}{2}\left\{\bar V^{(1)}_a B^{(1)}_b-
\bar V^{(1)}_b B^{(1)}_a + [B^{(1)}_a, B^{(1)}_b]\right\}
\bar\theta_{(1)}^a\wedge \bar\theta_{(1)}^b,
\eqno(7.10b)
$$
where the (0,1)-forms $\{\bar\theta_{1,2}^a\}$ were introduced in
\S\,\ref{3.3}, $\psi_1\in \cs(\U_1)$, $B^{(1)}\in \cb^{0,1}(\U_1)$.

The sheaf $\cs$ acts on the sheaves $\cb^{0,1}$ and $\hat\cb^{0,q}$
by means of the adjoint representation. In particular, for $\cb^{0,1}$
and $\hat\cb^{0,2}$ we have the same formulae (7.3) with replacement
$\hat\psi$ by $\psi\in \cs(\U)$,
$$
B\mapsto\mbox{Ad}(\psi, B)=
\psi^{-1}B\psi + \psi^{-1}\bar\p\psi ,
\eqno(7.11a)
$$
$$
\hat F\mapsto \mbox{Ad}(\psi,\hat F)=\psi^{-1}\hat F\psi ,
\eqno(7.11b)
$$
where $B\in\cb^{0,1}(\U)$, $\hat F\in \hat\cb^{0,2}(\U)$.

At last, let us denote by $\cb$ the subsheaf of $\cb^{0,1}$
consisting of germs of $sl(n,\C)$-valued (0,1)-forms
$B$ such that $\bar\d^1B=0$, i.e., sections $B$ of the sheaf
$\cb=\ker\bar\d^1$ satisfy the equations
$$
\bar\p B+B\wedge B=0.
\eqno(7.12a)
$$
In components for $B\in\cb^{0,1}(\U_1)$ on the open set $\U_1$
eqs.(7.12a) have the form
$$
\bar V^{(1)}_1B^{(1)}_2 - \bar V^{(1)}_2B^{(1)}_1+
[B^{(1)}_1,B^{(1)}_2]=0,\quad
\bar V^{(1)}_3B^{(1)}_1=0,\quad
\bar V^{(1)}_3B^{(1)}_2=0,
\eqno(7.12b)
$$
since $B^{(1)}_3=0$. We have analogous equations on $\U_2\subset\rp$.

\subsection{Exact sequences of sheaves}
\label{7.4}

Let us consider the sheaves $\hat\cs$, $\hat\cb^{0,1}$ and
$\hat\cb^{0,2}$.
The triple $\{\hat\cs, \hat\cb^{0,1}, \hat\cb^{0,2}\}$ with the maps
$\bar\d^0$ and $\bar\d^1$ is a resolution of the sheaf $\ch$, i.e.,
the sequence of sheaves
$$
{\bf 1}\lra\ch\stackrel{i}{\lra}\hat\cs\stackrel{\bar\d^0}{\lra}
\hat\cb^{0,1}
\stackrel{\bar\d^1}{\lra}\hat\cb^{0,2},
\eqno(7.13)
$$
where $i$ is an embedding, is exact.  For proof see~\cite{Oni}.
Restricting $\bar\d^0$  to $\cs\subset\hat\cs$
and $\bar\d^1$  to $\cb^{0,1}\subset\hat\cb^{0,1}$,
we obtain the exact sequence of sheaves
$$
{\bf 1}\lra\ch\stackrel{i}{\lra}\cs\stackrel{\bar\d^0}{\lra}\cb^{0,1}
\stackrel{\bar\d^1}{\lra}\hat\cb^{0,2},
\eqno(7.14)
$$
where ${\bf 1}$ is the identity of the sheaf $\ch$.

\smallskip

By virtue of the exactness of the sequence  (7.13), we have
$$
\bar\d^0\hat\cs= \ker\bar\d^1=\hat\cb .
\eqno(7.15a)
$$
Since $\bar\d^0$ is the projection, connected with the action (7.3a)
of the sheaf $\hat\cs$ on  $\hat\cb^{0,1}$, the sheaf $\hat\cs$  acts
transitively with the help of Ad on $\hat\cb$ and
$\hat\cb\simeq\hat\cs/\ch$.  Thus, we obtain the exact sequence of
sheaves
$$
{\bf 1}\lra\ch\stackrel{i}{\lra}\hat\cs\stackrel{\bar\d^0}{\lra}
\hat\cb\stackrel{\bar\d^1}{\lra}0.
\eqno(7.15b)
$$
{}For more details see~\cite{Oni}. Restricting the map
$\bar\d^0$ to $\cs$ and $\bar\d^1$  to $\cb$, we obtain the exact
sequence of sheaves
$$
{\bf 1}\lra\ch\stackrel{i}{\lra}\cs\stackrel{\bar\d^0}{\lra}
\cb\stackrel{\bar\d^1}{\lra}0,
\eqno(7.16)
$$
since $\bar\d^0\cs=\ker\bar\d^1$ (the exactness of the sequence (7.14)),
and $\cs$ acts on $\cb$ transitively ($\cb\simeq \cs/\ch$). For
sections of the sheaf $\cb$ over $\U_1$ and $\U_2$ we have
$$
B^{(1)}_1=-(\bar V^{(1)}_1\psi_1)\psi^{-1}_1,\quad
B^{(1)}_2=-(\bar V^{(1)}_2\psi_1)\psi^{-1}_1,\quad
B^{(1)}_3=-(\bar V^{(1)}_3\psi_1)\psi^{-1}_1\equiv 0,
\eqno(7.17a)
$$
$$
B^{(2)}_1=-(\bar V^{(2)}_1\psi_2)\psi^{-1}_2,\quad
B^{(2)}_2=-(\bar V^{(2)}_2\psi_2)\psi^{-1}_2,\quad
B^{(2)}_3=-(\bar V^{(2)}_3\psi_2)\psi^{-1}_2\equiv 0,
\eqno(7.17b)
$$
where $\psi_{1,2}\in \cs(\U_{1,2})$, $B^{(1,2)}\in\cb(\U_{1,2})$.

\subsection{The group $H^0(\rp, \cs)$ and the cohomology set
            $H^1(\rp, \cs)$}
\label{7.5}

Having the sheaf $\cs$ of partially holomorphic smooth maps from
$\rp$ into $SL(n,\C)$ and the two-set open cover $\fu=\{\U_1,\U_2\}$,
we consider the groups of cochains
$$
C^0(\fu, \cs)=\{\mbox{maps}\ \psi_1: \U_1\to \cs(\U_1), \psi_2:
\U_2\to \cs(\U_2)\}=\cs(\U_1)\times \cs(\U_2),
\eqno(7.18a)
$$
$$
C^1(\fu, \cs)=\{\mbox{maps}\ f_{12}: \U_{12}\to \cs(\U_{12}), f_{21}:
\U_{12}\to \cs(\U_{12})\}=\cs(\U_{12})\times \cs(\U_{12}),
\eqno(7.18b)
$$
where $\cs(\U)$ is a space of sections of the sheaf $\cs$ over an open set
$\U\subset\rp$.

\smallskip

{}For 0- and 1-cocycles we have
$$
Z^0(\rp, \cs)=\left\{\psi =\{\psi_1,\psi_2\}\in C^0(\fu, \cs):\ \psi_1=\psi_2
\ \mbox{on}\ \U_{12}\right\},
\eqno(7.19a)
$$
$$
Z^1(\fu, \cs)=\left\{f=\{f_{12},f_{21}\}\in C^1(\fu, \cs):\ f_{21}=f_{12}^{-1}
\right\}.
\eqno(7.19b)
$$
By definition, $H^0(\rp, \cs):=Z^0(\rp, \cs)=\Gamma (\rp, \cs)$. As usual,
two cocycles $\cf, \hat\cf\in Z^1(\fu,\cs)$ are called equivalent if
$\hat\cf_{12}=\psi_1 \cf_{12}\psi_2^{-1}$ for some $\psi =\{\psi_1,
\psi_2\}\in C^0(\fu, \cs)$. A set of equivalence classes of 1-cocycles
$\cf$ is the \v{C}ech 1-cohomology set $H^1(\fu, \cs)$. For the considered
cover $\fu$ we have $H^1(\rp, \cs)=H^1(\fu,\cs)$.

\smallskip

By replacing the sheaf $\ch$ by the sheaf $\cs$ in the formulae of
\S\,\ref{5.2}, one can define the action of the group $C^0(\fu,\cs)$ on
$C^1(\fu,\cs)$ by automorphisms $\s_0$,
$$
\s_0(\psi ,f)_{12}=\psi_2f_{12}\psi_2^{-1},\quad
\s_0(\psi ,f)_{21}=\psi_1f_{21}\psi_1^{-1},
$$
$$
\psi =\{\psi_1,\psi_2\}\in C^0(\fu,\cs),\quad
f =\{f_{12},f_{21}\}\in C^1(\fu,\cs),
\eqno(7.20)
$$
and define a twisted homomorphism $\d^0:\ C^0(\fu,\cs)\to C^1(\fu,\cs)$
by the formulae
$$
\d^0(\phi  )_{12}=\phi_1\phi_2^{-1},\quad
\d^0(\phi  )_{21}=\phi_2\phi_1^{-1},\quad
\d^0(h\phi  )=\d^0(h)\s_0 (h,\d^0(\phi  )),
\eqno(7.21)
$$
where $\phi  =\{\phi_1, \phi_2\}\in C^0(\fu,\cs)$,
$\d^0(\phi  )\in Z^1(\fu,\cs)\subset C^1(\fu,\cs)$. Then we have
$$
H^0(\rp, \cs)= \ker\d^0,
\eqno(7.22)
$$
and the image
$$
\im\d^0=\d^0(C^0(\fu,\cs))\subset  Z^1(\fu,\cs)
\eqno(7.23)
$$
of the map $\d^0$ corresponds to the marked element  $e\in H^1(\rp, \cs)$,
i.e., to the class of smoothly trivial bundles over $\rp$ which are
holomorphically trivial over $\cp\hra\rp$, $x\in U$.  Transition
matrices $\cf\in \im\d^0$ have the form (4.17):
$\cf_{12}=\psi_1^{-1}(x,\l )\psi_2(x,\l )$.

\smallskip

Finally, for $\psi\in C^0(\fu,\cs)$, $\cf\in Z^1(\fu,\cs)$, the formula
$$
\rho_0(\psi ,\cf):= \d^0(\psi )\s_0(\psi ,\cf )\ \Leftrightarrow\
\rho_0(\psi ,\cf)_{12}=\psi_1\cf_{12}\psi_2^{-1}
\eqno(7.24)
$$
defines the action of the group $C^0(\fu,\cs)$ on the set $Z^1(\fu,\cs)$,
and we obtain
$$
H^1(\fu,\cs)=\r_0(C^0(\fu,\cs))\bl Z^1(\fu,\cs).
\eqno(7.25)
$$
{}For the chosen cover $\fu$ we have $H^1(\rp,\cs)=H^1(\fu,\cs)$.

\subsection{Exact sequences of cohomology sets}
\label{7.6}

{}From (7.15b) we obtain the exact sequence of cohomology sets~\cite{Oni}
$$
e\lra H^0(\rp,\ch)\stackrel{i_*}{\lra}H^0(\rp,\hat\cs) \stackrel
{\bar\d^0_*}{\lra}H^0(\rp,\hat\cb)\stackrel{\bar\d^1_*}{\lra}
H^1(\rp,\ch)\stackrel{\hat\vp}{\lra} H^1(\rp,\hat\cs) ,
\eqno(7.26)
$$
where $e$ is a marked element (identity) of the considered sets, and
a homomorphism $\hat\vp$ coincides with the canonical embedding,
induced by the embedding of sheaves $i: \ch\to\hat\cs$.  The kernel
$\ker\hat\vp=\hat\vp^{-1}(e)$ of the map $\hat\vp$ coincides with a
subset of those elements from $H^1(\rp,\ch)$, which are mapped into
the class $e\in H^1(\rp,\hat\cs)$ of topologically (and smoothly)
trivial bundles.  This means that representatives of the subset
$\ker\hat\vp$ are those transition matrices $\cf\in Z^1(\fu,\ch)$ for
which there exists a splitting
$$
\cf_{12}=\psi_1^{-1}(x,\l ,\bar\l )\psi_2(x,\l ,\bar\l )
\eqno(7.27)
$$
with {\it smooth} matrix-valued functions $\psi_1, \psi_2 \in SL(n,\C)$.

\smallskip

Similarly, from (7.16) we obtain the exact cohomology sequence
$$
e\lra H^0(\rp, \ch)\stackrel{i_*}{\lra}H^0(\rp, \cs) \stackrel
{\bar\d^0_*}{\lra}H^0(\rp,\cb)\stackrel{\bar\d^1_*}{\lra}
H^1(\rp,\ch)\stackrel{\vp}{\lra} H^1(\rp, \cs) ,
\eqno(7.28)
$$
where a homomorphism $\vp$ is an embedding, induced by the
embedding of sheaves $i: \ch\to \cs$.  The kernel
$\ker\vp=\vp^{-1}(e)$ of the map $\vp$ coincides with a subset of
those elements from $H^1(\rp,\ch)$, which are mapped into the class
$e\in H^1(\rp,\cs)$ of smoothly trivial bundles over $\rp$, which are
holomorphically trivial on any projective line $\cp\hra\rp$, $x\in
U$. This means that representatives of the subset $\ker\vp$ of the
1-cohomology set $H^1(\rp,\ch)$ are those transition matrices $\cf\in
Z^1(\fu,\ch)$ for which there exists a Birkhoff decomposition
(cf.(4.17))
$$
\cf_{12}=\psi_1^{-1}(x,\l )\psi_2(x,\l )
\eqno(7.29)
$$
with smooth matrix-valued functions $\psi_1, \psi_2 \in SL(n,\C)$ that
are {\it holomorphic} in $\l$.

\smallskip

The map $\bar\d^0$ corresponds  a global section
$$
B=\left\{
B^{(1)}=-(\bar\p\psi_1)\psi_1^{-1}\ \mbox{on}\ \U_1,\quad
B^{(2)}=-(\bar\p\psi_2)\psi_2^{-1}\ \mbox{on}\ \U_2, \quad
B^{(1)}= B^{(2)} \ \mbox{on}\ \U_{12}\right\},
\eqno(7.30)
$$
of the sheaf $\cb$ over $\rp$ to $\{\psi_1, \psi_2\} \in C^0(\fu,\cs)$.
The equality $B^{(1)}= B^{(2)}$ on
$\U_{12}$, which means that the (0,1)-form $B\in H^0(\rp,\cb)$ is
defined globally, follows from the identity
$$
\bar\p\cf_{12}=\bar\p (\psi_1^{-1}\psi_2)= (\bar\p \psi_1^{-1})\psi_2
+\psi_1^{-1}\bar\p \psi_2 = \psi_1^{-1}\{- (\bar\p \psi_1)\psi_1^{-1} +
(\bar\p \psi_2)\psi_2^{-1}\}\psi_2=0.
\eqno(7.31)
$$
The group $\cs(\rp ):=H^0(\rp,\cs)=Z^0(\rp,\cs)=\Gamma (\rp,\cs)$ of global
sections of the sheaf $\cs$ acts on the set $H^0(\rp,\cb)$ with the help
of Ad$(g,\cdot )$ transformations
$$
\mbox{Ad}(g,B)=g^{-1}Bg + g^{-1}\bar\p g,
\eqno(7.32)
$$
where $g\in H^0(\rp,\cs)$, $B\in H^0(\rp,\cb)$. Notice that from the
definition (7.19a) of the group $H^0(\rp,\cs)$  and from the
Liouville theorem for $\cp\hra\rp$ it follows that the elements $g\in
H^0(\rp,\cs)$ do not depend on $\l$. Comparing (7.12) and (7.30) with
(4.20)--(4.23), we conclude that the 0-cohomology set $H^0(\rp,\cb)$
coincides with the space of (complex) local solutions to the SDYM
equations on $U\subset\R^4$, the group $H^0(\rp, \cs)$ coincides with
the group of (complex) gauge transformations, and the quotient space
$H^0(\rp,\cb)/H^0(\rp, \cs)$ coincides with the moduli space of
(complex) local solutions to the SDYM equations on $U$.

\smallskip

The space $\ker\vp$ is a representative of the germ {\bm} at the point
$e\in H^1(\rp,\ch)$ of the moduli space of bundles $E'$ over $\rp$,
holomorphically trivial on $\cp\hra\rp$, $x\in U$. We will denote it by
$\cm :=\ker\vp$; this set was described in detail in \S\,\ref{6}.
{}From the exactness of the sequence (7.28) it follows that the set
$\cm =\ker\vp\subset H^1(\rp,\ch)$ is bijective to the moduli space
$H^0(\rp,\cb)/H^0(\rp, \cs)$  of (complex) solutions to the SDYM equations,
$$
\cm\simeq H^0(\rp,\cb)/H^0(\rp, \cs).
\eqno(7.33)
$$
This correspondence is a non-Abelian analogue of the Dolbeault theorem
about the isomorphism of (Abelian) \v{C}ech and Dolbeault 1-cohomology
groups.

\smallskip

{\bf Remark.} Using the sheaves $\hat\cs$ and $\hat\cb$, considered in
\S\S\,7.2,7.4 and \S\,7.6, one can introduce a
Dolbeault 1-cohomology set $H^{0,1}_{\bar\p_{\hat B}}(\rp)$
as a set of orbits of the group $H^0(\rp,\hat\cs)$ in the set
$H^0(\rp,\hat\cb)$, i.e.,
$$
H^{0,1}_{\bar\p_{\hat B}}(\rp):= H^0(\rp,\hat\cb)/H^0(\rp,\hat\cs).
\eqno(7.34)
$$
The set $H^0(\rp,\cb)/H^0(\rp, \cs)$ considered
above is an open subset in the Dolbeault 1-cohomology set
$H^{0,1}_{\bar\p_{\hat B}}(\rp)$.  It follows from the exactness of the
sequence (7.26) that $H^{0,1}_{\bar\p_{\hat B}}(\rp)\simeq \ker\hat\vp$,
i.e., the moduli space $H^{0,1}_{\bar\p_{\hat B}}(\rp)$ of global
solutions of eqs.(7.4) on $\rp$ is bijective to the moduli space of
holomorphic bundles over $\rp$ which are trivial as smooth bundles.
Transition matrices of such bundles have the form (7.27).

\smallskip

Using the bijection (7.33), we will identify the spaces $\cm$ and
$H^0(\rp,\cb)/H^0(\rp, \cs)$ and denote them by the same letter $\cm$.
It also follows from (7.33) that $H^0(\rp,\cb)$ is a principal fibre
bundle
$$
H^0(\rp,\cb)=P(\cm , H^0(\rp, \cs))
\eqno(7.35)
$$
with the base space $\cm$ and the structure group $H^0(\rp, \cs)$.

\subsection{Unitarity conditions}
\label{7.7}

In \S\,\ref{6.6} we discussed the imposition of a unitarity condition
on transition matrices $\cf\in Z^1(\fu,\ch)$ and defined various subsets
of transition matrices and their moduli satisfying the unitarity condition.

\smallskip

As it has been discussed in \S\,\ref{4.4}, the matrices $\psi_1,
\psi_2\in SL(n,\C)$ corresponding to gauge fields with values in
the algebra $su(n)$ have to satisfy  the condition (4.16c). The conditions
(4.16a) for components of the gauge potential follow from (4.16c),
(4.22) and (4.23). To satisfy these conditions, consider the
{}following real subgroup $C^0_{\tau}(\fu,\cs)$ (a real form) of the
group $C^0(\fu,\cs)$:
$$
C^0_{\tau}(\fu,\cs):=\left\{\psi =\{\psi_1,\psi_2\}\in C^0(\fu,\cs):
\psi_1^{\dagger}(\tau (x,\l ))=\psi_2^{-1}(x,\l )\right\},
\eqno(7.36)
$$
compatible with the real form $\tau$ on $\rp$. Of course, one can also
define other real forms of the complex group $C^0(\fu,\cs)$ assuming
$$
\psi_1^{\dagger}(\tau (x,\l ))=\Pi\psi_2^{-1}(x,\l ),
\eqno(7.37)
$$
where $\Pi$ is a diagonal matrix with $m$ copies of $+1$ and
$n-m$ copies of $-1$. For all these subgroups the matrices
$\d^0(\psi^{-1} )=\psi_1^{-1}\psi_2\in Z^1(\fu,\ch)$ will
satisfy the unitarity condition (4.16a) and therefore
$\d^0(\psi^{-1} )\in Z^1_{\tau}(\fu,\ch)$.

\smallskip

The map $\d^0: C^0_{\tau}(\fu,\cs)\to Z^1_{\tau}(\fu,\cs)$ defines in
$Z^1_{\tau}(\fu,\cs)$ a subset of matrices $\psi_1^{-1}\psi_2$
with $\{\psi_1, \psi_2\}\in C^0_{\tau}(\fu,\cs)$ which corresponds to
the element $e\in H^1_{\tau}(\rp,\cs)$. The set $H^1_{\tau}(\rp,\cs)$
is defined analogously with the set  $H^1(\rp,\cs)$ (see \S\,\ref{7.5}).
The kernel $\ker\vp_{\tau}=\vp^{-1}_{\tau}(e)$ of the map
$$
\vp_{\tau}:=\vp\mid_{H^1_{\tau}(\rp,\ch)}:\quad
H^1_{\tau}(\rp,\ch)\to H^1_{\tau}(\rp,\cs)
\eqno(7.38)
$$
coincides with the moduli space $\cm_{\tau}$ of transition matrices
$\cf\in Z^1_{\tau}(\fu,\ch)$, for which there exists a Birkhoff
decomposition (7.29) with $\psi_1, \psi_2$ satisfying the unitarity
conditions (4.16c). The map $\bar\d^0$ associates with
$\psi_1, \psi_2$ the global section (7.17), (7.30) of the sheaf
$\cb$ satisfying the unitarity condition (4.16a). We denote the
space of all these solutions by $H^0_{\tau}(\rp,\cb)$. The matrices
$g\in SL(n,\C)$ from the group $H^0(\rp,\cs)$ do not depend on $\l$,
and the subgroup
$$
H^0_{\tau}(\rp,\cs)=\left\{g\in H^0(\rp,\cs):\quad g^{\dagger}=
g^{-1}\right\}
\eqno(7.39)
$$
of unitary matrices $g(x)\in SU(n)$ preserves the space
$H^0_{\tau}(\rp,\cb)$. So we have a one-to-one correspondence
between $\cm_{\tau}$ and the moduli space
$H^0_{\tau}(\rp,\cb)/H^0_{\tau}(\rp,\cs)$
of real local solutions to the SDYM equations,
$$
\cm_{\tau}\simeq H^0_{\tau}(\rp,\cb)/H^0_{\tau}(\rp,\cs).
\eqno(7.40)
$$

\section{Symmetries in terms of smooth sheaves}
\label{8}

\subsection{Riemann-Hilbert problems from the cohomological
            point of view}
\label{8.1}

In \S\,\ref{7.5} we described the twisted homomorphism
$\d^0: C^0(\fu,\cs)\to C^1(\fu,\cs)$, the image of which
$\im\d^0=\d^0 (C^0(\fu,\cs))$ belongs to the
set $Z^1(\fu,\cs)\subset C^1(\fu,\cs)$. More precisely, we have
$\im\d^0\simeq C^0(\fu,\cs)/H^0(\rp, \cs)$, where the group
$H^0(\rp, \cs)=\ker\d^0$ is a kernel of the map $\d^0$.
Hence $\d^0 (C^0(\fu,\cs))$ can be identified
with $C^0(\fu,\cs)/H^0(\rp, \cs)$, and
$$
\d^0:\ C^0(\fu,\cs)\to C^0(\fu,\cs)/H^0(\rp, \cs)
\eqno(8.1)
$$
is a projection of the group $C^0(\fu,\cs)$ onto the homogeneous space
$Q:=C^0(\fu,\cs)/H^0(\rp, \cs)$. So, the group  $C^0(\fu,\cs)$ can be
considered  as a principal fibre bundle
$$
C^0(\fu,\cs)=P(Q, H^0(\rp,\cs))
\eqno(8.2)
$$
with the structure group $H^0(\rp,\cs)$ and the base space
$Q\subset Z^1(\fu,\cs)$, points of which  correspond to smoothly
trivial bundles.

\smallskip

As described in detail in \S\S\,\ref{5} -\,\ref{7}, the space $Q$
contains as a subset the set $\cn$ of those holomorphic bundles which
are not only trivial as smooth bundles, but also holomorphically
trivial on $\cp\hra\rp$, $x\in U$. The group $C^0(\fu,\cs)$ acts on
$Q$ transitively by formula (7.24) and therefore  for any cocycle
$\cf\in\cn\subset Q$ there exists an element $\psi
=\{\psi_1,\psi_2\}\in C^0(\fu,\cs)$ such that the action $\r_0(\psi
,\cdot )$ transforms $\cf$ into $\cf^0=1$,

$$
\r_0(\psi ,\cf )_{12}=\psi_1\cf_{12}\psi^{-1}_2=1 \ \Rightarrow\
\cf_{12}= \psi_1^{-1}\psi_2,
\eqno(8.3)
$$
and to solve the Riemann-Hilbert problem means to find such an element
$\psi$ from the group $C^0(\fu,\cs)$. Of course, this element $\psi
\in C^0(\fu,\cs)$ is not unique; it is defined up to an element $g$
from the stability subgroup $H^0(\rp, \cs)$ of the point $\cf^0=1$.

Indeed, if $\psi_1\cf_{12}\psi^{-1}_2=1$, then
$(g^{-1}\psi_1)\cf_{12}(g^{-1}\psi_2)^{-1}=1$ for any
$g\in H^0(\rp, \cs)$. In other words, to solve the Riemann-Hilbert
problem means to define a section
$$
s:\ \cn\to  C^0(\fu,\cs)
\eqno(8.4)
$$
over $\cn\subset Q$ of the bundle (8.2). The section $s$ is {\it not
uniquely defined}, and the group $H^0(\rp,\cs)$  defines a
transformation $g$ of the section $s$ into an {\it equivalent} section
$s_g$.

\smallskip

{\bf Remark.} It should be stressed that the cohomological description
of the construction of solutions is applicable not only to the
SDYM equations, but also to all equations integrable with the help of
a Birkhoff decomposition of matrices on $\C\bp^1$ (the dressing
method~\cite{ZS}-\cite{FT}). For such equations, one can write an
exact sequence of sheaves like (7.16) and an exact sequence of
cohomology sets like (7.28). In many cases this can be done by
reductions of the sheaves $\ch, \cs$ and $\cb$, which explains the
known fact that (almost) all integrable equations in 2D can be
obtained by reductions of the SDYM equations (see e.g.~\cite{Wa,IP,MW,Leg}
and references therein).

\smallskip

Consider the restriction
$$
P(\cn, H^0(\rp,\cs)):=P(Q, H^0(\rp,\cs))\mid_\cn =(\d^0)^{-1}(\cn )
\eqno(8.5)
$$
of the principal fibre bundle $P(Q, H^0(\rp,\cs))$ to the subset
$\cn\subset Q$. As described in \S\,\ref{6.2}, the group $C^0(\fu,\ch)$
acts on the space $\cn$ on the left, and this action
can be lifted up to the action on $P(\cn, H^0(\rp,\cs))$, since
this (left) action commutes with the (right) action of the group
$H^0(\rp,\cs)$ on the space $P(\cn, H^0(\rp,\cs))$. Thus, we have the space
$P(\cm, H^0(\rp,\cs))$
as a space of orbits of the group $C^0(\fu,\ch)$  in the space
$P(\cn, H^0(\rp,\cs))$,
$$
P(\cm, H^0(\rp,\cs))=P(\r_0(C^0(\fu,\ch))\bl\cn, H^0(\rp,\cs))=
\r_0(C^0(\fu,\ch))\bl P(\cn, H^0(\rp,\cs)).
\eqno(8.6a)
$$
 At the same time, it follows from (7.35) that
this space coincides with the space
$$
H^0(\rp,\cb)=P(\cm, H^0(\rp,\cs))
\eqno(8.6b)
$$
of (complex) local solutions to the SDYM equations.

\smallskip

{}Finally, it follows from (8.6) that the moduli space of
(complex) local solutions to the SDYM equations is
$$
\cm\simeq \r_0(C^0(\fu,\ch))\bl P(\cn, H^0(\rp,\cs))/ H^0(\rp,\cs),
\eqno(8.7)
$$
i.e., $\cm$ is the biquotient space of the space
$P(\cn, H^0(\rp,\cs))$
under the action of the groups $C^0(\fu,\ch)$ and $H^0(\rp,\cs)$.

\smallskip

Using \S\,\ref{7.7}, where we discussed the unitarity conditions
in terms of $\cf_{12}$, $\psi\in C^0(\fu,\cs)$ etc., one can rewrite
all formulae of \S\,\ref{8.1} in a way compatible with the real structure
$\tau$ on $\rp$. In particular, for the moduli space $\cm_{\tau}$ of
(real) local solutions to the SDYM equations we have
$$
\cm_{\tau}\simeq \r_0(C^0_{\tau}(\fu,\ch))\bl P(\cn_{\tau},
H^0_{\tau}(\rp,\cs))/ H^0_{\tau}(\rp,\cs).
\eqno(8.8)
$$
Then  gauge fields take values in the Lie algebra $su(n)$.

\subsection{Action of the symmetry group $\frg_{\tau}$ on
           real solutions of the SDYM equations}
\label{8.2}

We consider the cover $\fu =\{\U_1,\U_2\}$ of the twistor space $\rp$
and holomorphic bundles  $E'\in\cn_{\tau}\subset Z^1_{\tau}(\fu,\ch)$.
In \S\S\,\ref{6.5},\,\ref{6.6}, the (local) action of the local group
$\frg_{\tau}=\fh_{\tau}\ltimes\fc_{\tau}$ on the space $\cn_{\tau}
\simeq \fc_{\tau}/\fc_{\tau\t}$ was described. Let us choose an
arbitrary transition matrix $\cf_{12}=\psi^{-1}_1\psi_2\in\cn_{\tau}$
and an element $h=\{\eta ,a\}\in \fh_{\tau}\ltimes\fc_{\tau}$.
Consider the action $\r (h,\cdot )$ of the element $h\in \frg_{\tau}$
given by formulae (6.19), (5.30) and (6.24). Then we have
$\r (h,\cdot ): \cf_{12}\mapsto \cf_{12}^h=\r(h,\cf)_{12}$. Since the local
action preserves $\cn_{\tau}$, then $\cf^h\in \cn_{\tau}$ and therefore
there exists an element $\psi^h=\{\psi^h_1,\psi^h_2\}\in C^0_{\tau}(\fu,\cs)$
such that
$$
\cf_{12}^h= (\psi^h_1)^{-1} \psi^h_2 .
\eqno(8.9)
$$
Let us introduce $\phi (h)=\{\phi_1(h), \phi_2(h)\}\in C^0_{\tau}(\fu,\cs)$
by the formulae
$$
\phi_1(h):= \psi^h_1 \psi^{-1}_1,\quad
\phi_2(h):= \psi^h_2 \psi^{-1}_2 .
\eqno(8.10)
$$
Then  we have a map
$$
\phi :\quad \frg_{\tau}\to C^0_{\tau}(\fu,\cs)
\eqno(8.11)
$$
of the group $\frg_{\tau}$ into the group $C^0_{\tau}(\fu,\cs)$.

\smallskip

The elements $\phi (h)=\{\phi_1(h), \phi_2(h)\}$ of the group
$C^0_{\tau}(\fu,\cs)$ act by definition  on $\psi $ $=\{\psi_1,\psi_2\}$
$\in P(\cn_{\tau},$ $ H^0_{\tau}(\rp,\cs))$ as follows:
$$
h:\quad \psi =\{\psi_1,\psi_2\}\ \mapsto \ \r (h,\psi ):=
\psi^h =\{\psi_1^h,\psi_2^h\}=\{\phi_1(h)\psi_1,\phi_2(h)\psi_2\}.
\eqno(8.12)
$$
From (7.11) it follows that $B=\{B^{(1)}, B^{(2)}\}$ is transformed
by the formulae
$$
h:\ B\mapsto \r (h,B)\equiv B^h:=
\phi (h)B\phi^{-1}(h) +\phi (h)\bar\p\phi^{-1}(h)
\Rightarrow
\eqno(8.13a)
$$
$$
B^{(1)}\mapsto \phi_1 (h)B^{(1)}\phi^{-1}_1(h) +
\phi_1 (h)\bar\p\phi_1^{-1}(h),\quad
B^{(2)}\mapsto \phi_2 (h)B^{(2)}\phi^{-1}_2(h) +
\phi_2 (h)\bar\p\phi_2^{-1}(h).
\eqno(8.13b)
$$
With the help of formulae (8.13), (4.22) and (4.23) it is not difficult
to write down explicit formulae for transformations of components
$A_\mu$ of the gauge potential $A$. We shall not do this.

\smallskip

Consider now a transformation
$$
\cf_{12}\stackrel{h}{\mapsto}
\cf_{12}^h\stackrel{f}{\mapsto}\cf_{12}^{fh}=
f_{12}h_{12}\cf_{12}h^\dagger_{12}f^\dagger_{12}.
\eqno(8.14)
$$
It is easy to see that
$$
B^{fh}=\phi (fh)B\phi^{-1}(fh)+\phi (fh)\bar\p\phi^{-1}(fh) =
\phi (f)B^h\phi^{-1}(f) +\phi (f)\bar\p\phi^{-1}(f)=
$$
$$
=\phi (f)\phi (h)B(\phi (f)\phi (h))^{-1} +
\phi (f)\phi (h)\bar\p(\phi (f)\phi (h))^{-1}.
\eqno(8.15)
$$
It follows from (8.14), (8.15) that
$$
\phi (fh)=\phi (f)\phi (h),
\eqno(8.16)
$$
i.e., the map (8.11) is a {\it homomorphism} of the local Lie group
$\frg_{\tau}$ into the group $C^0_{\tau}(\fu,\cs)$.

\subsection{Gauge fixing and some formulae}
\label{8.3}

The SDYM equations (4.7) for $A_\mu\in sl(n,\C)$ imply that the
components of the gauge potential can be written in the form
$$
A_{y^1}=\th^{-1}\p_{y^1}\th,\quad
A_{y^2}=\th^{-1}\p_{y^2}\th,\quad
A_{\bar y^1}=\tilde\th^{-1}\p_{\bar y^1}\tilde\th,\quad
A_{\bar y^2}=\tilde\th^{-1}\p_{\bar y^2}\tilde\th,
\eqno(8.17)
$$
where $\th$ and $\tilde\th$ are some $SL(n,\C)$-valued functions on
$U\subset\R^4$. One may perform the following gauge transformation:
$$
A_{\bar y^1}\mapsto A_{\bar y^1}^{\tilde\th}=
\tilde\th A_{\bar y^1}\tilde\th^{-1}
+ \tilde\th\p_{\bar y^1}\tilde\th^{-1}=0,\quad
A_{\bar y^2}\mapsto A_{\bar y^2}^{\tilde\th}=
\tilde\th A_{\bar y^2}\tilde\th^{-1}
+ \tilde\th\p_{\bar y^2}\tilde\th^{-1}=0,
\eqno(8.18a)
$$
$$
A_{y^1}\mapsto A_{y^1}^{\tilde\th}=\tilde\th A_{y^1}\tilde\th^{-1}
+ \tilde\th\p_{y^1}\tilde\th^{-1}=\Phi^{-1}\p_{y^1}\Phi ,\quad
A_{y^2}\mapsto A_{y^2}^{\tilde\th}=\tilde\th A_{y^2}\tilde\th^{-1}
+ \tilde\th\p_{y^2}\tilde\th^{-1}=\Phi^{-1}\p_{y^2}\Phi ,
\eqno(8.18b)
$$
where $\Phi :=\th\tilde\th^{-1}\in SL(n,\C)$, and thus fix the gauge
$A_{\bar y^1}^{\tilde\th}=A_{\bar y^2}^{\tilde\th}=0$
~\cite{Prasad}-\cite{Tak}. Then eqs.(4.7) are replaced by the
matrix equations
$$
\p_{\bar y^1}(\Phi^{-1}\p_{y^1}\Phi )+\p_{\bar y^2}(\Phi^{-1}
\p_{y^2}\Phi )=0,
\eqno(8.19)
$$
which are the SDYM equations in the Yang gauge. Equations (8.19)
are a 4D analogue of the 2D WZNW equations.

\smallskip

It is also possible to perform the gauge transformation
$$
A_{\bar y^1}\mapsto  \th A_{\bar y^1} \th^{-1}
+  \th\p_{\bar y^1} \th^{-1}=\Phi\p_{\bar y^1}\Phi^{-1} ,\quad
A_{\bar y^2}\mapsto  \th A_{\bar y^2} \th^{-1}
+  \th\p_{\bar y^2} \th^{-1}=\Phi\p_{\bar y^2}\Phi^{-1} ,
$$
$$
A_{y^1}\mapsto  \th A_{y^1} \th^{-1}+\th\p_{y^1} \th^{-1}=0,\quad
A_{y^2}\mapsto  \th A_{y^2} \th^{-1}+\th\p_{y^2} \th^{-1}=0,
\eqno(8.20)
$$
then eqs.(4.7) get converted into the equations
$$
\p_{y^1}(\Phi \p_{\bar y^1}\Phi^{-1} )+\p_{y^2}(\Phi \p_{\bar y^2}
\Phi^{-1} )=0.
\eqno(8.21)
$$

\smallskip

{}From the linear system (4.10) it is easy to see that
$$
\th =\psi^{-1}_2 (\zeta =0),\quad
\tilde\th =\psi^{-1}_1 (\l =0),
\eqno(8.22)
$$
where the $SL(n,\C)$-valued function $\psi_1$ is defined on $\U_1$, and
the $SL(n,\C)$-valued function $\psi_2$ is defined on $\U_2$. Equations
(8.19) are the compatibility conditions of the linear system
$$
\p_{\bar y^1}\tilde\psi_1 -\l (\p_{y^2} +\Phi^{-1}\p_{y^2}\Phi )
\tilde\psi_1=0,\quad
\p_{\bar y^2}\tilde\psi_1 +\l (\p_{y^1} +\Phi^{-1}\p_{y^1}\Phi )
\tilde\psi_1=0,
\eqno(8.23)
$$
obtained from (4.10) for $\psi_1$ by performing the gauge transformation
$\psi_1(x,\l )\mapsto \tilde\psi_1(x,\l ) = \psi^{-1}_1(x, 0)
\psi_1(x, \l ) =
\tilde\th (x)\psi_1(x,\l )$, $\ \l\in\Omega_1$. Analogously, eqs.(8.21) are
the compatibility conditions for the  linear system
$$
\zeta (\p_{\bar y^1}+\Phi\p_{\bar y^1}\Phi^{-1})\tilde\psi_2-
\p_{y^2}\tilde\psi_2=0,\quad
\zeta (\p_{\bar y^2}+\Phi\p_{\bar y^2}\Phi^{-1})\tilde\psi_2+
\p_{y^1}\tilde\psi_2=0,
\eqno(8.24)
$$
where $\tilde\psi_2(x,\zeta )=\psi_2^{-1}(x,0)\psi_2(x,\zeta )=
\th (x)\psi_2(x,\zeta )$ is well defined for $\zeta\in \Omega_2$.

\smallskip

We have  $\tilde\psi_1(x,\l =0 )=1$ and therefore
$$
\tilde\psi_1=1+\l\Psi +O(\l^2)
\eqno(8.25)
$$
for some Lie algebra valued function $\Psi\in sl(n,\C)$. By substituting
(8.25) into (8.23), we find that
$$
\Phi^{-1}\p_{y^2}\Phi =\p_{\bar y^1}\Psi ,\quad
\Phi^{-1}\p_{y^1}\Phi =-\p_{\bar y^2}\Psi .
\eqno(8.26)
$$
Then  after substitution (8.26) into (8.23), the compatibility
conditions of the linear system (8.23) will be
$$
\p_{y^1}\p_{\bar y^1}\Psi +\p_{y^2}\p_{\bar y^2}\Psi +
[\p_{\bar y^1}\Psi ,\p_{\bar y^2}\Psi ]=0.
\eqno(8.27)
$$
Equations (8.27) are the SDYM equations in the so-called
Leznov-Parkes form.

\smallskip

Notice that the condition
$$\psi_1(x,\l =0 )=1,
\eqno(8.28)
$$
leading to the gauge fixing $A_{\bar y^1}=A_{\bar y^2}=0$, can be
imposed from the very beginning. Then  the Birkhoff factorization (8.3) is
unique, which corresponds to the choice of the fixed section (8.4) of
the bundle (8.5). Nevertheless, the gauge (8.28) does not remove all
degrees of freedom related to holomorphic transformations of the
group $C^0(\fu ,\ch)$, and if we want to obtain the moduli space
$\cm$, we have to factorize $s(\cn )\simeq \cn$ w.r.t. the action of
the subgroup
in $C^0(\fu ,\ch)$ preserving the gauge (8.28).  The same gauge may
be used in the description of the moduli space $\cm_\tau$ discussed in
\S\,\ref{8.1}.

\subsection{Generalization to self-dual manifolds}
\label{8.4}

As has been mentioned in \S\,\ref{4.6},  the twistor correspondence
between self-dual gauge fields and holomorphic bundles exists not
only for the Euclidean space $\R^4$, but also for  4-manifolds $M$,
the Weyl tensor of which is self-dual. Twistor spaces $\cz\equiv\cz
(M)$ for such manifolds $M$ are three-dimensional complex spaces. The
description of symmetries of local solutions to the SDYM equations
can be easily generalized to this general case.

\smallskip

It can be done as follows. Fix an open set $U\subset M$ such that
$\cz|_U\simeq U\times \C\bp^1$  and choose coordinates $x^\mu$ on
$U$.  Consider the restriction of the twistor bundle $\pi :\cz\to M$
to $U$ and put $\rp :=\cz|_U$. The space $\rp$ is an open subset of
$\cz$, and, as a real manifold, $\rp$ is diffeomorphic to the direct
product $U\times\C\bp^1$. Now a metric on $U$ is not flat, and a
conformal structure on $U$ is coded into a complex structure $\cj$
on $\rp$ ~\cite{Pen, AHS}.  In this ``curved" case we again have a natural
one-to-one correspondence between solutions of the SDYM equations on
$U$ and holomorphic bundles $E'$ over $\rp$, holomorphically trivial
on (real) projective lines $\cp\hra\rp$, $\forall x\in U$.

\smallskip

In our group-theoretic analysis of the twistor correspondence we did
not use the explicit form of the complex structure $\cj$ on $\rp$
and therefore  did not use the explicit form of the metric on $U$. This
explicit form was used only in some illustrating formulae, which can
easily be generalized. That is why, all statements about local
solutions and symmetry groups are also true for the SDYM equations on
self-dual manifolds $M$. Thus, as the local symmetry group we again
obtain the group $\frg_\tau=\fh_\tau\ltimes\fc_\tau$ from \S\S\,6--8
acting on the space of local solutions to the SDYM equations defined
on a self-dual 4-manifold $M$.

\section{Discussion}
\label{9}

\subsection{What is integrability?}
\label{9.1}

In books and papers on soliton equations one often poses the
question:  What is integrability? There is no general answer to this
question, and usually one connects the integrability with the
existence of Lax or zero curvature representations. Then  non-Abelian
cohomology, local groups  and deformation theory of bundles with
holomorphic or flat
connections form the basis of integrability. In other words, there
are always exact sequences of sheaves and cohomology sets of type
(7.16), (7.28) hiding behind the integrability. This explains, in
particular, why almost all integrable equations in two dimensions
can be obtained by reductions of the SDYM equations (see
e.g.~\cite{Wa,IP,MW,Leg} and references therein).

\smallskip

In~\cite{CDFN}-\cite{BKS} generalized SDYM equations in dimension
D$>$4 and their solutions have been considered. Some of these
equations in dimension D=4n ~\cite{War,CGK} are integrable, since
with the help of the twistor approach these quaternionic-type SDYM
equations can be rewritten as holomorphy conditions of the Yang-Mills
bundle over an auxiliary (twistor) (4n+2)-space. The situation with
the integrability of other generalized SDYM equations in D$>$4 is
much less clear. Solutions of these equations, e.g.
octonionic-type SDYM equations in D=8~\cite{CDFN, IPo}, were used in
constructing solitonic solutions of  string theories ~\cite{HS}. The
modification of these generalized SDYM equations arising after
replacement  of commutators by Poisson brackets are considered in
supermembrane theory (see e.g.~\cite{CFZ}).  At the moment it is not
clear whether all these equations can be interpreted as an existence
condition of flat or holomorphic connections in bundles over some
auxiliary spaces. This interesting problem deserves further study.

\subsection{Holomorphic Chern-Simons-Witten theory}
\label{9.2}

Let us consider a smooth six-dimensional manifold $\cz$ with an
integrable almost complex structure $\cj$. Then $\cz$ is a complex
3-manifold, and one can introduce a cover $\{\U_\a\}$ of $\cz$ and
coordinates $z_\a :\U_\a\to \C^3$. Let $E'$ be a smooth complex
vector bundle of rank $n$ over $\cz$ and let $\hat B$ be the
(0,1)-component of a connection 1-form on the bundle $E'$.  Suppose
that $\hat B$ satisfies the equations
$$
\bar\p\hat B + \hat B\wedge\hat B =0,
\eqno(9.1)
$$
where $\bar\p$ is the (0,1) part of the exterior derivative
$d=\p+\bar\p$.  The special case of eqs.(9.1) on the twistor space
$\rp$ of $U\subset \R^4$ was considered in \S\,7. Equations (9.1)
mean that the (0,2) part of the curvature of the bundle $E'$ is equal to
zero: $F^{0,2}:=\bar \p^2_{\hat B}=(\bar\p +\hat B)^2=0$ and,
therefore, the bundle $E'$ is holomorphic. We shall call eqs.(9.1)
defined on a complex 3-manifold $\cz$ the field equations of
holomorphic Chern-Simons-Witten (CSW) theory.

\smallskip

Equations (9.1) were suggested by Witten~\cite{Wit2} for a special case
of bundles over Calabi-Yau (CY) 3-folds $\cz$ as equations of a
holomorphic analogue of the ordinary Chern-Simons theory. Witten obtained
eqs.(9.1) from open $N=2$ topological strings with a central charge
$\hat c=3$ (6D target space) and the CY restriction $c_1(\cz)=0$  arised
from $N=2$ superconformal invariance of a sigma model used in constructing
the topological string theory. The connection of eqs.(9.1) with
topological strings was also considered in~\cite{BCOV}.  Equations (9.1)
on CY 3-folds were considered by Donaldson and Thomas ~\cite{DT} in the
frames of program on extending the results of Casson, Floer, Jones
and Donaldson to manifolds of dimension D$>$4. Donaldson and Thomas
~\cite{DT} pointed out that one may try to consider a more general
situation with eqs.(9.1) on complex manifolds $\cz$ which are not
Calabi-Yau ($c_1(\cz)\ne 0$).  This is important since the CY
restriction cannot be imposed if one uses the twistor correspondence
between  4D and 6D theories.

\smallskip

In \S\,5 we considered the special case of the holomorphic CSW theory
when field equations are defined not on an arbitrary complex 3-manifold,
but on the twistor space $\rp$ of $U\subset\R^4$. The manifold $\rp$
can be covered by two charts, and in \S\,5 we described the moduli
space and symmetries of the holomorphic CSW theory in the \v{C}ech
approach.  In \S\,7 (see formulae (7.13), (7.26) and (7.34)) we gave
the Dolbeault description of this moduli space. This analysis of
the moduli space and symmetries of the holomorphic CSW theory can be
generalized without difficulties to an arbitrary complex 3-manifold $\cz$.

\subsection{$N=2$ and $N=4$ topological strings}
\label{9.3}

The coupling of topological sigma models and topological gravity
gives the above-mentioned $N=2$ topological strings~\cite{DVV} which
were further studied in~\cite{Wit2, BCOV, GV}.  They have critical
dimension D=6
and are related to topological sigma models with the 6D target space.
There are two classes of such models, called A- and B-models. In the
open string sector of the critical topological string theories there
are A and B versions of these theories. The A-model is related to the
ordinary Chern-Simons theory in 3 real dimensions and the B-model is
related to the holomorphic  Chern-Simons-Witten theory in 3 complex
dimensions. We discuss only the B-model, the field equations for
which concide with eqs.(9.1) on a CY 3-fold.

\smallskip

Besides $N=2$ topological strings with $\hat c=3$ (6D target space)
there are $N=4$ topological strings with a central charge $\hat c=2$
(4D target space)~\cite{BV} and nontopological $N=2$ strings (see
e.g.~\cite{OV1}-\cite{BKL} and references therein). In~\cite{BV} it was
shown that
$N=2$ strings are a special case of $N=4$ topological strings. The
$N=2$ string theories describe quantum SDYM fields on a self-dual
gravitational background~\cite{OV1, BV, OV2, BKL}.  For heterotic $N=2$
strings~\cite{OV3}  besides SDYM fields there are also matter fields
depending on the details of the construction.

\smallskip

Comparing the above-mentioned string theories and field theories
corresponding to them, one obtains the following ``commutative'' diagram

\smallskip

$$
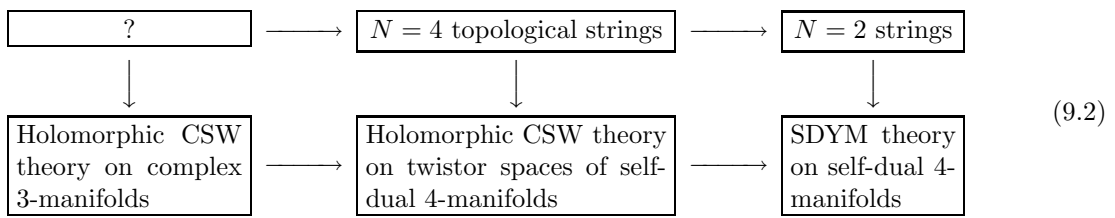
\begin{CD}
\fbox{\makebox[3cm]{?}}@>>>\fbox{\makebox[4cm]{$N=4$ topological strings}}@>>>
\fbox{\makebox[2.2cm]{$N=2$ strings}} \\
@VVV                        @VVV           @VVV\\
\fbox{\parbox{3cm}{Holomorphic CSW theory on complex 3-manifolds}}@>>>
\fbox{\parbox{4cm}{Holomorphic CSW theory on twistor spaces of self-dual
4-manifolds}}@>>>\fbox{\parbox{2.2cm}{SDYM theory on self-dual 4-manifolds}}
\end{CD}
\eqno(9.2)
$$
\medskip

\noindent
The arrows mean that one theory can be derived from another one.  The
difference between the holomorphic CSW theory on a general complex
3-manifold and the one defined on a twistor space $\cz$ is stipulated
by the existence in $\cz$ of a bundle structure $\pi : \cz\to M$ with a
self-dual 4D manifold $M$ as a base space and $\C\bp^1$ as a typical fibre.
In the general case, complex 3-spaces are arbitrary.

\smallskip

Into the box with the question-mark from (9.2) one cannot substitute
`$N=2$ topological strings', since they are obtained from sigma
models on CY 3-folds. One should substitute there some generalized
$N=2$ topological strings on a complex 3-manifold without the CY
restriction. The possibility of introducing such strings was pointed
out in the papers~\cite{BCOV, OV2}. Ooguri and Vafa~\cite{OV2}
gave reasons for
possible equivalence of $N=4$ topological strings and generalized
$N=2$ topological strings on the twistor space with a holomorphic
(2,0)-form turned on. It would be very interesting to study this
possibility.

\subsection{Integrable 4D conformal field theories}
\label{9.4}

It is well-known that the ordinary 3D Chern-Simons theory is
connected with 2D conformal field theories if one supposes that a
3-manifold has the form $\Sigma\times\R$, where $\Sigma$ is a
2-manifold with or without a boundary~\cite{Wit3, EMSS}. In particular,
if $\Sigma$ has a boundary, the quantum Hilbert space $H_{\Sigma}$ is
infinite-dimensional and is a representation space of the chiral
algebra of CFT on $\Sigma$.  Analogously, the holomorphic
Chern-Simons-Witten theory on a complex 3-manifold $\cz$ is connected
with integrable 4D CFT's on a self-dual 4-manifold $M$ if one
supposes that $\cz$ is the twistor space of $M$. This means that
$\cz$ is the bundle $\pi :\cz\to M$ over $M$ with $\C\bp^1$ as a
typical fibre.
On $M$ it is possible to consider a CFT of fields of an arbitrary spin.
Most of these CFT's will describe free fields in a fixed background.
By considering local solutions of field equations on
$M$ we take an open set $U\subset M$ and consider the twistor space
$\rp =\cz|_U$ of $U$ which is an open subset in $\cz$.

\smallskip

In this paper, we actually discuss how the concrete nonlinear 4D CFT --
the SDYM theory -- is connected with the holomorphic CSW theory on the
twistor space $\rp$ of $U$. The SDYM model on an open ball $U\subset\R^4$
is a generalization of the WZNW model on the complex plane $\C$, and
we mainly consider sets $U$ with the flat metric. We described symmetries
of the SDYM model and the moduli space of self-dual gauge fields
on $U$. Naturally, the following questions arise:
\begin{enumerate}
\item What is an analogue of affine Lie algebras of 2D CFT's?
\item What is an analogue of the Virasoro algebra?
\end{enumerate}
In this paper we have not discussed symmetry algebras yet.
But knowing the symmetry groups of the SDYM equations, described
in \S\S\,7,8, it is not difficult to write down the algebras
corresponding to them.

\smallskip

A symmetry algebra of integrable 4D CFT's is connected with the
algebra $\cg_h$ of functions that are holomorphic on
$\U_{12}=\U_1\cap\U_2\subset \rp$ and take values in the Lie algebra
$\bf g$ of a complex Lie group $\bf G$.  The algebra $\cg_h$ with
pointwise commutators generalizes affine Lie algebras. The symmetry
algebra is the algebra $$ C^1(\fu ,\co^{\bf g}_\rp )\simeq
\cg_h\oplus \cg_h
\eqno(9.3)
$$ of 1-cochains of the cover $\fu
=\{\U_1,\U_2\}$ of the space $\rp$ with values in the sheaf $\co^{\bf
g}_\rp$ of holomorphic maps from $\rp$ into the Lie algebra ${\bf
g}$. We mainly considered the case ${\bf g} =sl(n,\C)$.  The algebra
(9.3) was also considered by Ivanova~\cite{tai}.

\smallskip

Notice that the affine Lie algebra ${\bf g}\otimes\C[\l ,\l^{-1}]$ (without
a central term) is the algebra of ${\bf g}$-valued meromorphic functions
on $\C\bp^1\simeq\C^*\cup\{0\}\cup\{\infty\}$ with the poles at
$\l =0, \l =\infty$
and holomorphic on $\Omega_{12}=\Omega_1\cap\Omega_2\simeq\C^*$. Hence, it
is a subalgebra in the algebra
$$
C^1(\fo , \co^{\bf g}_{\C\bp^1})\simeq {\bf g}\otimes\C[\l ,\l^{-1}]\oplus
{\bf g}\otimes\C[\l ,\l^{-1}]
\eqno(9.4)
$$
of 1-cochains of the cover $\fo =\{\Omega_1,\Omega_2\}$ of  $\C\bp^1$
with values in the sheaf of holomorphic maps from $\C\bp^1$
into the Lie algebra ${\bf g}$. Thus, the algebra (9.3) is an analogue of
the 2D affine Lie algebra (9.4). Notice that (central) extensions of the
algebras (9.3) and (9.4) will appear after the transition to quantum theory.

\subsection{The \v{C}ech description of the Virasoro algebra}
\label{9.5}

Elements of the Virasoro algebra $Vir^0$ (with zero central charge) are
meromorphic vector fields on $\C\bp^1$ having poles at the points
$\l =0, \l =\infty$ and holomorphic on the overlap $\Omega_{12}=\Omega_1
\cap\Omega_2\simeq\C^*=\C\bp^1-\{0\}-\{\infty\}$. This algebra has the
following \v{C}ech description. Let us consider the sheaf $\cv_{\C\bp^1}$
of holomorphic vector fields  on $\C\bp^1$. Then  for the space of \v{C}ech
1-cochains with values in $\cv_{\C\bp^1}$ we have
$$
C^1(\fo , \cv_{\C\bp^1})\simeq  Vir^0\oplus Vir^0.
\eqno(9.5)
$$
Notice that for $\{v_{12},v_{21}\}\in C^1(\fo , \cv_{\C\bp^1})$ the
antisymmetry condition cannot be imposed on cohomology indices of the
holomorphic vector fields $v_{12},v_{21}$, since it is not preserved
under commutation. So we have $v_{21}\ne -v_{12}$ in the general case.

\smallskip

The space  $Z^1(\fo , \cv_{\C\bp^1})$ of 1-cocycles of the cover
$\fo =\{\Omega_1,\Omega_2\}$ of  $\C\bp^1$ with values in the sheaf
$\cv_{\C\bp^1}$ coincides with the algebra $Vir^0$ as a vector space,
since
$$
Z^1(\fo , \cv_{\C\bp^1})\simeq
(Vir^0\oplus Vir^0)/diag(Vir^0\oplus Vir^0).
\eqno(9.6)
$$
{}Further, by virtue of the equality
$$
H^1(\C\bp^1, \cv_{\C\bp^1})=0,
\eqno(9.7)
$$
which means the rigidity of the complex structure of $\C\bp^1$,
any element $v$ from $Vir^0\simeq Z^1$ can be represented
in the form
$$
v=v_1-v_2.
\eqno(9.8)
$$
Here, $v_1$ can be extended to a holomorphic vector field on $\Omega_1$,
and $v_2$ can be extended to a holomorphic vector field on $\Omega_2$.

\smallskip

It follows from (9.6)--(9.8) that the algebra $Vir^0$ is connected
with the algebra
$$
C^0(\fo , \cv_{\C\bp^1})
\eqno(9.9)
$$
of 0-cochains of the cover $\fo$ with values in the sheaf $\cv_{\C\bp^1}$
by the (twisted) homomorphism
$$
\dot\d^0: C^0(\fo , \cv_{\C\bp^1})\lra C^1(\fo , \cv_{\C\bp^1})
\Leftrightarrow
\eqno(9.10a)
$$
$$
\dot\d^0: \{v_1,v_2\}\mapsto \{v_1-v_2,v_2-v_1\}.
\eqno(9.10b)
$$
Just the cohomological nature of the algebra $Vir^0$ permits one to
define its local action on Riemann surfaces of arbitrary genus and on
the space of conformal structures of Riemann surfaces~\cite{BMS}. A
central extension arises under an action of the Virasoro algebra on
holomorphic sections of linear bundles over moduli spaces
(quantization).

\subsection{Infinitesimal deformations of self-dual conformal structures}
\label{9.6}

Here we briefly answer the question of \S\,9.4 about an analogue of the
Virasoro algebra (without a central term).

\smallskip

In \S\S 5.4, 5.5, 8.2 we described the local group $\fh$ of
biholomorphisms of the twistor space $\rp$ and its action on the
space of local solutions to the SDYM equations. To this group there
corresponds the algebra (cf.(9.9), (9.10))
$$
C^0(\fu , \cv_\rp )
\eqno(9.11)
$$
of 0-cochains of the cover $\fu=\{\U_1,\U_2\}$ of $\rp$ with values
in the sheaf $\cv_\rp$ (of germs) of holomorphic vector fields on
$\rp =\U_1\cup\U_2$.  However, this algebra is not a correct
generalization of the Virasoro algebra.

\smallskip

An analogue of the Virasoro algebra is the algebra $\cv_\rp (\U_{12})$
of holomorphic vector fields on $\U_{12}=\U_1\cap\U_2\subset\rp$.
It is a subalgebra of the algebra
$$
C^1(\fu , \cv_\rp )\simeq \cv_\rp (\U_{12})\oplus \cv_\rp (\U_{12})
\eqno(9.12)
$$
of 1-cochains of the cover $\fu$ with values in the sheaf $\cv_\rp$.
Elements of the algebra $C^1(\fu , \cv_\rp )$ are the collections of
vector fields
$$
\chi =\{\chi_{12}, \chi_{21}\}=\{\chi_{12}^a\frac{\p}{\p z^a_1},
\chi_{21}^a\frac{\p}{\p z^a_2}\}
\eqno(9.13)
$$
with ordered ``cohomology indices".

\smallskip

{}From the Kodaira-Spencer deformation theory~\cite{Kod} it follows that
the algebra (9.12) acts on the transition function $f_{12}$ of the space
$\rp$ (see \S\,3.3) by the formula
$$
\d f^a_{12}=\chi_{12}^a - \frac{\p f_{12}^a}{\p z^b_2}\chi^b_{21}\quad
\Leftrightarrow\quad
\d f_{12}:= \d f^a_{12}  \frac{\p}{\p z^a_1} = \chi_{12}-\chi_{21}.
\eqno(9.14)
$$
Accordingly, one may define the following action of the algebra
$C^1(\fu , \cv_\rp )$ on the transition matrices $\cf_{12}$ of holomorphic
bundles $E'$ over the twistor space $\rp$:
$$
\d_\chi \cf_{12}=\chi_{12}(\cf_{12})-\chi_{21}(\cf_{12}).
\eqno(9.15)
$$
The algebra  $C^0(\fu , \cv_\rp )$ acts on the transition function
$f_{12}$ of the space $\rp$ and on the transition matrices $\cf_{12}$
of bundles $E'$  over $\rp$ by formulae (9.14),(9.15) via the twisted
homomorphism
$$
\dot\d^0: C^0(\fu , \cv_\rp )\ni\{\chi_1,\chi_2\}\mapsto\{\chi_1-\chi_2,
\chi_2-\chi_1\}\in C^1(\fu , \cv_\rp )
\eqno(9.16)
$$
of the algebra $C^0(\fu , \cv_\rp )$ into the algebra $C^1(\fu ,
\cv_\rp)$.

\smallskip

Notice that $\d f:=\{\d f_{12}, \d f_{21}\}\in
Z^1(\fu , \cv_\rp)$, and the quotient space
$$
H^1(\fu , \cv_\rp):= Z^1(\fu , \cv_\rp)/ \dot\d^0(C^0(\fu , \cv_\rp))
\eqno(9.17)
$$
describes nontrivial infinitesimal deformations of the complex structure
of $\rp$. For a cover $\fu =\{\U_1,\U_2\}$, where $\U_1,\U_2$ are Stein
manifolds, we have $H^1(\rp , \cv_\rp)=H^1(\fu , \cv_\rp)$. In contrast
with the 2D case (9.7) now we have $H^1(\rp , \cv_\rp)\ne 0$. Hence, the
transformations (9.14) of the transition function in general change the
complex structure of $\rp$ and therefore  change the conformal structure
on $U$. Recall that a conformal structure $[g]$ is called self-dual if
the Weyl tensor for any metric $g$ in the conformal equivalence class
$[g]$ is self-dual~\cite{AHS}. In virtue of the twistor
correspondence~\cite{Pen, AHS} the moduli space of self-dual conformal
structures on a 4-manifold $M$ is bijective to the moduli space of complex
structures on the twistor space of
$M$.

\smallskip

All algebras of infinitesimal symmetries of the self-dual gravity
equations known by now (see e.g.~\cite{PBR} and references therein)
are subalgebras in the algebra $C^1(\fu ,\cv_\rp)$. The action of the
algebra  $C^0(\fu ,\cv_\rp)$ (and the group $\fh (\rp)$ corresponding
to it) transforms $f_{12}$ into an equivalent transition function
and therefore  preserves the conformal structure on $U$. At the same
time, the action of the algebra $C^0(\fu ,\cv_\rp)$ on transition
matrices of holomorphic bundles $E'\to\rp$ is not trivial.

\smallskip

If we want to define an action of the algebra $C^1(\fu ,\cv_\rp)$ on
the coordinates  $\{z^a_1\}$, $\{z^a_2\}$, q-forms etc, we should
define:
1) a sheaf $\ct^{1,0}$ of (1,0) vector fields on $\rp$,
holomorphic along fibres $\cp$ of the bundle $\rp\to U$;
2) a sheaf $\cw$ of (0,1)-forms $W$ on $\rp$ with values in $\ct^{1,0}$,
vanishing on the distribution $V^{0,1}$ (see \S\,7.3) and satisfying
the equations
$$
\bar\p W =0
\eqno(9.18)
$$
on any open set $\U\subset\rp$, where $W\in \cw (\U)$. Then  we have
the exact sequence of sheaves
$$
0\lra\cv_\rp\lra\ct^{1,0}\lra\cw\lra 0
\eqno(9.19)
$$
and the corresponding exact sequence of cohomology spaces
$$
0\lra H^0(\rp,\cv_\rp)\lra H^0(\rp,\ct^{1,0})\lra H^0(\rp,\cw)
\lra H^1(\rp,\cv_\rp)\lra 0,
\eqno(9.20)
$$
describing infinitesimal deformations of the complex structure of the
twistor space $\rp$.

\smallskip

{}From (9.20) it follows that for any element $\d f\in Z^1(\fu ,\cv_\rp )
\subset Z^1(\fu ,\ct^{1,0})$ there exists an element $\{\vp_1,\vp_2\}
\in  C^0 (\fu ,\ct^{1,0})$ such that
$$
\d f=\{\chi_{12}-\chi_{21}, \chi_{21}-\chi_{12}\}=\{\vp_1-\vp_2,
\vp_2-\vp_1\}\in \dot\d^0(C^0 (\fu ,\ct^{1,0})).
\eqno(9.21)
$$
Then  for infinitesimal transformations of coordinates on
$\rp =\U_1\cup\U_2$ we have
$$
\d z^a_1:=\vp^a_1(z_1,\bar z_1),\quad
\d z^a_2:=\vp^a_2(z_2,\bar z_2).
\eqno(9.22)
$$
To preserve the reality of the conformal structure on $U$, one should
define real subalgebras of the algebras $C^1(\fu ,\cv_{\rp})$  and
$C^0 (\fu ,\ct^{1,0})$ by analogy with \S\S\,6.6,\,7.7. We shall not
write down transformations of the metric and conformal structure on $U$,
since this will
require a lot of additional explanations. Details will be published
elsewhere.

\subsection{Quantization}
\label{9.7}

Some problems related to the quantization of the SDYM model were
discussed in~\cite{NS, CY, LMNS}. The quantization was carried out in
four dimensions in terms of $\fg$-valued  fields $A_\mu$ or in terms
of a $G$-valued scalar field by using the Yang gauge. But the
obtained results are fragmentary; the picture is not complete and far
from what we have in 2D CFT's. Remembering the connection between 2D
CFT's and the ordinary 3D CS theory, one may come to the reasonable
conclusion that the quantization of integrable 4D CFT's may be much
more successful if we use the 6D holomorphic CSW theory.

\smallskip

When quantizing the holomorphic CSW theory on the twistor space $\rp$
one may use the results on the quantization of the ordinary CS theory
(see e.g.~\cite{Wit3, EMSS} and references therein) after a proper
generalization. We are mainly interested in quantizing the SDYM
model. As such, we have to put $\hat B_3=0$ in eqs.(9.1), which leads
to the equations (cf.(7.12)) $$ \bar\p B + B\wedge B=0 \eqno(9.23) $$
equivalent to the SDYM equations, as has been discussed in this
paper. The comparison with the ordinary CS theory in the Hamiltonian
approach shows that $\bar\l$ may be considered as (complex) time of
the holomorphic CSW theory.

\smallskip

Further, one can use two standard approaches to the quantization of
constrained systems:
1) one first solves the constraints and then performs the quantization
of the moduli space;
2) one first quantizes the free theory and then imposes (quantum)
constraints. The first approach will mainly be discussed. We shall
write down the list of questions and open problems whose solutions are
necessary to give the holomorphic CSW and the SDYM theories a status
of quantum field theories.

\smallskip

1. One should rewrite a symplectic structure $\tilde\o$ on the space
of gauge potentials or their relatives~\cite{NS, LMNS, MW, CY} in
terms of fields on the twistor space $\rp$. This 2-form $\tilde\o$
induces a symplectic structure $\o$ on the moduli space $\cm$ of
solutions to eqs.(9.23), and the cohomology class $[\o ]\in H^2(\cm,
\R)$ has to be integral.

\smallskip

2. Over the moduli space $\cm$ one should define a complex line bundle
$\cl$ with the Chern class $c_1(\cl)=[\o]$. Then $\cl$ admits a connection
with the curvature  2-form equal to $\o$.

\smallskip

3. A choice of a complex structure $\cj$ on the twistor space $\rp$
endows the moduli space $\cm$ with a complex structure which we shall
denote by the same letter $\cj$. Then the bundle $\cl$ over
$(\cm,\cj)$ has a holomorphic structure, and a quantum Hilbert space
of the SDYM theory can be introduced as the space $H_\cj$ of (global)
holomorphic sections of $\cl$.

\smallskip

4. Is it possible to introduce the bundle $\cl\to\cm$ as the
holomorphic determinant line bundle $Det\bar\p_B$ of the operator
$\bar\p_B=\bar\p + B$ on $\rp$?

\smallskip

5. The action functional of the holomorphic CSW theory on a Calabi-Yau
3-fold has a simple form~\cite{Wit2, BCOV} analogous to the action of the
standard CS theory. How should one modify this action if we go over
to the case of an arbitrary complex 3-manifold?

\smallskip

6. One should lift the action of the symmetry groups and algebras
described in this paper up to an action on the space $H_\cj$ of
holomorphic sections of the bundle $\cl$ over $\cm$. What is an
extension (central or not) of these groups and algebras? Finding of
an extension of the algebra $C^1(\fu,\co^{\bf g}_\rp)$ is equivalent
to finding a curvature of the bundle $\cl$ since this curvature
represents a local anomaly.

\smallskip

7. What can be said about representations of the algebras
$C^1(\fu,\cv_\rp)$ and $C^1(\fu,\co^{\bf g}_\rp)$? Which of these
representations are connected with the Hilbert space $H_\cj$?

\smallskip

8. In the quantum holomorphic CSW and SDYM theories there exist
Sugawara-type formulae, i.e., generators of the algebra
$C^1(\fu,\cv_\rp)$ can be quadratically expressed in terms of
generators of the algebra $C^1(\fu,\co^{\bf g}_\rp)$. This follows
from the fact that any transformation of transition matrices of a
holomorphic bundle $E'\to\rp$ under the action of the algebra
$C^1(\fu,\cv_\rp)$ can be compensated by an action of the algebra
$C^1(\fu,\co^{\bf g}_\rp)$. What are the explicit formulae
connecting the generators of these algebras?

\smallskip

9. One should write down Ward identities resulting from the symmetry
algebra  $C^1(\fu,\cv_\rp)\dotplus C^1(\fu,\co^{\bf g}_\rp)$. To what
extent do these identities define correlation functions?

\smallskip

\noindent Clearly, to carry out this quantization program,
it will be necessary to overcome a number of technical difficulties.

\smallskip

The general picture arising as a result of quantization of the SDYM
model on a self-dual 4-manifold $M$ and the holomorphic CSW theory on
the twistor space $\cz$ of $M$ resembles the one that arises in
the quantization of the ordinary CS theory and is as follows: Let $[g]$ be a
self-dual conformal structure on a 4-manifold $M$ and let $\cj$ be a
complex structure on the twistor space $\cz$ of $M$. As has already
been noted, there exists a bijection ~\cite{Pen, AHS} between the moduli
space of self-dual conformal structures on $M$ and the moduli space
$\fx$ of complex structures on $\cz$. Let $\cm$ be a moduli space of
solutions to the SDYM equations on $M$ and let $H_\cj$ be the quantum
Hilbert space of holomorphic sections of the line bundle $\cl$ over
$(\cm,\cj)$. The space $H_\cj$ depends on $\cj\in \fx$ and one can
introduce a holomorphic vector bundle
$$
p: \tilde H\lra \fx
\eqno(9.24)
$$
with fibres $H_\cj$ at the points $\cj\in\fx$. Then one may put a question
about the existence of a (projectively) flat connection in the bundle (9.24).
If such a connection exists, then as a quantum Hilbert space one may take a
space of covariantly constant sections of the vector bundle $\tilde H$.

\section{Conclusion}

In this paper, the group-theoretic analysis of the Penrose-Ward
correspondence was undertaken. Having used sheaves of non-Abelian
 groups and cohomology sets we have described the symmetry group
acting on the
space of local solutions to the SDYM equations and the moduli space
$\cm$ of local solutions. It has been shown that $\cm$ is a double
coset space. The full  algebra of infinitesimal deformations of
self-dual conformal structures on a 4-space $M$ has also been described.
We have discussed the program of quantization of the SDYM model on $M$
based on the equivalence of this model to a subsector of the holomorphic
CSW model on the twistor space $\cz$ of $M$.  There are a lot of open
problems, which deserve further study.

\section*{Acknowledgements}
\addtocontents{toc}{\medskip}
\addcontentsline{toc}{appe}{\noindent\bf Acknowledgements\hfill
\medskip
}

The author is grateful to Yu.I.Manin and I.T.Todorov for helpful
discussions. I also thank for its hospitality the Max-Planck-Institut
f\"ur Mathematik in Bonn, where part of this work was done, and the
Alexander von Humboldt Foundation for support.  This work is supported
in part by the grant RFBR-98-01-00173.

\section*{Appendix A. Actions of groups on sets}
\addcontentsline{toc}{appe}{\\
\medskip
\bf Appendix A. Actions of groups  on sets\hfill}

The left action of a group $\cg$ on a set $\up$  is a map
$\r :\cg\times\up\to\up$ with the following properties:
$$
\r (e,x)=x,
\eqno(A.1a)
$$
$$
\r (a,\r (b,x))=\r (ab, x),
\eqno(A.1b)
$$
for any $x\in\up , a,b,e\in\cg$. Here $e$ is the identity in the
group $\cg$. If we are given an action $\r$ on a set $\up$, to any
$a\in \cg$ we can correspond a bijective transformation
$\r_a:x\mapsto\r (a,x)$ of the set $\up$ such that a map $\g :
a\mapsto\r_a$ is a homomorphism of the group $\cg$ into the group
$S_\up$ of all permutations (bijective transformations) of the set
$\up$. Conversely, any homomorphism $\g :\cg\to S_\up$ defines the
action of the group $\cg$ on $\up$ by the formula $$ \r (a,x):=\g
(a)(x) \eqno(A.2) $$ for any $a\in\cg , x\in\up$. If $\up$ is a
smooth manifold, then to define an action of $\cg$ on $\up$ is
equivalent to assigning a homomorphism $\g :\cg\to\di(\up)$ of the
group $\cg$ into the group of diffeomorphisms of the manifold $\up$.

\smallskip

Usually the left action of the group $\cg$ is represented as a
multiplication of elements from $\up$ by elements of the group $\cg$
and written as $\r (a,x)=ax, a\in\cg, x\in\up$. One also considers
the right action of the group $\cg$ on $\up$ in the definition of which
the condition (A.1b) is replaced by the condition
$$
\r (a,\r (b,x))=\r (ba,x).
\eqno(A.1c)
$$
Then the notation $\r (a,x)=xa$ is used.

\smallskip

Recall that a space $\cg$ is called a {\it local   group}, if for
elements $a,b$ sufficiently close to the identity $e$ (marked
element) the multiplication $ab$ is defined, the inverse elements
$a^{-1}, b^{-1}$ exist and all group axioms are fulfilled every time
the objects participating in these axioms are defined. More
precisely, a space $\cg$ is called a local group if:  1) some
element $e$ (identity)  of $\cg$ is chosen; 2) a neighbourhood
{\vv}$\subset\cg$ of the element $e$ is chosen; 3) there is a map
{\vv}$\times${\vv}$\to\cg$, $(a,b)\mapsto ab$ (multiplication)
satisfying the conditions $ea=ae=a$ and $(ab)c=a(bc)$ for
$a,b,c,ab,bc\in${\vv}.\  From these conditions it follows that
there  exists a neighbourhood {\ww}$\subset \cg$ of the identity and
a map  $\imath :$ {\ww}$\to${\ww}, $a\mapsto a^{-1}$
(inversion) such that $aa^{-1}=a^{-1}a=e$. Choosing {\vv}$\,=\,${\ww}
$=\cg$, one can consider any group $\cg$ as a local group; this is
why we use the same letter $\cg$ for groups and for local groups.

\smallskip

If one replaces $\cg$ and {\vv}\  by open subsets $\cg'\subset\cg$,
\vv$'\subset\ $\vv$\cap\cg'$ satisfying the condition
\vv$'$\vv$'\subset\cg'$, one obtains a local group $\cg'$,
called a restriction or a part of the initial one.
Two local groups are called {\it equivalent}, if some of their
parts coincide. The equivalence class of the local group $\cg$
is called the {\it germ} of the group $\cg$ at the point $e\in \cg$
and denoted by \bg.

\smallskip

An action of a group $\cg$ on a set $\up$ can be {\it localized} if
one considers $\cg$ as a local   group.  Namely, let $\r$ be an
action of the   group $\cg$ on the set $\up$ and let $\cn$ be an open
subset in $\up$. The action $\r$, generally speaking, does not map
$\cn$ into itself and therefore  does not define an action of the whole
group $\cg$ on $\cn$.  However, an action of $\cg$ as a local group
is defined, i.e., a map $\r :\ $\ww$\to\cn$ is defined, where
\ww$=\{(a,x)\in\cg\times\cn :  \r (a,x)\in\cn\}$ is an open subset in
$\cg\times\cn$ containing $\{e\}\times\cn$. Moreover, for any fixed
point $x\in\cn$ there exists a neighbourhood {\vv}\ of the identity
in $\cg$ and a neighbourhood $\cn'$ of the point $x$ in $\cn$ such
that $\r ($\vv$\times\cn')\subset\cn$.

\smallskip

In a more general situation, a {\it local action of a local group}
$\cg$ on a set $\cn$ is a map $\r :\ $\ww$\to\cn$, where {\ww}\ is an open
set in $\cg\times\cn$ containing  $\{e\}\times\cn$, and the
properties (A.1) are satisfied for all $a,b\in \cg$, $x\in\cn$
for which both parts of the equality (A.1b) are defined. A local action $\r$
of the local group $\cg$ on the set $\cn$ generates a local action of
$\cg$ on any open subset $\cn'\subset\cn$. This action is called a
restriction of the action $\r$ to the subset $\cn'$. A local action
of the group $\cg$ is called {\it globalizable} if it is a localization
of some global action of the group.

\section*{Appendix B. Sheaves of (non-Abelian) groups}
\addcontentsline{toc}{appe}{\\
\medskip
\bf Appendix B. Sheaves of (non-Abelian) groups\hfill}

Let us consider a topological space $X$ and recall the definitions of a
presheaf and a sheaf of groups over $X$ (see e.g.~\cite{GR, Hir}).

\smallskip

One has  a {\it presheaf} $\{\fs (U), r^U_V\}$ of groups over a
topological space $X$ if with any nonempty
open set $U$ of the space $X$ one associates a group $\fs(U)$
and  with any two open sets $U$ and $V$ with $V\subset U$
one associates a homomorphism $r^U_V: \fs(U)\to\fs(V)$ satisfying
the following conditions:
(i) the homomorphism  $r^U_U: \fs(U)\to\fs(U)$ is the identity
map id$_U$;
(ii) if $W\subset V\subset U$, then $r^U_W=r^V_W\circ r^U_V$.

\smallskip

A {\it sheaf} of groups over a topological space $X$ is a topological
space $\fs$ with a {\it local homeomorphism} $\pi : \fs\to X$. This
means that any point $s\in\fs$ has an open neighbourhood $V$ in $\fs$
such that $\pi (V)$ is open in $X$ and $\pi : V\to\pi (V)$ is a
homeomorphism.  A set $\fs_x=\pi^{-1}(x)$ is called a {\it stalk} of
the sheaf $\fs$ over $x\in X$, and the map $\pi$ is called the
projection.  For any point $x\in X$ the stalk $\fs_x$ is a group, and
the group operations are continuous.

\smallskip

A {\it section}  of a sheaf $\fs$ over an open set $U$ of the space $X$
is a continuous map $s: U\to\fs$ such that $\pi\circ s=$id$_U$. A set
$\fs (U):=\Gamma (U,\fs)$ of all sections of the sheaf $\fs$ of groups
over $U$ is a group. Corresponding to any open set $U$ of the space
$X$ the group $\fs(U)$ of sections of the sheaf $\fs$ over $U$ and to
any two open sets $U,V$ with $V\subset U$ the restriction homomorphism
$r^U_V: \fs(U)\to\fs(V)$, we obtain the presheaf $\{\fs(U),r^U_V\}$ over
$X$. This presheaf is called the canonical presheaf.

\smallskip

On the other hand, one can associate a sheaf with any presheaf
$\{\fs(U),r^U_V\}.$
Let
$$
\fs_x={\mathop{\mbox{lim}}\limits_{\stackrel{\lra}{x\in U}}}\fs(U)
$$
be a direct limit of sets $\fs(U)$. There exists a natural map
$r^U_x : \fs(U)\to\fs_x$, $x\in U$, sending elements from
$\fs(U)$ into their equivalence classes in the direct limit.
If $s\in\fs(U)$, then
${\bf s}_x:=r^U_x(s)$ is called a {\it germ} of the section $s$ at the
point $x$, and $s$ is called a {\it representative} of the germ ${\bf s}_x$.
In other terms, two sections $s,s'\in \fs(U)$ are called {\it equivalent}
at the point $x\in U$ if there exists an open neighbourhood $V\subset U$
such that $s|_V=s'|_V$; the equivalence class of such sections is called
the germ ${\bf s}_x$ of section $s$ at the point $x$. Put
$$
\fs={\mathop\cup\limits_{x\in X}}\fs_x
$$
and let $\pi : \fs\to X$ be a projection mapping points from $\fs_x$
into $x$. The set $\fs$ is
equipped with a topology, the basis of open sets of which consists of sets
$\{{\bf s}_x, x\in U\}$ for all possible $s\in\fs(U), U\subset X$.
In this topology $\pi$ is a local homeomorphism, and we
obtain the sheaf $\fs$.

\smallskip

Let $X$ be a smooth manifold. Consider a complex (non-Abelian) Lie group
${\bf G}=G^\C$ and define a presheaf $\{\hat\cs(U),r^U_V\}$ of groups by
putting
$$
 \hat\cs(U):=\{C^\infty\mbox{-maps}\ f: U\to {\bf G}\},
\eqno(B.1)
$$
and using the canonical restriction homomorphisms $r^U_V$ when for
$f\in\hat\cs(U)$ its image $r^U_V(f)$ equals $f|_V\in\hat\cs(V)$,
$V\subset U$. To each elements $\a_x$ and $\b_x$ from
$\hat\cs_x:=r^U_x(\hat\cs(U))$ one can correspond their pointwise
multiplication $\a_x\b_x$.  To this presheaf $\{\hat\cs(U), r^U_V\}$
there corresponds the sheaf $\hat\cs$ of germs of smooth maps of the
space $X$ into the group $\bf G$.

\smallskip

Suppose now that $X$ is a complex manifold. Then one can define a
presheaf $\{\ch(U), r^U_V\}$ of groups assuming that
$$
\ch(U)\equiv\co^{\bf G}(U):=\{\mbox{holomorphic maps}\ h: U\to {\bf G}\},
\eqno(B.2)
$$
and associate with it the sheaf $\ch\equiv\co^{\bf G}$ of germs
of holomorphic maps of the space $X$ into the complex Lie group $\bf G$.

\section*{Appendix C. Cohomology sets and vector bundles}
\addcontentsline{toc}{appe}{\\
\medskip
\bf Appendix C. Cohomology sets and vector bundles\hfill}

We shall consider a complex manifold $X$ and a sheaf $\fs$ coinciding
with either the sheaf $\hat\cs$ or the sheaf $\ch$ introduced in
Appendix B. So $\fs$ is the sheaf of germs of smooth or holomorphic maps
of the space $X$ into the complex Lie group $\bf G$.

\smallskip

\v{C}ech cohomology sets $H^0(X,\fs)$ and $H^1(X,\fs)$ of the space $X$
with values in the sheaf $\fs$ of groups are defined as follows~\cite{GR,
Hir, Ma}.

\smallskip

Let there be given an open cover $\fu=\{\U_\a\}$, $\a\in I$, of the
manifold $X$.
The family $\la\U_0,...,\U_q\ra$ of elements of the cover such that
$\U_0\cap...\cap\U_q\ne \varnothing$ is called a q-simplex. The support
of this simplex
is $\U_0\cap...\cap\U_q$. Define a {\it 0-cochain with coefficients
in} $\fs$ as a map $f$ associating with $\a\in I$ a section $f_\a$
of the sheaf $\fs$ over $\U_\a$:
$$
f_\a\in\fs(\U_\a):=\Gamma (\U_\a, \fs).
\eqno(C.1)
$$
A set of 0-cochains is denoted by $C^0(\fu,\fs)$ and is a group
under the pointwise multiplication.

\smallskip

Consider now the ordered set of two indices $\la\a,\b\ra$
such that $\a,\b\in I$
and $\U_\a\cap\U_\b\ne\varnothing$. Define a {\it 1-cochain} with
coefficients in $\fs$ as a map $f$ associating with $\la\a,\b\ra$
a section of the sheaf $\fs$ over $\U_\a\cap\U_\b$:
$$
f_{\a\b}\in\fs(\U_\a\cap\U_\b):=\Gamma (\U_\a\cap\U_\b,\fs).
\eqno(C.2)
$$
A set of 1-cochains is denoted by $C^1(\fu,\fs)$ and is a group under the
pointwise multiplication.

\smallskip

Subsets of {\it cocycles} $Z^q(\fu,\fs)\subset C^q(\fu,\fs)$ for $q=0,1$
are defined by the formulae
$$
Z^0(\fu,\fs)=\{f\in C^0(\fu,\fs): f_\a f_\b^{-1}=1\ \mbox{on}\
\U_\a\cap\U_\b\ne\varnothing\},
\eqno(C.3)
$$
$$
Z^1(\fu,\fs)=\{f\in C^1(\fu,\fs): f_{\b\a}=f_{\a\b}^{-1} \ \mbox{on}\
\U_\a\cap\U_\b\ne\varnothing ,\ f_{\a\b}f_{\b\g}f_{\g\a}=1\ \mbox{on}\
\U_\a\cap\U_\b\cap\U_\g\ne\varnothing\}.
\eqno(C.4)
$$
It follows from (C.3) that $Z^0(\fu,\fs)$  coincides with the group
$H^0(X,\fs):=\fs(X)\equiv\Gamma (X,\fs )$ of global sections of the
sheaf $\fs$. The set $Z^1(\fu,\fs)$ is not in general a subgroup
of the group $C^1(\fu,\fs)$. It contains the marked element $\bf 1$,
represented by the 1-cocycle $f_{\a\b}= 1$ for any $\a,\b$ such that
$\U_\a\cap\U_\b\ne\varnothing$.

\smallskip

{}For $h\in C^0(\fu,\fs)$, $f\in Z^1(\fu,\fs)$ let us define an action
$\r_0$ of the group $C^0(\fu,\fs)$ on the set $Z^1(\fu,\fs)$ by
the formula
$$
\r_0(h,f)_{\a\b}=h_\a f_{\a\b}h_{\b}^{-1}.
\eqno(C.5)
$$
So we have a map $\r_0: C^0\times Z^1\ni (h,f)\mapsto\r_0(h,f)\in Z^1$.
A set of orbits of the group $C^0$ in $Z^1$ is called a {\it 1-cohomology
set} and denoted by $H^1(\fu,\fs)$. In other words, two cocycles
$f,\tilde f\in Z^1$ are called equivalent, $f\sim\tilde f$, if
$$
\tilde f= \r_0(h,f)
\eqno(C.6)
$$
{}for some $h\in C^0$, and by the 1-cohomology set $H^1=\r_0(C^0)\bl Z^1$
one calls a set  of equivalence classes of 1-cocycles. Finally, we
should take the direct limit of these sets $H^1(\fu,\fs)$ over
successive refinement of the cover $\fu$ of $X$ to obtain
$H^1(X,\fs)$, the 1-cohomology set of $X$ with coefficients in $\fs$.
In fact, one can always choose a cover $\fu=\{\U_\a\}$ such that it
will be $H^1(\fu,\fs)=H^1(X,\fs)$ and therefore  it will not be
necessary to take the direct limit of sets.  This is realized, for
instance, when the coordinate charts $\U_\a$ are Stein manifolds (see
e.g.~\cite{GR}).

\smallskip

Recall that $\fs$ is the sheaf of germs of  (smooth or holomorphic)
{}functions with values in the complex Lie group $\bf G$.
Suppose we are given a representation of $\bf G$ in $\C^n$. It is
well-known that any
1-cocycle $\{f_{\a\b}\}$ from $Z^1(\fu,\fs)$ defines a unique complex
vector bundle $E'$ over $X$, obtained from the direct products
$\U_{\a}\times\C^n$  by glueing with the help of $f_{\a\b}\in {\bf
G}$. Moreover, two 1-cocycles define isomorphic complex vector
bundles over $X$ if and only if the same element from $H^1(X,\fs)$
corresponds to them. Thus, we have a one-to-one correspondence
between the set $H^1(X,\fs)$ and the set of equivalence classes of
complex vector bundles of the rank $n$ over $X$. Smooth bundles are
parametrized by the set $H^1(X,\hat\cs)$ and holomorphic bundles are
parametrized by the set $H^1(X,\ch)$, where the sheaves $\hat\cs$ and
$\ch$ were described in Appendix B. For more details see
e.g.~\cite{GR, Hir}.

\end{document}